\documentclass[aps,preprint]{revtex4}%
\usepackage{amsfonts}
\usepackage{amsmath}
\usepackage{amssymb}
\usepackage{graphicx}%
\providecommand{\U}[1]{\protect \rule{.1in}{.1in}}
\newtheorem{theorem}{Theorem}
\newtheorem{acknowledgement}[theorem]{Acknowledgement}

\begin{document}
\preprint{ \emph{LPHE-MS Preprint: April 2012}\bigskip \bigskip \bigskip \bigskip
\bigskip \bigskip \bigskip \bigskip}
\title[Fermions on hyperdiamond]{\bigskip \textbf{Topological\ Aspects of Fermions on Hyperdiamond}\bigskip}
\author{E.H Saidi$^{1,2}$, O. Fassi-Fehri$^{1}$, M. Bousmina$^{1}$\bigskip \bigskip}
\affiliation{1. Hassan II Academy of Science and Technology, Avenue Mohammed VI, KM 4,
Souissi, Rabat, Morocco.\bigskip \bigskip}
\affiliation{2. Lab of High Energy Physics, Modeling and Simulations, Faculty of
Science,University Mohammed V-Agdal, Rabat, Morocco,}
\affiliation{Centre Of Physics and Mathematics, CPM-CNESTEN, Rabat, Morocco\bigskip \bigskip}
\affiliation{}
\pacs{PACS number\bigskip}

\begin{abstract}
Motivated by recent results on the index of the Dirac operator $D=\gamma^{\mu
}D_{\mu}$ of QCD on lattice and also by results on topological features of
electrons and holes of 2-dimensional graphene, we compute in this paper the
Index of D for fermions living on a family of even dimensional lattices
denoted as $\mathbb{L}_{2N}$ and describing the 2N-dimensional generalization
of the graphene honeycomb. The calculation of this topological Index is done
by using the direct method based on solving explicitly the gauged Dirac
equation and also by using specific properties of the lattices $\mathbb{L}%
_{2N}$ which are shown to be intimately linked with the weight lattices of
$SU\left(  2N+1\right)  $. The Index associated with the two leading $N=1$ and
$N=2$ elements of this family describe precisely the chiral anomalies of
graphene and QCD$_{4}$.\ Comments on the method using the spectral flow
approach as well as the computation of the topological charges on 2-cycles of
2N-dimensional compact supercell in $\mathbb{L}_{2N}$ and applications to
\emph{QCD}$_{4}$ are also given.\newline \textbf{Key words: }%
{\small Topological index, Lattice QCD}$_{4}${\small , Graphene, Chiral
anomaly, Root and Weight lattices of }$SU\left(  N\right)  ${\small , Adam's
spectral flow}.\emph{ \  \  \ }

\end{abstract}
\maketitle




\section{Introduction}

It is quite well known that the spectrum of the gauged Dirac operator
$D=\gamma^{\mu}\left(  \partial_{\mu}-iA_{\mu}\right)  $ $\equiv \gamma^{\mu
}D_{\mu}$ of \emph{2}- and \emph{4}- dimensional Dirac theories describing the
dynamics of fermionic waves $\Psi \left(  x\right)  $ in a uniform background
field with topological charge $Q_{tot}$ obeys the famous \emph{Atiyah-Singer
index theorem }\textrm{\cite{1A,1B}; }one of the most substantial achievements
of modern mathematics. The index of the Dirac operator $D$, to which we refer
below to as $Ind\left(  \emph{D}\right)  $, relates the topological charge
$Q_{tot}$ to the net numbers $N_{+}$ and $N_{-}$ of chiral and antichiral zero
modes of $D$ as
\begin{equation}
Q_{tot}=N_{+}-N_{-} \label{01}%
\end{equation}
showing in turns that the background field breaks implicitly the left-right
parity symmetry. The $Ind\left(  \emph{D}\right)  $ is a powerful topological
quantity that has been used for diverse purposes \textrm{\cite{2A}-\cite{2G};
}see also\textrm{\  \cite{AL} }and refs therein for other applications; it
gives a rigorous explanation of the origin of chiral anomalies and constitutes
an alternative approach to the perturbative method based on computing
radiative corrections to describe $\bar{\Psi}\gamma^{\mu}\Psi A_{\mu}$
interactions in \emph{QFT}$_{1+2}$ where the topological Chern-Simons gauge
theory emerges as a 1-loop correction of the gauge field propagator
\textrm{\cite{3A,3B}}. In 2-dimensions, the computation of $Ind\left(
\emph{D}\right)  $ shows that the anomalous quantum Hall effect (QHE) of
graphene \textrm{\cite{4A,4B,4C}} is precisely due to the chiral anomaly of
zero modes; a basic result that is expected to be valid as well for higher
dimensional Dirac fermions in background fields including light quarks on
\emph{4D} hyperdiamond \textrm{\cite{5A}-\cite{5G}} and fermions on higher 2N-
dimensional honeycombs. Recall by the way that in graphene the quantum Hall
effect is a very special effect in the sense it can be observed \emph{at room
temperature}; the gap between the $n=0$ and $n=1$ Landau levels is around
\emph{1300K} at \emph{10} Tesla compared to around \emph{100K} in an ordinary
2D electronic gaz \textrm{\cite{RT,TR}; see also \cite{17,18,19,20} }for other
related aspects.

\  \  \  \  \  \newline Motivated by recent developments on the index theorem on
lattice \emph{QCD}$_{4}$ \textrm{\cite{6A,6B,6C,6D}}, we compute in this paper
the $Ind\left(  \emph{D}_{4}\right)  $, and in general the $Ind\left(
\emph{D}_{2N}\right)  $, of fermions in background fields living on a class of
even- dimensional lattices $\mathbb{L}_{2N}$ describing the 2N- dimensional
generalization of the honeycomb $\mathbb{L}_{2}$. The calculation of
$Ind\left(  \emph{D}_{2N}\right)  $ will be done as follows:

\begin{itemize}
\item use known results on the index to bring the lattice analysis to the
spectrum of the fermionic operator near the Fermi level; i.e the spectrum near
the Dirac points of lattice QCD$_{2N}$ where live \emph{fermionic} \emph{zero
modes }contributing to $Ind\left(  \emph{D}_{2N}\right)  $.

\item then determine these fermionic zero modes by using the direct method
based on the two following: \newline$\left(  a\right)  $ using specific
features of the $SU\left(  2N+1\right)  $ roots and weights to deal with the
symmetry properties of the lattices $\mathbb{L}_{2N}$ with $N=1,2,...$;
\newline$\left(  b\right)  $ solving explicitly the gauged Dirac equation
$D\Psi=\epsilon \Psi$, with $\epsilon$ standing for massive deformations around
the Fermi level.
\end{itemize}

\  \  \newline Recall that there are two basic ways to compute $Ind\left(
\emph{D}_{4}\right)  $; $\left(  i\right)  $ the direct method which we will
be considering here; and $\left(  ii\right)  $ the so called\textrm{\ }%
spectral flow approach of Adams \textrm{\cite{6A,6B,6C,6D}} based on the
introduction of a hermitian version $H_{sp}$ of the Dirac operator. In
\emph{QCD}$_{4}$ on hyperdiamond, this spectral hamiltonian has the form
$H_{sp}=\gamma_{5}\left(  D-m\right)  $ showing that any zero mode of $D$ with
$\pm$\ chirality corresponds to some eigen-modes of $H_{sp}$ with eigenvalues
$\lambda \left(  m\right)  =\mp m$; for details see
\textrm{\cite{6A,6B,6C,6D,6DD,D6,5G}}.

\  \  \  \  \  \  \  \newline Our interest into the study of the index of the Dirac
operator of fermions on $\mathbb{L}_{2N}$ has been also motivated by the two following:

\  \  \  \  \newline \textbf{(1)} the \emph{2} leading lattices $\mathbb{L}_{2}$,
$\mathbb{L}_{4}$ of the family $\mathbb{L}_{2N}$ are precisely given by the
honeycomb of graphene, thought of here as lattice QCD$_{2}$, and the 4D
hyperdiamond used in lattice \emph{QCD}$_{4}$. So, one expects the members of
this family to share some basic features; in particular methods to approach
the physical properties of fermions on $\mathbb{L}_{2N}$. This link opens a
window for new ways to modeling and simulating in euclidian relativistic
theory by borrowing ideas from graphene as done by M. Creutz in
\textrm{\cite{5A}};\textrm{\ }see also\textrm{\  \cite{5B,5C,5E}}.

\  \  \  \  \  \newline \textbf{(2)} the existence of a remarkable relation between
the $\mathbb{L}_{2N}$ honeycomb's class and the family of weight lattices of
the $SU\left(  2N+1\right)  $ Lie algebras. This extraordinary relation allows
a unified description of the tight binding description of fermions on
$\mathbb{L}_{2N}$ with $N$ a generic integer; and permits moreover to take
advantage of the power of the $SU\left(  2N+1\right)  $ representations to
work out explicit configurations for fermions and gauge fields on
$\mathbb{L}_{2N}$. For example boundary conditions of the fields on supercells
in $\mathbb{L}_{2N}$ as well as the determination of the Dirac points
$\left \{  P_{i}\right \}  $ regarding fermions on $\mathbb{L}_{2N}$ get mapped
to manageable equations on the $SU\left(  2N+1\right)  $ weight lattice where
the \emph{duality} between simple roots $\mathbf{\alpha}_{i}$ of $SU\left(
2N+1\right)  $ and its fundamental weight vectors $\mathbf{\omega}_{i}$ plays
a crucial role \textrm{\cite{5E,5F,5H,5I}}.

\  \  \  \  \  \newline Our explicit analysis for computing the index of the Dirac
operator allows us as well to get more insight into the structure of the
topological index on lattice; in particular into the two following things: (a)
the relation between the various possible fluxes $Q_{i}$ through the 2-cycles
$C_{i}$ of the supercell compactification and the total charge $Q_{tot}$ of
the background fields. (b) the role played by the different matrices
$\Gamma_{_{2N+1}}$ one can construct from the 2D Pauli ones namely
\begin{equation}
\Gamma_{_{2N+1}}=\tau_{_{3}}\otimes \left(  \sigma_{_{3}}\right)  ^{n_{_{1}}%
}\otimes...\otimes \left(  \sigma_{_{3}}\right)  ^{n_{_{N-1}}} \label{20}%
\end{equation}
with $n_{i}=0,1$ and $\tau_{3}=diag\left(  1,-1\right)  =\sigma_{3}$.

\  \  \  \  \  \newline Regarding the topological charges $Q_{i}$ and $Q_{tot}$,
\textrm{notice} that the lattice $\mathbb{L}_{2N}$ is recovered by considering
a 2N- dimensional compact supercell with homology classes as those of the real
2N-torus $\mathbb{T}^{2N}$. While in 2-dimensions the homology of the
supercell has one 2-cycle given by the parallelogram of fig \ref{PARA}, the
situation is richer in higher dimensions. In the case of 4D hyperdiamond for
instance, the homology of the 4D\ compact supercell has, in addition to the
real 4-cycle $C_{4}\sim \mathbb{T}^{4}$, a basis of six 2-cycle $C_{i}%
\sim \mathbb{T}_{i}^{2}$ leading to the following topological charges%
\[
Q_{tot}=\frac{1}{8\pi^{2}}%
{\displaystyle \int \nolimits_{C_{4}}}
F\wedge F,\qquad Q_{i}=\frac{1}{2\pi}%
{\displaystyle \int \nolimits_{C_{i}}}
F
\]
related as in eq(\ref{B}) and playing a central role in computing the index of
the Dirac operator.

\  \  \  \  \newline Concerning the matrices $\Gamma_{_{2N+1}}$ for fermions on
2N-honeycombs, one has in general $2^{N-1}$ possible representations depending
on the values of the $n_{i}$'s of (\ref{20}). For the case of the 4D
hyperdiamond, there are two kinds of such matrices and are as follows
\[
\gamma_{5}=\sigma_{3}\otimes I,\qquad \sigma_{3}\otimes \tau_{3}.
\]
The explicit computation of the Dirac index given in sections 5 and 6 shows
that these matrices lead to different relations between the zero modes and the
topological charges. We will show that the right index of the Dirac operator
that recovers (\ref{01}) is given by $Tr\left(  \bar{\Psi}\sigma_{3}%
\otimes \tau_{3}\Psi \right)  =Q_{top}$.

\  \  \  \  \  \newline The presentation is as follows: In section 2, we review the
usual approach to honeycomb; but now by using the root and weight lattices of
$SU\left(  3\right)  $; the latter is a hidden symmetry of the honeycomb. In
section 3, we study the topological aspects of fermions in graphene in
presence of an external magnetic field $B$ and develop the explicit
computation of the zero modes of the gauged Dirac operator. As we will show,
the basic properties of the ground state are encoded in the sign $\frac
{B}{\left \vert B\right \vert }$ of the background field. In section 4, we use
the results on graphene to approach the fermions on hyperdiamond in presence
of two background fields $B $ and $E$. In this study, we also use roots and
weight lattices of $SU\left(  5\right)  $ that appears as a hidden symmetry of
the 4D crystal. In section 5, we compute the spectrum of the Dirac operator in
continuum, and in section 6 we work out the complete spectrum of fermions on
the hyperdiamond lattice. We also compute the topological index giving the
relation between the zero modes and the various topological charges of the
background fields. Here also, we show that the ground state features are
encoded in the sign of $\frac{B}{\left \vert B\right \vert }$ and $\frac
{E}{\left \vert E\right \vert }$. In section 7, we give a conclusion and make
comments on the extension to fermions on 2N- dimensional honeycomb in presence
of N background fields $B_{i}$ whose signs $\frac{B_{i}}{\left \vert
B_{i}\right \vert }$ characterize completely the ground states.

\  \  \  \  \newline Before proceeding, we would like to notice that we will omit
certain technical details on lattice calculations from the core of the paper;
the essential of these details is reported in appendix 8; and related
extensions can be found in \textrm{\cite{5A,5B,5C,5D,5E,TB,PSM}.}

\section{Fermions on 2D honeycomb}

In this section, we develop the so called primitive compactification of
fermion on honeycomb used in \textrm{\cite{4C}} to study topological aspects
of fermions in graphene. To get more insight into our analysis, we first
describe the canonical frame formulation, generally used to deal with fermions
on square lattice; then we turn back to study the primitive frame (P-frame)
and exhibit its link with the root and the weight lattices of the $SU\left(
3\right)  $ Lie algebra. This study can be also viewed as a step to fix the
ideas before moving to the case of QCD$_{4}$ fermions on hyperdiamond to be
considered in sections 4, 5 and 6.

\subsection{2D honeycomb and $SU\left(  3\right)  $ weight lattice}

\begin{figure}[ptbh]
\begin{center}
\hspace{0cm} \includegraphics[width=16cm]{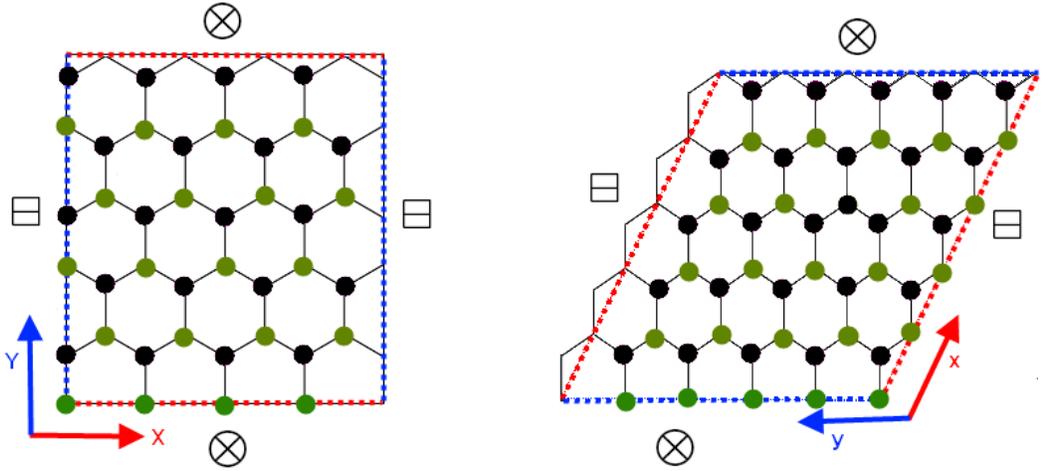}
\end{center}
\par
\vspace{-0.8cm}\caption{{\protect \small On left the perpendicular
compactification. On right the primitive compactification}}%
\label{PC}%
\end{figure}

Honeycomb is a 2D lattice where each site has \emph{3} first nearest neighbors
and \emph{6} second nearest ones as shown on fig \ref{PC}; the \emph{3} first
neighbors will be viewed here as the vertices of an equilateral triangle
transforming as a $SU\left(  3\right)  $ triplet; and the second \emph{6} ones
the vertices of an hexagon. The lattice parametrization of the honeycomb can
be obtained from the real plane $\mathbb{R}^{2}$ by restricting the local real
coordinate $\mathbf{r}=\left(  X,Y\right)  $ to $\mathbf{r}_{n,m}=\left(
X_{n},Y_{m}\right)  $ with $n,m$ integers. As a 2D vector, one may expand the
position vector $\mathbf{r}$ in diverse ways; in particular into 2 particular
frames used in \textrm{\cite{4C}} and named there as \emph{perpendicular}
frame and \emph{primitive} one; see fig \ref{PC} for illustration. Here, we
will refer to the perpendicular frame as the \emph{canonical }one.

\begin{itemize}
\item \emph{Canonical compactification}
\end{itemize}

In the canonical compactification (C-frame), one uses the usual
\emph{orthogonal} cartesian basis $\left \{  \mathbf{e}_{X},\mathbf{e}%
_{Y}\right \}  $ of $\mathbb{R}^{2}$ to decompose vectors like $\mathbf{r}%
=X\mathbf{e}_{X}+Y\mathbf{e}_{Y}$,
\begin{equation}%
\begin{tabular}
[c]{llllll}%
$\mathbf{e}_{\mu}.\mathbf{e}_{\mu}=\delta_{\mu \nu}$ & $,$ & $X=\mathbf{e}%
_{X}.\mathbf{r}$ & $,$ & $Y=\mathbf{e}_{Y}.\mathbf{r}$ & .
\end{tabular}
\  \  \  \label{1}%
\end{equation}
The restriction of local fields $F\left(  \mathbf{r}\right)  $ to the
honeycomb is obtained by using the Dirac delta function $\delta \left(
\mathbf{r-r}_{n,m}\right)  $ as
\begin{equation}
F_{\mathbf{r}_{n,m}}=\int_{\mathbb{R}^{2}}d^{2}\mathbf{r}\text{ }\delta \left(
\mathbf{r-r}_{n,m}\right)  F\left(  \mathbf{r}\right)  . \label{FN}%
\end{equation}
The dynamics of fermionic waves $\Psi \left(  T,X,Y\right)  $ describing the
delocalized electrons near the Dirac points is given by the Dirac equation\
\begin{equation}
\left(  \gamma^{0}\frac{\partial}{\partial T}+\gamma^{1}\frac{\partial
}{\partial X}+\gamma^{2}\frac{\partial}{\partial Y}\right)  \Psi=0, \label{DR}%
\end{equation}
with%
\begin{equation}%
\begin{tabular}
[c]{lll}%
$\gamma^{0}=\left(
\begin{array}
[c]{cc}%
1 & 0\\
0 & -1
\end{array}
\right)  ,$ & $\gamma^{1}=\left(
\begin{array}
[c]{cc}%
0 & i\\
i & 0
\end{array}
\right)  ,$ & $\gamma^{2}=\left(
\begin{array}
[c]{cc}%
0 & 1\\
-1 & 0
\end{array}
\right)  ,$%
\end{tabular}
\  \  \  \label{GM}%
\end{equation}
satisfying the Clifford algebra in $\left(  2+1\right)  $-dimensions. The
solutions of this equation are given by plane waves $e^{ik_{T}T-i\left(
k_{X}X+k_{Y}Y\right)  }$ with $k_{T}^{2}=k_{X}^{2}+k_{Y}^{2}$. Boundary
conditions such as $\Psi \left(  T,X+L,Y\right)  $ $=\Psi \left(  T,X,Y\right)
$ leads to discrete $K_{X}=\frac{2\pi n}{L}$, $n\in \mathbb{Z}$.\  \  \ 

\begin{itemize}
\item \emph{Primitive compactification}
\end{itemize}

In the primitive frame (P-frame) of honeycomb, one uses the non orthogonal
vector basis%
\begin{equation}%
\begin{tabular}
[c]{llll}%
$\mathbf{e}_{x}\mathbf{.e}_{x}=1$ & , & $\mathbf{e}_{y}\mathbf{.e}_{y}=1$ & \\
$\mathbf{e}_{x}\mathbf{.e}_{y}=-\frac{1}{2}$ & , & $\widehat{\mathbf{e}%
_{x}\mathbf{,e}}_{y}=\frac{2\pi}{3}$ &
\end{tabular}
\label{02}%
\end{equation}
allowing to take advantage of the two following remarkable features of the honeycomb:

\textbf{(i) }\emph{honeycomb contains the root lattice of SU}$\left(
3\right)  $\newline The basis vectors $\mathbf{e}_{x}\mathbf{,}$
$\mathbf{e}_{y}$ are, up to a scale factor, precisely the simple roots
$\mathbf{\alpha}_{1}$ and $\mathbf{\alpha}_{2}$\ of the $SU\left(  3\right)  $
Lie algebra
\[
\mathbf{e}_{x}=\frac{\sqrt{2}}{2}\mathbf{\alpha}_{1},\qquad \mathbf{e}%
_{y}=\frac{\sqrt{2}}{2}\mathbf{\alpha}_{2}%
\]
so that space vectors like positions can be decomposed as%
\begin{equation}%
\begin{tabular}
[c]{lll}%
$\mathbf{r}$ & $=x^{1}\mathbf{\alpha}_{1}+x^{2}\mathbf{\alpha}_{2}$ & .
\end{tabular}
\  \label{2}%
\end{equation}
The relation between the C- and P- frames is given by
\begin{equation}%
\begin{tabular}
[c]{lll}%
$X^{\mu}$ & $=\frac{\sqrt{2}}{2}\alpha_{1}^{\mu}x^{1}+\frac{\sqrt{2}}{2}%
\alpha_{2}^{\mu}x^{2}$ & ,\\
$x^{i}$ & $=\sqrt{2}\beta_{1}^{i}X^{1}+\sqrt{2}\beta_{2}^{i}X^{2}$ & ,
\end{tabular}
\label{3}%
\end{equation}
or by using matrices%
\begin{equation}
\left(
\begin{array}
[c]{c}%
X\\
Y
\end{array}
\right)  =\frac{\sqrt{2}}{2}\left(
\begin{array}
[c]{cc}%
\alpha_{1}^{1} & \alpha_{1}^{2}\\
\alpha_{2}^{1} & \alpha_{2}^{2}%
\end{array}
\right)  \left(
\begin{array}
[c]{c}%
x\\
y
\end{array}
\right)  .
\end{equation}
The inverse transformation is
\begin{equation}
\left(
\begin{array}
[c]{c}%
x\\
y
\end{array}
\right)  =\sqrt{2}\left(
\begin{array}
[c]{cc}%
\beta_{1}^{1} & \beta_{1}^{2}\\
\beta_{2}^{1} & \beta_{2}^{2}%
\end{array}
\right)  \left(
\begin{array}
[c]{c}%
X\\
Y
\end{array}
\right)  ,
\end{equation}
with the constraint $\beta_{\mu}^{i}\alpha_{i}^{\nu}=\delta_{\mu}^{\nu}$ which
is solved in terms of the $\alpha_{i}^{\mu}$'s like
\[
\beta_{\mu}^{i}=\frac{1}{\Delta}\varepsilon_{\mu \nu}\varepsilon^{ji}\alpha
_{j}^{\nu}%
\]
with $\Delta=\det \left(  \alpha_{i}^{\mu}\right)  =\frac{1}{2}\varepsilon
_{\mu \nu}\varepsilon^{ji}\alpha_{j}^{\nu}\alpha_{i}^{\mu}$.

\textbf{(ii) }\emph{honeycomb is the weight lattice of SU}$\left(  3\right)
$\newline The matrix $\beta_{\mu}^{i}$ is precisely the entries of the
fundamental weight vectors of $SU\left(  3\right)  $%
\[
\mathbf{\omega}^{1}=\left(  \beta_{\mu}^{1}\right)  ,\qquad \mathbf{\omega}%
^{2}=\left(  \beta_{\mu}^{2}\right)
\]
These vectors are dual to the simple roots
\[
\mathbf{\alpha}_{i}.\mathbf{\omega}^{j}=\delta_{i}^{j}.
\]
Notice that the roots of Lie algebras can be expanded in terms of the weight
vectors and vice versa. For the example of SU$\left(  3\right)  $, the simple
roots decompose as
\[
\mathbf{\alpha}_{1}=2\mathbf{\omega}^{1}-\mathbf{\omega}^{2},\qquad
\mathbf{\alpha}_{2}=2\mathbf{\omega}^{2}-\mathbf{\omega}^{1}%
\]
This features teaches as that honeycomb sites are given by $n_{1}%
\mathbf{\omega}^{1}+n_{2}\mathbf{\omega}^{2}$ with $n_{i}$ arbitrary integers.
Moreover sites in the sublattices $\mathbb{A}$ and $\mathbb{B}$ of honeycomb
are rather given by $n_{1}\mathbf{\alpha}_{1}+n_{2}\mathbf{\alpha}_{2}.$

\begin{figure}[ptbh]
\begin{center}
\hspace{0cm} \includegraphics[width=10cm]{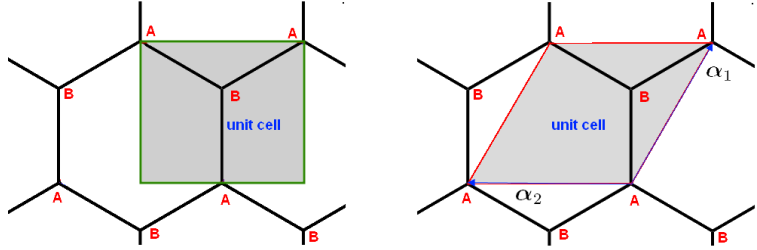}
\end{center}
\par
\vspace{-0.5cm}\caption{{\protect \small The real unit cell of the 2D honeycomb
in C- and P- frames. }Sites $A$ (resp $B$) belong to sublattice $\mathbb{A}$
(resp $\mathbb{B}$) of the honeycomb.}%
\label{ARA}%
\end{figure}

\textbf{(iii) }\emph{from 2D honeycomb to 4D hyperdiamond}\newline The link
between honeycomb and the root lattice of $SU\left(  3\right)  $ is very
suggestive. It allows to extend results on 2D graphene to fermions on higher
2N- dimensional lattices. The latters are isomorphic to the weight lattice of
$SU\left(  2N+1\right)  $ Lie algebras and permits the following
correspondence%
\[%
\begin{tabular}
[c]{|l|l|l|}\hline
\ {\small N \  \ } & $\ {\small SU}\left(  2N+1\right)  $ \  \  &
\  \  \  \  \  \  \  \  \  \  \  \  \ {\small Lattice}\\ \hline
\ {\small 1} & $\ {\small SU}\left(  3\right)  $ & \ {\small 2D honeycomb of
graphene}\\ \hline
\ {\small 2} & $\ {\small SU}\left(  5\right)  $ & \ {\small 4D honeycomb }%
$=${\small \ 4D hyperdiamond of QCD}$_{4}$ \  \  \  \\ \hline
\ {\small 3} & $\ {\small SU}\left(  7\right)  $ & \ {\small 6D honeycomb}%
\\ \hline
\end{tabular}
\  \
\]
In this view, 2D graphene appears as the leading term of a family of lattice
models.\ The second element of this family is remarkably given by the 4D
hyperdiamond that is used in dealing with quarks in lattice QCD$_{4}$
\textrm{\cite{5A}-\cite{5G}}.

\  \  \  \textrm{\newline}To make contact with the study of \textrm{\cite{4C}},
we take the components of the simple roots in the C-frame as,%
\begin{equation}%
\begin{tabular}
[c]{ll}%
$\mathbf{\alpha}_{1}=\left(  \frac{\sqrt{2}}{2},\frac{\sqrt{6}}{2}\right)  ,$
& $\mathbf{\alpha}_{2}=\left(  -\sqrt{2},0\right)  ,$%
\end{tabular}
\  \  \  \  \  \  \label{AL}%
\end{equation}
leading to
\begin{equation}%
\begin{tabular}
[c]{ll}%
$\alpha_{i}^{\mu}=\left(
\begin{array}
[c]{cc}%
\frac{\sqrt{2}}{2} & \frac{\sqrt{6}}{2}\\
-\sqrt{2} & 0
\end{array}
\right)  ,$ & $\beta_{\mu}^{i}=\left(
\begin{array}
[c]{cc}%
0 & -\frac{\sqrt{2}}{2}\\
\frac{\sqrt{6}}{3} & \frac{\sqrt{6}}{6}%
\end{array}
\right)  $%
\end{tabular}
\end{equation}
with $\mathbf{\alpha}_{1}\wedge \mathbf{\alpha}_{2}=$ $\Delta=\sqrt{3}$; and
\begin{equation}%
\begin{tabular}
[c]{lll}%
$\mathbf{\omega}^{1}=\left(  0,\frac{\sqrt{6}}{3}\right)  $ & $,$ &
$\mathbf{\omega}^{2}=\left(  -\frac{\sqrt{2}}{2},\frac{\sqrt{6}}{6}\right)  ,$%
\end{tabular}
\  \  \  \  \  \  \label{OM}%
\end{equation}
with $\mathbf{\omega}_{1}\wedge \mathbf{\omega}_{2}=\frac{\sqrt{3}}{3}$ which
is $\frac{1}{3}$ smaller compared to $\mathbf{\alpha}_{1}\wedge \mathbf{\alpha
}_{2}$. Notice that the transpose vectors
\begin{equation}%
\begin{tabular}
[c]{lll}%
$\mathfrak{a}^{1}=\left(  \alpha_{i}^{1}\right)  =\left(  \frac{\sqrt{2}}%
{2},-\sqrt{2}\right)  $ & $,$ & $\mathfrak{b}_{1}=\left(  \beta_{1}%
^{i}\right)  =\left(  0,-\frac{\sqrt{2}}{2}\right)  ,$\\
$\mathfrak{a}^{2}=\left(  \alpha_{i}^{2}\right)  =\left(  \frac{\sqrt{6}}%
{2},0\right)  $ & $,$ & $\mathfrak{b}_{2}=\left(  \beta_{2}^{i}\right)
=\left(  \frac{\sqrt{6}}{3},\frac{\sqrt{6}}{6}\right)  ,$%
\end{tabular}
\end{equation}
satisfy as well the duality duality relation $\mathfrak{a}^{\mu}%
.\mathfrak{b}_{\nu}=\delta_{\nu}^{\mu}$. Notice also that the vectors
$\mathbf{\hat{0}}$, $\mathbf{\hat{1}}$, $\mathbf{\hat{2}}$ used in \cite{4C}
are nothing but the three \emph{3} roots SU$\left(  3\right)  $ namely
$\mathbf{\alpha}_{1},$ $\mathbf{\alpha}_{2},$ $\mathbf{\alpha}_{3}%
=\mathbf{\alpha}_{1}+\mathbf{\alpha}_{2}$; see also fig \ref{ARA} for
illustration. The area of the unit cell in the C-frame is given by $L_{1}L_{2}$
and in the P-frame is $l_{1}l_{2}\frac{\sqrt{3}}{2}$. The equality of these
areas can checked by using eqs(\ref{3},\ref{AL}).

\subsection{Dirac equation in primitive frame}

One of the lessons we have learned from above is that simple roots
$\mathbf{\alpha}_{i}$ and fundamental weight vectors of $\mathbf{\omega}^{i}$
are appropriate tools to deal with fermionic waves on honeycomb and reciprocal
space. Real positions $\mathbf{r}$ and wave vectors $\mathbf{k}$ in the
momentum space can be decomposed as,
\begin{equation}%
\begin{tabular}
[c]{lll}%
$\mathbf{r}=x^{i}\mathbf{\alpha}_{i}$ & , \  \  \  \  \  & $x^{i}=\mathbf{\omega
}^{i}.\mathbf{r}$\\
$\mathbf{k}=k_{i}\mathbf{\omega}^{i}$ & , \  \  \  & $k_{i}=\mathbf{\alpha}%
_{i}.\mathbf{k}$%
\end{tabular}
\  \  \label{RX}%
\end{equation}
The respective norms read as $\mathbf{r}^{2}=C_{ij}x^{i}x^{j}$, $\mathbf{k}%
^{2}=G^{ij}k_{i}k_{j}$ with metrics $C_{ij}=\mathbf{\alpha}_{i}.\mathbf{\alpha
}_{i}$ and $G^{ij}=\mathbf{\omega}^{i}.\mathbf{\omega}^{j}$ given by
\begin{equation}%
\begin{tabular}
[c]{llll}%
$C_{ij}=\left(
\begin{array}
[c]{cc}%
2 & -1\\
-1 & 2
\end{array}
\right)  $ & , & $G^{ij}=\left(
\begin{array}
[c]{cc}%
\frac{2}{3} & \frac{1}{3}\\
\frac{1}{3} & \frac{2}{3}%
\end{array}
\right)  $ &
\end{tabular}
\  \  \label{ME}%
\end{equation}
Plane waves $e^{i\mathbf{k.r}}$ which read in C-frame as $e^{iK_{\mu}X^{\mu}}
$ takes also the form $e^{ik_{i}x^{i}}$ in the P-frame because of the duality
relation $\mathbf{\omega}^{i}.\mathbf{\alpha}_{j}=\delta_{j}^{i}$ appearing in
$\exp i\left[  \mathbf{\omega}^{i}.\mathbf{\alpha}_{j}\text{ }k_{i}%
x^{j}\right]  $. The Fourier transform reads as usual
\begin{equation}
f\left(  \mathbf{r}\right)  =\int \frac{d^{2}\mathbf{k}}{\left(  2\pi \right)
^{2}}F\left(  \mathbf{k}\right)  e^{i\mathbf{k.r}},
\end{equation}
and the gradients are related as $\frac{\partial}{\partial X^{\mu}}%
=\omega_{\mu}^{i}\frac{\partial}{\partial x^{i}}$ and $\frac{\partial
}{\partial x^{i}}=\alpha_{i}^{\mu}\frac{\partial}{\partial X^{\mu}}$. The two
sublattices $\mathbb{A}$ and $\mathbb{B}$ of the 2D honeycomb are as follows%
\[%
\begin{tabular}
[c]{ll}%
$\mathbb{A}:$ & $\mathbf{r}_{n}=n_{1}\mathbf{a}_{1}+n_{2}\mathbf{a}_{2}$\\
$\mathbb{B}:$ & $\mathbf{r}_{n}=\left(  n_{1}-\frac{1}{3}\right)
\mathbf{a}_{1}+\left(  n_{2}-\frac{2}{3}\right)  \mathbf{a}_{2}$%
\end{tabular}
\  \
\]
where $\mathbf{a}_{i}=d\sqrt{\frac{3}{2}}\mathbf{\alpha}_{i}\equiv$ $\frac
{a}{\sqrt{2}}\mathbf{\alpha}_{i}$ with $d$ standing for the lattice parameter
and $n=\left(  n_{1}\mathbf{,}n_{2}\right)  \in \mathbb{Z}^{2}$. The area of
the unit cells in real and momentum spaces are given by
\begin{equation}%
\begin{tabular}
[c]{lll}%
$\left \vert \mathbf{a}_{1}\mathbf{\wedge a}_{2}\right \vert $ & $=\frac
{3d^{2}\sqrt{3}}{4}=a^{2}\frac{\sqrt{3}}{4}$ & \\
$\left \vert \mathbf{b}_{1}\mathbf{\wedge b}_{2}\right \vert $ & $=\frac
{8\pi^{2}}{3d^{2}}\frac{\sqrt{3}}{3}=\frac{8\pi^{2}}{a^{2}}\frac{\sqrt{3}}{3}$
&
\end{tabular}
\end{equation}
with $\mathbf{b}_{i}=\frac{2\pi}{d}\sqrt{\frac{2}{3}}\mathbf{\omega}^{i}%
\equiv$ $\frac{2\pi \sqrt{2}}{a}\mathbf{\omega}^{i}$ the dual vectors of
$\mathbf{a}_{i}$. For a real supercell
\[
SC_{2}=\left(  N_{1}\mathbf{a}_{1}\right)  \times \left(  N_{2}\mathbf{a}%
_{2}\right)
\]
described by a parallelogram with edges $\mathbf{l}_{i}=N_{i}\mathbf{a}_{i}$,
the real area $\mathcal{S}$ is given by $\mathcal{S}=N_{1}N_{2}a^{2}%
\frac{\sqrt{3}}{4}$. \newline The dynamics of fermionic waves in the C-frame
is given by the Dirac equation $\gamma^{\mu}\frac{\partial}{\partial X^{\mu}%
}\Psi=0$; it reads in the P-frame as,\
\begin{equation}
\left(  \Upsilon^{0}\frac{\partial}{\partial t}+\Upsilon^{1}\frac{\partial
}{\partial x}+\Upsilon^{2}\frac{\partial}{\partial y}\right)  \Psi=0
\label{DQ}%
\end{equation}
with the new $\Upsilon^{i}$ Dirac matrices related to the $\gamma^{\mu}$'s
like
\begin{equation}%
\begin{tabular}
[c]{lllll}%
$\Upsilon^{i}=\omega_{\mu}^{i}\gamma^{\mu}$ & , & $\  \gamma^{\mu}=\alpha
_{i}^{\mu}\Upsilon^{i}$ & , & $\Upsilon^{0}=\gamma^{0}$%
\end{tabular}
\  \  \label{GA}%
\end{equation}
and obeying the following Clifford algebra%
\begin{equation}%
\begin{tabular}
[c]{llll}%
$\left \{  \Upsilon^{i},\Upsilon^{j}\right \}  =2G^{ij}$ & , & $\left \{
\Upsilon^{0},\Upsilon^{i}\right \}  =0$ & \\
$\left[  \Upsilon^{1},\Upsilon^{2}\right]  =-2i\frac{\sqrt{3}}{3}\sigma^{3}$ &
&  & .
\end{tabular}
\  \  \label{CL}%
\end{equation}

\section{Topological aspects of fermion on honeycomb}

First we give some general features regarding fermion on supercell; then we
study the spectrum of the Dirac equation of these fermions in a uniform
background field $B$. After that, we compute the index of the Dirac operator
in presence of $B$.

\subsection{Fermions on supercell}

To study the topological aspects of fermions on honeycomb, one has to
transcribe the expressions of the usual fields $\Psi \left(  \mathbf{x}\right)
$ and $A_{\mu}\left(  \mathbf{x}\right)  $ in the continuum to the case of the
lattice. This extension is not straightforward because the field expressions
seem to be in conflict with the lattice periodicity. Below, we study this
issue by working in the P-frame of the honeycomb and by using its underlying
$SU\left(  3\right)  $ symmetry.

\subsubsection{Supercell in honeycomb}

The sites of a honeycomb supercell $SC_{2}$ are of two types: A-type and
B-type; and are defined as,%
\begin{equation}
SC_{2}=\left \{  \mathbf{r}_{n}^{\mathbb{A}}=n_{1}\mathbf{a}_{1}+n_{2}%
\mathbf{a}_{2}%
\begin{tabular}
[c]{l}%
\\
\end{tabular}
\right \}  \cup \left \{  \mathbf{r}_{n}^{\mathbb{B}}=(n_{1}+\frac{2}%
{3})\mathbf{a}_{1}+(n_{2}+\frac{1}{3})\mathbf{a}_{2}%
\begin{tabular}
[c]{l}%
\\
\end{tabular}
\right \}
\end{equation}
with $n=\left(  n_{1},n_{2}\right)  $ integers restricted as
\begin{equation}%
\begin{tabular}
[c]{llll}%
$0\leq n_{i}\leq N_{i}-1$ & , & $L_{i}=d\sqrt{3}N_{i}\equiv aN_{i}$ & .
\end{tabular}
\end{equation}
The vectors $\mathbf{r}_{\left(  n_{1},0\right)  }^{\mathbb{A}}$ and
$\mathbf{r}_{\left(  0,n_{2}\right)  }^{\mathbb{A}}$ are the lattice positions
along the $\mathbf{a}_{1}$ and $\mathbf{a}_{2}$ directions with $N_{1}$,
$N_{2}$ the respective number of sites in these directions. So the total
number of the A-type sites in the supercell is $N_{1}N_{2}$ and then the total
area $\mathcal{S}$ of the supercell is equal to $N_{1}N_{2}\frac{a^{2}\sqrt
{3}}{2}$. Notice also that in the supercell $SC_{2}$ we also have B-type sites
$\mathbf{r}_{\left(  n_{1},n_{2}\right)  }^{\mathbb{B}}$ with same number as
the A-type ones so that the total sites in the supercell is
\begin{equation}
2N_{1}N_{2}.
\end{equation}
Notice that the $\mathbf{r}_{\left(  n_{1},n_{2}\right)  }^{\mathbb{B}}$ sites
are shifted with respect to the $\mathbf{r}_{\left(  n_{1},n_{2}\right)
}^{\mathbb{A}}$ sites by a constant vector as $\mathbf{r}_{n}^{\mathbb{B}%
}=\mathbf{r}_{n}^{\mathbb{A}}+\mathbf{s}$ with $\mathbf{s=}\frac{2}%
{3}\mathbf{a}_{1}+\frac{1}{3}\mathbf{a}_{2}$. \begin{figure}[ptbh]
\begin{center}
\hspace{0cm} \includegraphics[width=10cm]{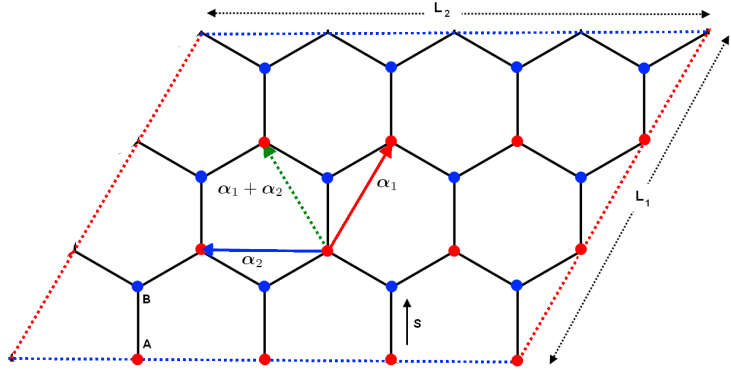}
\end{center}
\par
\vspace{-0.5cm}\caption{{\protect \small A supercell in primitive
compactification; it has \emph{24} sites: \emph{12} sites are A-type (balls in
red) and the 12 others of B-type (balls in blue). The origin of the primitive
frame belongs to }$\mathbb{A}${\protect \small -sublattice. A shift by the
vector }$\mathbf{s}${\protect \small \ maps A-type sites into B-type ones.}}%
\label{PARA}%
\end{figure}This shift vector $\mathbf{s}$ has two remarkable features: first
it is independent of the site positions $\left(  n_{1},n_{2}\right)  $; and
second it is proportional to $\mathbf{\omega}^{1}=\frac{2}{3}\mathbf{\alpha
}_{1}+\frac{1}{3}\mathbf{\alpha}_{2}$; we have $\mathbf{s}=d\sqrt{\frac{3}{2}%
}\mathbf{\omega}^{1}$. Using the parametrization of \textrm{\cite{4C}}, one
can check that $\mathbf{\omega}^{1}=(0,\frac{\sqrt{6}}{3})$ and by
substituting back into the expression of the shift vector, we get%
\begin{equation}
\mathbf{s}=\left(
\begin{array}
[c]{c}%
0\\
d
\end{array}
\right)
\end{equation}

\subsubsection{Gauge field and fermionic waves on supercell}

To deal with the Dirac equation on honeycomb supercell in presence of the
background field $B$, one has to worry about the boundary conditions of the
gauge field and the fermionic waves under lattice periodicity%
\begin{equation}
\mathbf{r}_{n}\rightarrow \mathbf{r}_{n}+nN_{1}\mathbf{a}_{1}+mN_{2}%
\mathbf{a}_{2},\qquad n,m\in \mathbb{Z}%
\end{equation}
Constructions regarding this issue has been first done by \emph{Smit, J.C.
Vink} in \textrm{\cite{B2}} by working in the C-frame of a square lattice; see
also \textrm{\cite{4C,5G}}. Below we extend this analysis to the P-frame; this
is helpful when we consider the extension to \emph{QCD}$_{4}$ on 4D
hyperdiamond and extension to higher dimensional lattices.

\begin{itemize}
\item \emph{Periodicity of the gauge field }
\end{itemize}

In the canonical frame, the gauge field is given by the vector $A_{\mu}$ while
in the primitive frame, it is denoted as $\mathcal{A}_{i}$. These two fields
capture the same physical property; they are related as $A_{\mu}=\omega_{\mu
}^{i}\mathcal{A}_{i}$ or inversely $\mathcal{A}_{i}=\alpha_{i}^{\mu}A_{\mu}$.
\newline Here, we use the P-frame coordinates $x^{i}=\left(  x,y\right)  $ and
work with the gauge covariant derivatives $\mathcal{D}_{i}=\frac{\partial
}{\partial x^{i}}-i\mathcal{A}_{i}$ with curvature%
\begin{equation}
\left[  \mathcal{D}_{i},\mathcal{D}_{j}\right]  =-i\mathcal{F}_{ij}%
=i\varepsilon_{ij}B \label{CU}%
\end{equation}
related to the C-frame one $\left[  D_{\mu},D_{\nu}\right]  =-iF_{\mu \nu}$
like
\[
\mathcal{F}_{ij}=\alpha_{i}^{\mu}\alpha_{j}^{\nu}F_{\mu \nu}%
\]
As usual, the gauge field $\mathcal{A}_{i}$ is defined up to gauge
transformations $\mathcal{A}_{i}^{\Omega}=$ $\mathcal{A}_{i}+i\Omega
\partial_{i}\Omega^{-1}$ where $\Omega=e^{-i\lambda \left(  \mathbf{x}\right)
}$ is a $U\left(  1\right)  $ gauge group element with gauge parameter
$\lambda \left(  \mathbf{x}\right)  $. We also have:%
\begin{equation}%
\begin{tabular}
[c]{ll}%
$\Psi^{\Omega}\left(  \mathbf{x}\right)  $ & $=e^{-i\lambda \left(
\mathbf{x}\right)  }\Psi \left(  \mathbf{x}\right)  $\\
$\mathcal{A}_{i}^{\Omega}$ & $=\mathcal{A}_{i}+\partial_{i}\lambda$%
\end{tabular}
\  \  \  \label{m}%
\end{equation}
Since $B$ is constant, a typical gauge configuration of $\mathcal{A}_{i}$ that
solve (\ref{CU}) in the continuum is as follows,%
\begin{equation}%
\begin{tabular}
[c]{llll}%
$\mathcal{A}_{x}\left(  x,y\right)  =\frac{1}{2}By$ & , & $\mathcal{A}%
_{y}\left(  x,y\right)  =-\frac{1}{2}Bx$ & .
\end{tabular}
\  \  \  \label{GF}%
\end{equation}
This not unique since a different, but equivalent, gauge field representation
with curvature $B$ is given by the following simpler one used by \emph{Smit,
J.C. Vink} in \textrm{\cite{B2},}
\begin{equation}%
\begin{tabular}
[c]{llll}%
$\mathcal{A}_{x}\left(  x,y\right)  =0$ & , & $\mathcal{A}_{y}\left(
x,y\right)  =-Bx$ & .
\end{tabular}
\  \  \  \label{SV}%
\end{equation}
However, this gauge field configuration, which is valid in the continuous
case, has to be adapted to the lattice; neither (\ref{GF}) nor (\ref{SV}) is
periodic on the honeycomb since%
\begin{equation}%
\begin{tabular}
[c]{llll}%
$\mathcal{A}_{i}\left(  x+L_{1},y\right)  $ & $\neq$ & $\mathcal{A}_{i}\left(
x,y\right)  $ & ,\\
$\mathcal{A}_{i}\left(  x,y+L_{2}\right)  $ & $\neq$ & $\mathcal{A}_{i}\left(
x,y\right)  $ & .
\end{tabular}
\end{equation}
Explicitly, by using the expression (\ref{GF}) for the gauge field
$\mathcal{A}_{i}$, we have%
\begin{equation}%
\begin{tabular}
[c]{llll}%
$\mathcal{A}_{x}\left(  x,y\right)  |_{y=L_{2}}=\frac{BL_{2}}{2}$ & $,$ &
$\mathcal{A}_{x}\left(  x,y\right)  |_{y=0}=0$ & ,\\
$\mathcal{A}_{y}\left(  x,y\right)  |_{x=L_{1}}=\frac{BL_{1}}{2}$ & $,$ &
$\mathcal{A}_{y}\left(  x,y\right)  |_{x=0}=0$ & .
\end{tabular}
\end{equation}
If using the gauge configuration (\ref{SV}), we also have%
\begin{equation}%
\begin{tabular}
[c]{llll}%
$\mathcal{A}_{x}\left(  x,y\right)  |_{y=L_{2}}=0$ & $,$ & $\mathcal{A}%
_{x}\left(  x,y\right)  |_{y=0}=0$ & ,\\
$\mathcal{A}_{y}\left(  x,y\right)  |_{x=L_{1}}=-BL_{1}$ & $,$ &
$\mathcal{A}_{y}\left(  x,y\right)  |_{x=0}=0$ & .
\end{tabular}
\end{equation}
To overcome this difficulty, we use the gauge symmetry freedom to ensure
field's periodicity. This is done by requiring%
\begin{equation}%
\begin{tabular}
[c]{lll}%
$\mathcal{A}_{x}\left(  x,y\right)  |_{y=0}$ & $=\mathcal{A}_{x}\left(
x,y\right)  |_{y=L_{2}}+\partial_{x}\lambda$ & ,\\
$\mathcal{A}_{y}\left(  x,y\right)  |_{x=0}$ & $=\mathcal{A}_{y}\left(
x,y\right)  |_{x=L_{1}}+\partial_{y}\lambda$ & ,
\end{tabular}
\end{equation}
where now the gauge group parameter $\lambda \left(  x,y\right)  $ is no longer
an arbitrary function as in the continuum; this is the price to pay to
implement lattice periodicity.\newline In what follows, we choose the gauge
configuration (\ref{SV}) used by Smit and Vink for the case of gauge fields on
square lattice. This choice is tricky and allows tremendous simplifications
when computing the spectrum of the Dirac operator; especially the degeneracy
of the ground state. Using this choice the gauge covariant derivatives reduce
to,%
\[%
\begin{tabular}
[c]{llll}%
$\mathcal{D}_{x}=\frac{\partial}{\partial x}$ & $,$ & $\mathcal{D}_{y}%
=\frac{\partial}{\partial y}-i\mathcal{A}_{y}$ & ,
\end{tabular}
\  \  \
\]
with no gauge component along the x-direction. In this case, the periodicity
condition on the gauge field namely
\[
\mathcal{G}_{y}\left(  x,y\right)  |_{x=0}=\mathcal{G}_{y}\left(  x,y\right)
|_{x=L_{1}}+\partial_{y}\lambda
\]
leads to $\partial_{y}\lambda=BL_{1}$ and then to $\lambda=BL_{1}%
y+\varphi \left(  x\right)  \equiv \vartheta \left(  y\right)  +\varphi \left(
x\right)  $ with $\varphi$ a real function in the $x$-variable that we shall
ignore below as it doesn't affect the analysis \textrm{\footnote{By using
periodicity property of the honeycomb supercell, one can show that
$\varphi \left(  x\right)  =\frac{2\pi x}{L_{1}}Q$.}}. Notice that as a
function on lattice, the gauge symmetry element $\Omega$, thought of as a
function of the y-variable only, i.e: $\Omega=\Omega \left(  y\right)  $,
should be also periodic $\Omega \left(  y+L_{2}\right)  =\Omega \left(
y\right)  $. The solving of this condition requires the quantization of the
background field as
\begin{equation}%
\begin{tabular}
[c]{llll}%
$BL_{1}L_{2}=2\pi Q$ & , & $Q\in \mathbb{Z}$ & .
\end{tabular}
\  \  \  \label{QD}%
\end{equation}
So the gauge group element reads as
\begin{equation}%
\begin{tabular}
[c]{ll}%
$\Omega \left(  y\right)  =\exp \left(  -i\frac{2\pi Q}{L_{2}}y\right)  $ & ,
\end{tabular}
\  \  \  \label{TE}%
\end{equation}
and can be thought of as describing a wave plane propagating along the
y-direction with the momentum $K_{y}=\frac{2\pi Q}{L_{2}}$. This property
teaches us that gauge transformations are generated by the shifts
\begin{equation}
p_{y}\rightarrow p_{y}+n\frac{2\pi Q}{L_{2}},\qquad n\in \mathbb{Z}%
\end{equation}
of the momenta $p_{y}=\frac{2\pi l}{L_{2}}$ along the y-axis.

\begin{itemize}
\item \emph{Periodicity of fermionic waves}
\end{itemize}

Fermionic waves $\Psi \left(  x,y\right)  $ on honeycomb supercell have to obey
consistent boundary conditions. Because of gauge freedom, the boundary
condition should be as
\[
\Psi \left(  x+nL_{1},y+mL_{2}\right)  =\mathcal{R}_{n,m}\Psi \left(
x,y\right)  ,
\]
where $\mathcal{R}_{n,m}$ is some representation of the gauge symmetry which
is given by some local phase $e^{i\vartheta_{n,m}\left(  y\right)  }$.
However, with the gauge choice (\ref{SV}), these condition reduces to
\begin{equation}%
\begin{tabular}
[c]{lll}%
$\Psi \left(  x,y=L_{2}\right)  $ & $=\Psi \left(  x,y=0\right)  $ & ,\\
$\Psi \left(  x=L_{1},y\right)  $ & $=e^{i\vartheta_{1}\left(  y\right)  }%
\Psi \left(  x=0,y\right)  $ & .
\end{tabular}
\  \  \  \label{PER}%
\end{equation}
To get the explicit expression of the fermionic waves, one has to solve the
Dirac equation on the lattice; this will be done in next subsection; for the
moment we use general arguments to derive some useful information on these
waves. \newline First, because of the choice (\ref{SV}) where the gauge field
$\mathcal{A}_{i}$ has no dependence in the y- variable, the periodicity
condition of fermionic waves namely $\Psi \left(  x,y+L_{2}\right)
=\Psi \left(  x,y\right)  $ can be solved in terms of plane waves propagating
along the y-direction as follows%
\begin{equation}
\Psi \left(  x,y\right)  =%
{\displaystyle \sum \limits_{l\in \mathbb{Z}}}
e^{i\frac{2\pi l}{L_{2}}y}\Psi_{l}\left(  x\right)  \label{XX}%
\end{equation}
The Fourier modes $\Psi_{l}\left(  x\right)  $, which depend on $x$, carry an
integer charge $l$ and because of the second condition of (\ref{PER}) are
expected to be not completely independent fields. \newline Second, to solve
the boundary condition along the x-direction, we expand the wave function at
$x+L_{1}$ in a similar manner as in (\ref{XX})
\begin{equation}
\Psi \left(  x+L_{1},y\right)  =%
{\displaystyle \sum \limits_{l\in \mathbb{Z}}}
e^{i\frac{2\pi l}{L_{2}}y}\Psi_{l}\left(  x+L_{1}\right)
\end{equation}
and then require that the value of the fermionic waves at the equivalent
positions $x$ and $x+L_{1}$\ to be related by a gauge transformation like
$\Psi \left(  x,y\right)  =$ $e^{-iBL_{1}y}$ $\Psi \left(  x+L_{1},y\right)  $.
This leads to the following relation between the Fourier modes,%
\begin{equation}
\Psi_{l}\left(  x+L_{1}\right)  =\Psi_{l-Q}\left(  x\right)  \label{SH}%
\end{equation}
\  \ This property can be exhibited by help of the quantization relation $2\pi
Q=BL_{1}L_{2}$ allowing to express the wave plane basis $e^{i\frac{2\pi
l}{L_{2}}y}$ in terms of the magnetic field and the topological charge like
$e^{iBL_{1}\frac{l}{Q}y}$. Substituting this expression back into (\ref{XX}),
one ends with (\ref{SH}). Notice that eq(\ref{SH}) is solved by taking
\begin{equation}
\Psi_{l}\left(  x\right)  =\Phi \left(  x+\frac{l}{Q}L_{1}\right)  \label{HS}%
\end{equation}
showing that a translation $y\rightarrow y+L_{2}$ along the y-direction by one
period induces the shift $x\rightarrow x+\frac{1}{Q}L_{1}$ along the x-axis.
The explicit expression of the function $\Phi \left(  \xi \right)  $ will be
determined below.

\subsection{Index of Dirac operator and chiral anomaly}

In the P-frame of the honeycomb, the Dirac equation of the two component
fermionic wave $\Psi=\left(  \psi,\chi \right)  $ in the background field $B$
is given by the following $2\times2$ matrix equation
\begin{equation}
\Upsilon^{i}\mathcal{D}_{i}\left(
\begin{array}
[c]{c}%
\psi \\
\chi
\end{array}
\right)  =iE\Upsilon^{0}\left(
\begin{array}
[c]{c}%
\psi \\
\chi
\end{array}
\right)  , \label{MAG}%
\end{equation}
where $\Upsilon^{0}$, $\Upsilon^{1}$, $\Upsilon^{2}$ are $2\times2$ gamma
matrices which are related to the usual gamma matrices $\gamma^{\mu}$ as in
(\ref{GA}).

\subsubsection{Solving Dirac equation on supercell}

The gauge covariant derivatives $\mathcal{D}_{i}=\partial_{i}-i\mathcal{A}%
_{i}$ with $\mathcal{A}_{i}$ taken the SV gauge are as follows%
\begin{equation}%
\begin{tabular}
[c]{llll}%
$\mathcal{D}_{1}=\frac{\partial}{\partial x}$ & $,$ & $\mathcal{D}_{2}%
=\frac{\partial}{\partial y}+iBx$ & ,
\end{tabular}
\end{equation}
with no component $\mathcal{A}_{x}$ for the gauge vector. Using the expression
of $\gamma^{0}=\sigma^{3}$, $\gamma^{1}=i\sigma^{1}$, $\gamma^{2}=i\sigma^{2}$
with the Pauli matrices $\sigma^{\mu}$ as in eq(\ref{GM}), the matrices
$\Upsilon^{i}$ read as:
\begin{equation}%
\begin{tabular}
[c]{lll}%
$\Upsilon^{1}=i\left(
\begin{array}
[c]{cc}%
0 & \xi \\
\bar{\xi} & 0
\end{array}
\right)  ,$ & $\Upsilon^{2}=i\left(
\begin{array}
[c]{cc}%
0 & \zeta \\
\bar{\zeta} & 0
\end{array}
\right)  ,$ & $\Upsilon^{0}=\left(
\begin{array}
[c]{cc}%
1 & 0\\
0 & -1
\end{array}
\right)  ,$%
\end{tabular}
\end{equation}
where we have set $\xi=\omega_{1}^{1}-i\omega_{2}^{1}$, $\zeta=\omega_{1}%
^{2}-i\omega_{2}^{2}$ with the useful relations%
\begin{equation}%
\begin{tabular}
[c]{lllll}%
$\xi \bar{\xi}=\left \Vert \mathbf{\omega}^{1}\right \Vert ^{2}$ & , &
$\  \  \  \  \  \  \  \zeta \bar{\xi}$ & $=$ & $\mathbf{\omega}^{1}$ $.$
$\mathbf{\omega}^{2}-i$ $\mathbf{\omega}^{1}\wedge \mathbf{\omega}^{2}$\\
$\zeta \bar{\zeta}=\left \Vert \mathbf{\omega}^{2}\right \Vert ^{2}$ & , &
$\xi \bar{\zeta}-\zeta \bar{\xi}$ & $=$ & $2i$ $\mathbf{\omega}^{1}%
\wedge \mathbf{\omega}^{2}$%
\end{tabular}
\end{equation}
By replacing the weight vectors $\mathbf{\omega}^{i}$ by their expression
(\ref{OM}), reads as
\begin{equation}%
\begin{tabular}
[c]{lllll}%
$\xi \bar{\xi}=\frac{2}{3}$ & $,$ & $\  \  \  \  \  \  \  \zeta \bar{\xi}$ & $=$ &
$\frac{1}{3}-i\frac{\sqrt{3}}{3}$\\
$\zeta \bar{\zeta}=\frac{2}{3}$ & $,$ & $\xi \bar{\zeta}-\zeta \bar{\xi}$ & $=$ &
$2i\frac{\sqrt{3}}{3}$%
\end{tabular}
\end{equation}
Notice that like in the C-frame, the Dirac operator $\Upsilon^{i}%
\mathcal{D}_{i}$ in presence of the background field B can be put as well into
the form%
\begin{equation}%
\begin{tabular}
[c]{ll}%
$\Upsilon^{i}\mathcal{D}_{i}=i\left(
\begin{array}
[c]{cc}%
0 & D\\
\bar{D} & 0
\end{array}
\right)  $ & ,
\end{tabular}
\end{equation}
with $D=\xi \mathcal{D}_{1}+\zeta \mathcal{D}_{2}$ and $\bar{D}=\bar{\xi
}\mathcal{D}_{1}+\bar{\zeta}\mathcal{D}_{2}$. Its square, which is needed for
solving the Dirac equation, is diagonal%
\begin{equation}%
\begin{tabular}
[c]{ll}%
$\left(  \Upsilon^{i}\mathcal{D}_{i}\right)  ^{2}=-\left(
\begin{array}
[c]{cc}%
D\bar{D} & 0\\
0 & \bar{D}D
\end{array}
\right)  $ & ,
\end{tabular}
\label{D2}%
\end{equation}
and has the same eigen wave functions as $\Upsilon^{i}\mathcal{D}_{i}$. Notice
also that the explicit expression of $D$ and $\bar{D}$\ are as
\begin{equation}%
\begin{tabular}
[c]{lll}%
$D$ & $=\xi \frac{\partial}{\partial x}+\zeta \left(  \frac{\partial}{\partial
y}+iBx\right)  $ & ,\\
$\bar{D}$ & $=\bar{\xi}\frac{\partial}{\partial x}+\bar{\zeta}\left(
\frac{\partial}{\partial y}+iBx\right)  $ & ,
\end{tabular}
\end{equation}
and satisfy the commutation relation%
\begin{equation}%
\begin{tabular}
[c]{ll}%
$\left[  D,\bar{D}\right]  =-iB\left(  \xi \bar{\zeta}-\bar{\xi}\zeta \right)
=-\frac{2B\sqrt{3}}{3}$ & ,
\end{tabular}
\  \label{C}%
\end{equation}
where the sign of the right hand side depends on the sign of $B$; it is
positive if sign$\left(  B\right)  $ is negative and vice versa.

\subsubsection{The explicit spectrum}

Setting
\begin{equation}%
\begin{tabular}
[c]{lll}%
$D=\sqrt{\frac{2\left \vert B\right \vert \sqrt{3}}{3}}A^{-}$ & , & $\bar
{D}=\sqrt{\frac{2\sqrt{3}\left \vert B\right \vert }{3}}A^{\dagger}$%
\end{tabular}
\  \  \label{EH}%
\end{equation}
the commutation relation (\ref{C}) becomes%
\begin{equation}%
\begin{tabular}
[c]{llll}%
$\left[  A^{-},A^{\dagger}\right]  =-\frac{B}{\left \vert B\right \vert }I$ &
, & $A^{\dagger}A^{-}=A^{-}A^{\dagger}+\frac{B}{\left \vert B\right \vert }I$ &
\end{tabular}
\  \  \label{HE}%
\end{equation}
with the remarkable dependence in $\frac{B}{\left \vert B\right \vert }$
capturing the sign of the background field $B$. By substituting back into
(\ref{D2}), the squared Dirac operator reads as
\begin{equation}%
\begin{tabular}
[c]{ll}%
$\left(
\begin{array}
[c]{cc}%
D\bar{D} & 0\\
0 & \bar{D}D
\end{array}
\right)  =\frac{2\left \vert B\right \vert \sqrt{3}}{3}\left(
\begin{array}
[c]{cc}%
A^{-}A^{\dagger} & 0\\
0 & A^{\dagger}A^{-}%
\end{array}
\right)  $ & ,
\end{tabular}
\end{equation}
To get the solution of the Dirac equation, we first use the periodicity
condition along the y-axis to expand the fermionic wave $\Psi \left(
x,y\right)  $ in Fourier series like%
\begin{equation}
\Psi \left(  x,y\right)  =%
{\displaystyle \sum \limits_{l\in \mathbb{Z}}}
e^{i\frac{2\pi l}{L_{2}}y}\Psi_{l}\left(  x\right)
\end{equation}
with
\begin{equation}
\Psi_{l}\left(  x\right)  =\left(
\begin{array}
[c]{c}%
\psi_{l}\left(  x\right) \\
\chi_{l}\left(  x\right)
\end{array}
\right)  ,\qquad l\in \mathbb{Z}.
\end{equation}
Substituting this expansion back into the Dirac equation, we obtain%
\begin{equation}
\left(
\begin{array}
[c]{cc}%
D\bar{D} & 0\\
0 & \bar{D}D
\end{array}
\right)  \left(
\begin{array}
[c]{c}%
\psi_{l}\left(  x\right) \\
\chi_{l}\left(  x\right)
\end{array}
\right)  =E^{2}\left(
\begin{array}
[c]{c}%
\psi_{l}\left(  x\right) \\
\chi_{l}\left(  x\right)
\end{array}
\right)
\end{equation}
where now the action of the operators $D$ and $\bar{D}$ on the waves is
restricted to the Fourier modes $\Psi_{l}\left(  x\right)  $; this leads to
\begin{equation}%
\begin{tabular}
[c]{lll}%
$D$ & $=\xi \frac{\partial}{\partial x}+i\zeta B\left(  x+\frac{2\pi l}{BL_{2}%
}\right)  $ & ,\\
$\bar{D}$ & $=\bar{\xi}\frac{\partial}{\partial x}+i\bar{\zeta}B\left(
x+\frac{2\pi l}{BL_{2}}\right)  $ & .
\end{tabular}
\end{equation}
Moreover, using the quantization property of the background field in terms of
the area of the supercell namely $BL_{1}L_{2}=2\pi Q$, we can rewrite the
above operators as%
\begin{equation}%
\begin{tabular}
[c]{lll}%
$D$ & $=\xi \frac{\partial}{\partial x}+i\zeta B\left(  x+\frac{l}{Q}%
L_{1}\right)  $ & ,\\
$\bar{D}$ & $=\bar{\xi}\frac{\partial}{\partial x}+i\bar{\zeta}B\left(
x+\frac{l}{Q}L_{1}\right)  $ & .
\end{tabular}
\end{equation}

\begin{itemize}
\item \emph{Case }$\frac{B}{\left \vert B\right \vert }=-1$
\end{itemize}

In this case the operators $A^{\dagger}=\sqrt{\frac{3}{2\left \vert
B\right \vert \sqrt{3}}}\bar{D}$ and $A=\sqrt{\frac{3}{2\left \vert B\right \vert
\sqrt{3}}}D$ in the Heisenberg algebra (\ref{HE}) are respectively the
creation operator and the annihilation one. So, we have%
\begin{equation}
\left(
\begin{array}
[c]{cc}%
D\bar{D} & 0\\
0 & \bar{D}D
\end{array}
\right)  =\frac{2\left \vert B\right \vert \sqrt{3}}{3}\left(
\begin{array}
[c]{cc}%
A^{\dagger}A+1 & 0\\
0 & A^{\dagger}A
\end{array}
\right)
\end{equation}
Thus, the wave functions $\Psi_{l}=\left(  \psi_{l},\chi_{l}\right)  $ solving
the Dirac equation are given by%
\begin{equation}%
\begin{tabular}
[c]{ll}%
$\Psi_{l,n}\left(  x\right)  =\left(
\begin{array}
[c]{c}%
\phi_{l,n-1}\left(  x\right) \\
\phi_{l,n}\left(  x\right)
\end{array}
\right)  $ & ,
\end{tabular}
\  \  \label{A}%
\end{equation}
for the integer $n\geq1$; and%
\begin{equation}
\Psi_{l,0}\left(  x\right)  =\left(
\begin{array}
[c]{c}%
0\\
\phi_{l,0}\left(  x\right)
\end{array}
\right)  . \label{AB}%
\end{equation}
We also have,%
\begin{equation}
\phi_{l,n}\left(  x\right)  =\frac{1}{n!}\left(  A^{\dagger}\right)  ^{n}%
\phi_{l,0}\left(  x\right)  \label{AA}%
\end{equation}
with the fundamental state $\phi_{l,0}\left(  x\right)  $ obeying the
condition%
\begin{equation}
\left[  \xi \frac{\partial}{\partial x}+i\zeta B\left(  x+\frac{l}{Q}%
L_{1}\right)  \right]  \phi_{l,0}\left(  x\right)  =0
\end{equation}
whose solution is given by%
\begin{equation}
\phi_{l,0}\left(  x\right)  =\mathcal{N}_{0}e^{-i\frac{\left \vert B\right \vert
\zeta \bar{\xi}}{2\xi \bar{\xi}}\left(  x+\frac{l}{Q}L_{1}\right)  ^{2}}
\label{AAA}%
\end{equation}
with $\zeta \bar{\xi}=\frac{1}{3}-i\frac{\sqrt{3}}{3}$ and $\xi \bar{\xi}%
=\frac{2}{3}$. We also have $\left \vert \phi_{l,0}\right \vert \sim
e^{-\left \vert B\right \vert \frac{\sqrt{3}}{4}\left(  x+\frac{l}{Q}%
L_{1}\right)  ^{2}}$.

\begin{itemize}
\item \emph{Case }$\frac{B}{\left \vert B\right \vert }>0$
\end{itemize}

In this case, the creation and annihilation are no longer as above; the new
creation and annihilation operators obeying $\left[  C^{-},C^{\dagger}\right]
=I$ are
\begin{equation}%
\begin{tabular}
[c]{llll}%
$C^{\dagger}=\sqrt{\frac{3}{2\left \vert B\right \vert \sqrt{3}}}D$ & , &
$C^{-}=\sqrt{\frac{3}{2\left \vert B\right \vert \sqrt{3}}}\bar{D}$ & ,
\end{tabular}
\end{equation}
so that%
\begin{equation}%
\begin{tabular}
[c]{lll}%
$\left(
\begin{array}
[c]{cc}%
D\bar{D} & 0\\
0 & \bar{D}D
\end{array}
\right)  $ & $=$ & $\frac{2\left \vert B\right \vert \sqrt{3}}{3}\left(
\begin{array}
[c]{cc}%
C^{\dagger}C^{-} & 0\\
0 & C^{\dagger}C^{-}+1
\end{array}
\right)  $%
\end{tabular}
\end{equation}
The wave functions $\Psi_{l}^{\prime}=\left(  \psi_{l}^{\prime},\chi
_{l}^{\prime}\right)  $ solving the Dirac equation with $\frac{B}{\left \vert
B\right \vert }>0$ are given by%
\begin{equation}%
\begin{tabular}
[c]{ll}%
$\Psi_{l,n}^{\prime}\left(  x\right)  =\left(
\begin{array}
[c]{c}%
\phi_{l,n}^{\prime}\left(  x\right) \\
\phi_{l,n-1}^{\prime}\left(  x\right)
\end{array}
\right)  $ & ,
\end{tabular}
\end{equation}
for the integer $n\geq1$; and%
\begin{equation}
\Psi_{l,0}^{\prime}\left(  x\right)  =\left(
\begin{array}
[c]{c}%
\phi_{l,0}\left(  x\right) \\
0
\end{array}
\right)  . \label{70}%
\end{equation}
The excited states are given by
\begin{equation}
\phi_{l,n}^{\prime}\left(  x\right)  =\frac{1}{n!}\left(  C^{\dagger}\right)
^{n}\phi_{l,0}^{\prime}\left(  x\right)
\end{equation}
with ground state%
\begin{equation}
\phi_{l,0}^{\prime}\left(  x\right)  =\mathcal{N}_{0}e^{-\frac{\left \vert
B\right \vert }{4}\left(  \sqrt{3}+i\right)  \left(  x+\frac{l}{Q}L_{1}\right)
^{2}}.
\end{equation}

\subsubsection{Computing the index of Dirac operator}

Here, we focus on the case $\frac{B}{\left \vert B\right \vert }<0$ and compute
the index of the Dirac operator of fermions on a honeycomb supercell in
presence of the background field $B$. Similar calculation can be done for the
case $\frac{B}{\left \vert B\right \vert }>0$.

\begin{itemize}
\item \emph{Case} $\frac{B}{\left \vert B\right \vert }<0:$ $Q_{top}<0$\  \ 
\end{itemize}

\ From eqs(\ref{A}-\ref{AB}), we learn that the ground state $\Phi_{0}\left(
x,y\right)  $ describing the fermionic wave with zero energy is antichiral%
\begin{equation}%
\begin{tabular}
[c]{llll}%
$\Psi_{0}\left(  x,y\right)  =\left(
\begin{array}
[c]{c}%
0\\
\mathrm{\phi}_{_{0}}\left(  x,y\right)
\end{array}
\right)  $ & , & $\sigma^{3}\Psi_{0}=-\Psi_{0}$ & .
\end{tabular}
\label{73}%
\end{equation}
The function $\mathrm{\phi}_{_{0}}\left(  x,y\right)  $ is given by the
following linear combination
\begin{equation}
\mathrm{\phi}_{_{0}}\left(  x,y\right)  =%
{\displaystyle \sum \limits_{l\in \mathbb{Z}}}
c_{l}e^{i\frac{2\pi l}{L_{2}}y}e^{-\eta \left(  x+\frac{l}{Q}L_{1}\right)
^{2}}%
\end{equation}
where we have set $\eta=\left \vert B\right \vert \frac{i\zeta \bar{\xi}}%
{2\xi \bar{\xi}}$ and where, a priori, the coefficients $c_{l}$ are arbitrary
moduli. As a wave function on the supercell, the function $\mathrm{\phi}%
_{_{0}}\left(  x,y\right)  $ has to obey the periodic boundary conditions%
\begin{equation}%
\begin{tabular}
[c]{llll}%
$\mathrm{\phi}_{_{0}}\left(  x,y+L_{2}\right)  $ & $=$ & $\mathrm{\phi}_{_{0}%
}\left(  x,y\right)  $ & \\
$\mathrm{\phi}_{_{0}}\left(  x+L_{1},y\right)  $ & $=$ & $e^{\frac{2i\pi
Q}{L_{2}}y}\mathrm{\phi}_{_{0}}\left(  x,y\right)  $ &
\end{tabular}
\end{equation}
The first condition giving the periodicity along the y-axis is manifestly
satisfied due to $e^{i\frac{2\pi l}{L_{2}}\left(  y+L_{2}\right)  }=e^{2i\pi
l}e^{i\frac{2\pi l}{L_{2}}y}$. The second condition giving the periodicity
along the $x$-direction requires the following identifications
\begin{equation}%
\begin{tabular}
[c]{llll}%
$c_{l}=c_{l-Q}$ & , & $l\in \mathbb{Z}$ &
\end{tabular}
\end{equation}
leaving afterward $\left \vert Q\right \vert $ free moduli
\begin{equation}%
\begin{tabular}
[c]{lllll}%
$c_{1},$ & $c_{2},$ & \ldots, & $c_{\left \vert Q\right \vert }$ & .
\end{tabular}
\end{equation}
This feature shows that the ground state $\mathrm{\phi}_{_{0}}\left(
x,y\right)  $ can be spanned as%
\begin{equation}
\mathrm{\phi}_{_{0}}\left(  x,y\right)  =%
{\displaystyle \sum \limits_{s=1}^{\left \vert Q\right \vert }}
c_{s}e^{i\frac{2\pi s}{L_{2}}y}\mathrm{\varphi}_{_{0,s}}\left(  x,y\right)
\label{fg}%
\end{equation}
with
\begin{equation}
\mathrm{\varphi}_{_{0,s}}\left(  x,y\right)  \sim%
{\displaystyle \sum \limits_{l\in \mathbb{Z}}}
e^{i\frac{2\pi l}{L_{2}}y}e^{-\eta \left(  x+\frac{s}{Q}+\frac{l}{Q}%
L_{1}\right)  ^{2}} \label{gf}%
\end{equation}
Therefore the degree of degeneracy of the fermionic waves is as follows
\begin{equation}%
\begin{tabular}
[c]{llllll}%
ground state & : & $N_{+}=0$ & , & $\  \ N_{-}=\left \vert Q\right \vert $ & \\
excited states & : & $N_{+}=\left \vert Q\right \vert $ & , & $\  \ N_{-}%
=\left \vert Q\right \vert $ &
\end{tabular}
\  \  \label{Q}%
\end{equation}
So the number $N_{\pm}^{\left(  \epsilon>0\right)  }$ of degeneracy of the
excited states is twice the number $N_{\pm}^{\left(  \epsilon=0\right)  }$ of
degeneracy of the ground state. The index of the Dirac operator which is given
by $N_{+}-N_{-}$ reduces to $N_{+}^{\left(  \epsilon=0\right)  }%
-N_{-}^{\left(  \epsilon=0\right)  }=-\left \vert Q\right \vert =Q_{top}$.

\begin{itemize}
\item \emph{anomalous QHE} \  \  \ 
\end{itemize}

To conclude this section, notice that applying the above results to the
\emph{2} Dirac points of graphene, we recover the well known value of the
anomalous quantum Hall effect in continuum. There, the transverse conductivity
is given by $\sigma_{XY}=\nu_{_{AQHE}}\frac{e^{2}}{h}$ with the filling factor
$\nu_{_{AQHE}}$, defined as the number of electrons per flux quantum, reads
for the case of $n$ Landau levels filled, as follows%
\begin{equation}
\nu_{_{AQHE}}=2\left(  2n+1\right)  =4\left(  n+\frac{1}{2}\right)  .
\end{equation}
Notice that the ground state corresponding to $n=0$ is half filled. From our
analysis, we learn moreover that in the case $\frac{B}{\left \vert B\right \vert
}<0$ this ground state is filled by the zero modes $\Psi_{0}$ eqs(\ref{AB}%
,\ref{73}) having negative chirality; they obey the chirality property
$\sigma^{3}\Psi_{0}=-\Psi_{0}$ with $\sigma^{3}=diag\left(  1,-1\right)  $. In
the case $\frac{B}{\left \vert B\right \vert }>0$, the ground state is filled by
the zero modes $\Psi_{0}^{\prime}$ eq(\ref{70}) with positive chirality
satisfying $\sigma^{3}\Psi_{0}^{\prime}=+\Psi_{0}^{\prime}$.

\section{Fermions on hyperdiamond}

In this section, we describe the supercell compactification in 4D hyperdiamond
and study the solutions of the boundary conditions on fields required by
periodicity properties. These tools will be used later on to derive the
spectrum of lattice fermions on hyperdiamond in presence of uniform background
fields $\mathcal{B}$ and $\mathcal{E}$ and to compute the topological index of
the Dirac operator.

\subsection{Hyperdiamond and $SU\left(  5\right)  $ weight lattice}

Like 2D honeycomb, the 4-dimensional hyperdiamond can be described either by
using the perpendicular compactification or the primitive compactification of
the supercell. As the first compactification is a standard method in (hyper)
cubic lattice QFT, let us focus below on the primitive one.

\begin{itemize}
\item \emph{Hyperdiamond in P-frame}
\end{itemize}

The hyperdiamond $\mathbb{L}_{4}$ is a 4-dimensional extension of the 2D
honeycomb; it is the world of fermions (light quarks) in lattice QCD$_{4}$
\textrm{\cite{5A}-\cite{5G}; }see also\textrm{\  \cite{7A}}. In the primitive
compactification, the hyperdiamond is generated by the basis vectors
\begin{equation}%
\begin{tabular}
[c]{llll}%
$\mathbf{a}_{1}=\frac{a\sqrt{2}}{2}\mathbf{\alpha}_{1}\mathbf{,}$ &
$\mathbf{\mathbf{a}_{2}}=\frac{a\sqrt{2}}{2}\mathbf{\alpha}_{2}\mathbf{,}$ &
$\mathbf{a}_{3}=\frac{a\sqrt{2}}{2}\mathbf{\alpha}_{3}\mathbf{,}$ &
$\mathbf{\mathbf{a}_{4}=\frac{a\sqrt{2}}{2}\alpha}_{4}$%
\end{tabular}
\  \  \label{41}%
\end{equation}
where $a=d\sqrt{\frac{5}{2}}$ with $d$ standing for the distance between sites
and where $\mathbf{\alpha}_{1}\mathbf{,}$ $\mathbf{\alpha}_{2}\mathbf{,}$
$\mathbf{\alpha}_{3}\mathbf{,}$ $\mathbf{\alpha}_{4}$ are the \emph{4} simple
roots of $SU\left(  5\right)  $. The reciprocal space $\mathbb{\tilde{L}}_{4}$
of the hyperdiamond is generated by
\begin{equation}%
\begin{tabular}
[c]{llll}%
$\mathbf{b}_{1}=\frac{2\pi \sqrt{2}}{a}\mathbf{\omega}_{1}\mathbf{,}$ &
$\mathbf{\mathbf{b}_{2}}=\frac{2\pi \sqrt{2}}{a}\mathbf{\omega}_{2}\mathbf{,}$
& $\mathbf{b}_{3}=\frac{2\pi \sqrt{2}}{a}\mathbf{\omega}_{3}\mathbf{,}$ &
$\mathbf{\mathbf{b}_{4}=\frac{2\pi \sqrt{2}}{a}\omega}_{4}$%
\end{tabular}
\end{equation}
where $\mathbf{\omega}_{1}\mathbf{,}$ $\mathbf{\omega}_{2}\mathbf{,}$
$\mathbf{\omega}_{3}\mathbf{,}$ $\mathbf{\omega}_{4}$ are the \emph{4}
fundamental weight vectors of $SU\left(  5\right)  $. We also have%
\begin{equation}%
\begin{tabular}
[c]{ll}%
$\mathbf{\alpha}_{1}$ & $=2\mathbf{\omega}_{1}-\mathbf{\omega}_{2}$\\
$\mathbf{\alpha}_{2}$ & $=2\mathbf{\omega}_{2}-\mathbf{\omega}_{1}%
-\mathbf{\omega}_{2}$\\
$\mathbf{\alpha}_{3}$ & $=2\mathbf{\omega}_{3}-\mathbf{\omega}_{2}%
-\mathbf{\omega}_{4}$\\
$\mathbf{\alpha}_{4}$ & $=2\mathbf{\omega}_{4}-\mathbf{\omega}_{3}$%
\end{tabular}
\end{equation}

\begin{itemize}
\item \emph{Unit cell of }$\mathbb{L}_{4}$
\end{itemize}

\begin{figure}[ptbh]
\begin{center}
\hspace{0cm} \includegraphics[width=8cm]{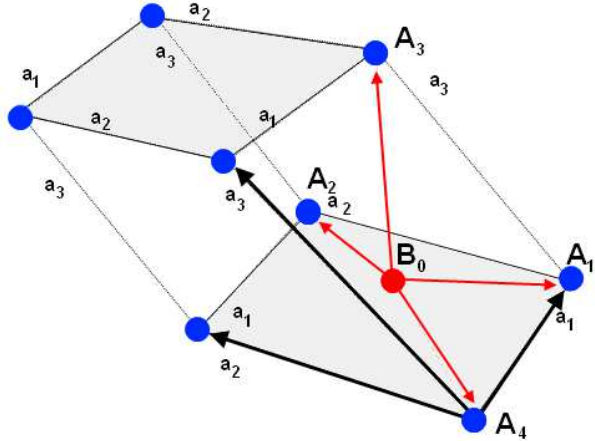}
\end{center}
\par
\vspace{-0.5cm}\caption{ 3-dimensional projection of the real unit cell of
hyperdiamond in the primitive frame. It is generated by the \textbf{4 }vectors
$\mathbf{a}_{1},\mathbf{a}_{2},\mathbf{a}_{3},$ $\mathbf{a}_{4}$ ( or up to
scale factor by the 4 simple roots $\mathbf{\alpha}_{1},\mathbf{\alpha}%
_{2},\mathbf{\alpha}_{3},$ $\mathbf{\alpha}_{4}$ of $SU\left(
{\protect \small 5}\right)  $). Sites $A_{i}$ (resp $B_{i}$) belong to
sublattice $\mathbb{A}$ (resp $\mathbb{B}$) of the hyperdiamond. B$_{0}$ has 5
nearest neighbors of A-type.}%
\label{UC}%
\end{figure}

A unit cell in the hyperdiamond is given by the fundamental region made by the
\emph{4} basis vectors. In P-frame the basis generators are given by the
vectors $\mathbf{a}_{i}$ and the hyper-volume reads as%
\begin{equation}
\mathbf{a}_{1}\wedge \mathbf{a}_{2}\wedge \mathbf{a}_{3}\wedge \mathbf{a}%
_{4}=\frac{a^{4}}{4}\det \left(  \alpha_{i}^{\mu}\right)  \label{hv}%
\end{equation}
Like in the case of the 2D honeycomb, a unit cell contains two particle sites:
one site inside the unit cell (\emph{the red ball in fig} \ref{UC}) and
\emph{5} first nearest neighbors (\emph{sites in blue in fig} \ref{UC})
contributing each by a fraction $\frac{1}{5}$ so that the total number is
$1+5\times \frac{1}{5}=2$. In the figure \ref{UC}, we give the projection of a
4D hyperdiamond unit cell on the 3D sub-space generated by $\left \{
\mathbf{a}_{1},\mathbf{a}_{2},\mathbf{a}_{3}\right \}  $; see also fig
\ref{PARA} for comparison with graphene. The first Brillouin zone of hyperdiamond is given by the unit cell
in the momentum space generated by the $\mathbf{b}_{i}$'s; its hyper-volume
is
\begin{equation}
\mathbf{b}_{1}\wedge \mathbf{b}_{2}\wedge \mathbf{b}_{3}\wedge \mathbf{b}%
_{4}=\frac{64\pi^{4}}{a^{4}}\det \left(  \omega_{\mu}^{i}\right)  ,
\end{equation}
and is equal to $\frac{\left(  2\pi \right)  ^{4}}{V_{uc}}$ with $V_{uc}$ the
volume of the real cell given by (\ref{hv}). Notice that the \emph{4} vectors
$\mathbf{b}_{i}$ (or up to a scale factor $\mathbf{\omega}^{i}$) that generate
the reciprocal space are dual to the $\mathbf{a}_{i}$ vectors (simple roots
$\mathbf{\alpha}_{i}$); they obey
\begin{equation}%
\begin{tabular}
[c]{llll}%
$\mathbf{b}_{i}.\mathbf{a}_{j}=2\pi \delta_{ij}$ & , & $\mathbf{\alpha}%
_{i}.\mathbf{\omega}^{j}=\alpha_{i}^{\mu}\omega_{\mu}^{j}=\delta_{i}^{j}$ & .
\end{tabular}
\end{equation}
and give a nice group theoretical interpretation of the relation between the
real and momentum spaces.

\begin{itemize}
\item \emph{More on hyperdiamond and} $SU\left(  5\right)  $
\end{itemize}

Hyperdiamond $\mathbb{L}_{4}$ is a 4-dimensional lattice contained in
$\mathbb{R}^{4}$; it can be thought of as made of the superposition of two
4-dimensional sublattices $\mathbb{A}$ and $\mathbb{B}$. Each site
$\mathbf{r}_{n}$ in $\mathbb{A}$ has $5$ nearest neighbors $\left(
\mathbf{r}_{n}+d\mathbf{e}_{i}\right)  $ belonging to $\mathbb{B},$
\[
\mathbf{r}_{n}\in \mathbb{A\rightarrow}\left \{
\begin{array}
[c]{c}%
\mathbf{r}_{n}+d\mathbf{e}_{1}\\
\mathbf{r}_{n}+d\mathbf{e}_{2}\\
\mathbf{r}_{n}+d\mathbf{e}_{3}\\
\mathbf{r}_{n}+d\mathbf{e}_{4}\\
\mathbf{r}_{n}+d\mathbf{e}_{5}%
\end{array}
\right.  \in \mathbb{B}%
\]
and \emph{20} second nearest ones $\mathbf{r}_{n}+d\left(  \mathbf{e}%
_{i}-\mathbf{e}_{j}\right)  $ belonging to $\mathbb{A}$. The 5 vectors
$\mathbf{e}_{i}$, giving the relative positions of the neighbors, obey the
traceless property
\begin{equation}
\mathbf{e}_{1}+\mathbf{e}_{2}+\mathbf{e}_{3}+\mathbf{e}_{4}+\mathbf{e}_{5}=0,
\end{equation}
capturing the hidden $SU\left(  5\right)  $ symmetry of the hyperdiamond.
\begin{figure}[ptbh]
\begin{center}
\hspace{0cm} \includegraphics[width=12cm]{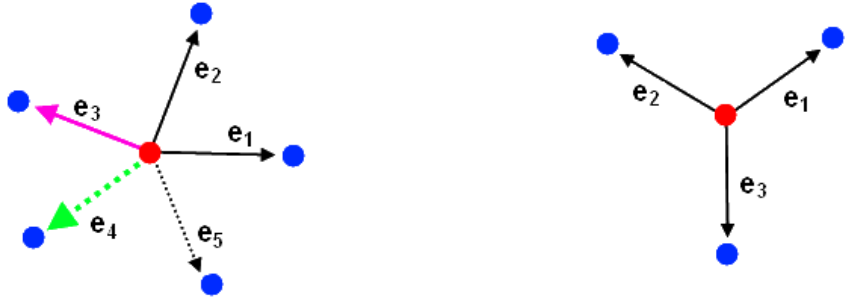}
\end{center}
\par
\vspace{-0.5 cm}\caption{On left the \emph{5} first nearest neighbors in the
\emph{4D} hyperdiamond with the property $\mathbf{e}_{1}+\mathbf{e}%
_{2}+\mathbf{e}_{3}+\mathbf{e}_{4}+\mathbf{e}_{5}=0$. On right, the \emph{3}
first nearest in \emph{2D} graphene with $\mathbf{e}_{1}+\mathbf{e}%
_{2}+\mathbf{e}_{3}=0$. }%
\label{5NE}%
\end{figure}To parameterize the site positions $\mathbf{r}_{n}$ of
$\mathbb{L}_{4}$, it is useful to use $\mathbf{e}_{1}\mathbf{,}$
$\mathbf{e}_{2}\mathbf{,}$ $\mathbf{e}_{3}\mathbf{,}$ $\mathbf{e}_{4}$ to
generate vectors in hyperdiamond. These \emph{4} basis vectors $\mathbf{e}%
_{i}$ have the following intersection matrix%
\begin{equation}%
\begin{tabular}
[c]{ll}%
$\mathbf{e}_{i}.\mathbf{e}_{j}=\left(
\begin{array}
[c]{cccc}%
1 & -\frac{1}{4} & 0 & 0\\
-\frac{1}{4} & 1 & -\frac{1}{4} & 0\\
0 & -\frac{1}{4} & 1 & -\frac{1}{4}\\
0 & 0 & -\frac{1}{4} & 1
\end{array}
\right)  $ & ,
\end{tabular}
\end{equation}
and can be realized as%
\begin{equation}%
\begin{tabular}
[c]{ll}%
$\mathbf{e}_{1}^{\mu}$ & $=\left(  \frac{\sqrt{5}}{4},\frac{\sqrt{5}}{4}%
,\frac{\sqrt{5}}{4},\frac{1}{4}\right)  $\\
$\mathbf{e}_{2}^{\mu}$ & $=\left(  \frac{\sqrt{5}}{4},\frac{-\sqrt{5}}%
{4},\frac{-\sqrt{5}}{4},\frac{1}{4}\right)  $\\
$\mathbf{e}_{3}^{\mu}$ & $=\left(  \frac{-\sqrt{5}}{4},\frac{-\sqrt{5}}%
{4},\frac{\sqrt{5}}{4},\frac{1}{4}\right)  $\\
$\mathbf{e}_{4}^{\mu}$ & $=\left(  \frac{-\sqrt{5}}{4},\frac{\sqrt{5}}%
{4},\frac{-\sqrt{5}}{4},\frac{1}{4}\right)  $\\
$\mathbf{e}_{5}^{\mu}$ & $=\left(  0,0,0,-1\right)  $\\
&
\end{tabular}
\  \  \  \  \  \label{Z2}%
\end{equation}
These \emph{5} vectors $\mathbf{e}_{1}\mathbf{,}$ $\mathbf{e}_{2}\mathbf{,}$
$\mathbf{e}_{3}\mathbf{,}$ $\mathbf{e}_{4},$ $\mathbf{e}_{5}$ define the first
nearest neighbors to $\left(  0,0,0,0\right)  $ as in fig \ref{5NE}; they
satisfy the relations $\mathbf{e}_{i}.\mathbf{e}_{i}=1$ and $\mathbf{e}%
_{i}.\mathbf{e}_{j}=-\frac{1}{4}$ $\  \forall$ $i\neq j$ showing that they are
distributed in a symmetric way as required by the hidden SU$\left(  5\right)
$ symmetry of $\mathbb{L}_{4}$. The \emph{20} second nearest neighbors are
given by
\begin{equation}%
\begin{tabular}
[c]{llll}%
$\mathbf{e}_{ij}=\mathbf{e}_{i}-\mathbf{e}_{j}$ & , & $\left(  \mathbf{e}%
_{ij}\right)  ^{2}=\frac{5}{2}$ & ,
\end{tabular}
\end{equation}
and are remarkably generated by the \emph{4} basis vectors%
\begin{equation}%
\begin{tabular}
[c]{lll}%
$\mathbf{\alpha}_{1}=\frac{2\sqrt{5}}{5}\left(  \mathbf{e}_{1}-\mathbf{e}%
_{2}\right)  $ & , & $\mathbf{\alpha}_{2}=\frac{2\sqrt{5}}{5}\left(
\mathbf{e}_{2}-\mathbf{e}_{3}\right)  $\\
$\mathbf{\alpha}_{3}=\frac{2\sqrt{5}}{5}\left(  \mathbf{e}_{3}-\mathbf{e}%
_{4}\right)  $ & , & $\mathbf{\alpha}_{4}=\frac{2\sqrt{5}}{5}\left(
\mathbf{e}_{4}-\mathbf{e}_{5}\right)  $%
\end{tabular}
\  \  \  \  \label{Z8}%
\end{equation}
By using (\ref{Z2}), we have%
\begin{equation}%
\begin{tabular}
[c]{lll}%
$\mathbf{\alpha}_{1}=\left(  0,1,1,0\right)  $ & , & $\mathbf{\alpha}%
_{2}=\left(  1,0,-1,0\right)  $\\
$\mathbf{\alpha}_{3}=\left(  0,-1,1,0\right)  $ & , & $\mathbf{\alpha}%
_{4}=\left(  \frac{-1}{2},\frac{1}{2},\frac{-1}{2},\frac{\sqrt{5}}{2}\right)
.$%
\end{tabular}
\  \  \  \  \label{SR}%
\end{equation}
from which we learn that $\mathbf{\alpha}_{i}^{2}=2$ and $\det \left(
\mathbf{\alpha}_{1},\mathbf{\alpha}_{2},\mathbf{\alpha}_{3},\mathbf{\alpha
}_{4}\right)  =\sqrt{5}$. We also have the intersection matrix $C_{ij}%
=\mathbf{\alpha}_{i}.\mathbf{\alpha}_{j}$%
\begin{equation}%
\begin{tabular}
[c]{ll}%
$C_{ij}=\left(
\begin{array}
[c]{cccc}%
2 & -1 & 0 & 0\\
-1 & 2 & -1 & 0\\
0 & -1 & 2 & -1\\
0 & 0 & -1 & 2
\end{array}
\right)  $ & ,
\end{tabular}
\end{equation}
indicating that the $\mathbf{\alpha}_{i}$'s are nothing but the \emph{4
}simple roots of $SU\left(  5\right)  $ and $C_{ij}$ the corresponding Cartan
matrix with inverse $G^{ij}$ as%
\begin{equation}%
\begin{tabular}
[c]{ll}%
$G^{ij}=\left(
\begin{array}
[c]{cccc}%
\frac{4}{5} & \frac{3}{5} & \frac{2}{5} & \frac{1}{5}\\
\frac{3}{5} & \frac{6}{5} & \frac{4}{5} & \frac{2}{5}\\
\frac{2}{5} & \frac{4}{5} & \frac{6}{5} & \frac{3}{5}\\
\frac{1}{5} & \frac{2}{5} & \frac{3}{5} & \frac{4}{5}%
\end{array}
\right)  $ & .
\end{tabular}
\end{equation}
This matrix $G^{ij}$ is just the intersection matrix of the fundamental weight
vectors $\mathbf{\omega}^{i}$ of SU$\left(  5\right)  $
\begin{equation}%
\begin{tabular}
[c]{ll}%
$G^{ij}=\mathbf{\omega}^{i}.\mathbf{\omega}^{j}$ & ,
\end{tabular}
\end{equation}
with
\begin{equation}%
\begin{tabular}
[c]{llll}%
$\left \Vert \mathbf{\omega}^{1}\right \Vert ^{2}=\left \Vert \mathbf{\omega}%
^{4}\right \Vert ^{2}=\frac{4}{5}$ & , & $\left \Vert \mathbf{\omega}%
^{1}\right \Vert ^{2}=\left \Vert \mathbf{\omega}^{4}\right \Vert ^{2}=\frac
{6}{5}$ & .
\end{tabular}
\end{equation}
The $\mathbf{\omega}^{i}$'s span the momentum space of the hyperdiamond; their
components in the canonical basis are
\begin{equation}%
\begin{tabular}
[c]{lll}%
$\mathbf{\omega}^{1}=\left(  \frac{1}{2},\frac{1}{2},\frac{1}{2},\frac
{\sqrt{5}}{10}\right)  $ & $,$ & $\mathbf{\omega}^{2}=\left(  1,0,0,\frac
{\sqrt{5}}{5}\right)  $\\
$\mathbf{\omega}^{3}=\left(  \frac{1}{2},\frac{-1}{2},\frac{1}{2},\frac
{3\sqrt{5}}{10}\right)  $ & $,$ & $\mathbf{\omega}^{4}=\left(  0,0,0,\frac
{2\sqrt{5}}{5}\right)  $%
\end{tabular}
\end{equation}
with $\det \left(  \mathbf{\omega}^{1},\mathbf{\omega}^{2},\mathbf{\omega}%
^{3},\mathbf{\omega}^{4}\right)  =-\frac{1}{5}\sqrt{5}$. Therefore, generic
positions $\mathbf{r}$ of $\mathbb{R}^{4}$ and generic wave vectors
$\mathbf{k}$ of the 4-dimensional momentum space decompose as follows,%
\begin{equation}%
\begin{tabular}
[c]{lll}%
$\mathbf{r}=%
{\displaystyle \sum \limits_{i=1}^{4}}
x^{i}\mathbf{\alpha}_{i}$ & , & $\  \  \ x^{i}=\mathbf{\omega}^{i}.\mathbf{r}$\\
$\mathbf{k}=%
{\displaystyle \sum \limits_{i=1}^{4}}
k_{i}\mathbf{\omega}^{i}$ & , & $\  \  \ k_{i}=\mathbf{\alpha}_{i}.\mathbf{k}$%
\end{tabular}
\end{equation}
With these tools, we can determine several features on the crystallographic
structure of the hyperdiamond; for instance the volume $V_{uc}$ of the real
unit cell is $V_{uc}=\frac{a^{4}}{4}\sqrt{5}$;\ and the volume of first
Brillouin zone $V_{BZ}$ reads as $V_{BZ}=\frac{64\pi^{4}}{5a^{4}}\sqrt{5}$.
Recall also that the SU$\left(  5\right)  $ Lie algebra has \emph{10}
\emph{positive} roots and \emph{10} \emph{negative }ones (opposites); the
positive ones consists of the \emph{4} above simple and the \emph{6} following%
\begin{equation}%
\begin{tabular}
[c]{lll}%
$\mathbf{\alpha}_{1}+\mathbf{\alpha}_{2},$ &  & $\mathbf{\alpha}%
_{1}+\mathbf{\alpha}_{2}+\mathbf{\alpha}_{3},$\\
$\mathbf{\alpha}_{2}+\mathbf{\alpha}_{3},$ &  & $\mathbf{\alpha}%
_{2}+\mathbf{\alpha}_{3}+\mathbf{\alpha}_{4},$\\
$\mathbf{\alpha}_{3}+\mathbf{\alpha}_{4},$ &  & $\mathbf{\alpha}%
_{1}+\mathbf{\alpha}_{2}+\mathbf{\alpha}_{3}+\mathbf{\alpha}_{4}$%
\end{tabular}
\end{equation}
Notice moreover that being also 4-dimensional vectors, the $\mathbf{\omega
}^{i}$'s may be decomposed as well like a sum over simple roots
\begin{equation}%
\begin{tabular}
[c]{ll}%
$\mathbf{\omega}_{1}=\frac{4}{5}\mathbf{\alpha}_{1}+\frac{3}{5}\mathbf{\alpha
}_{2}+\frac{2}{5}\mathbf{\alpha}_{3}+\frac{1}{5}\mathbf{\alpha}_{4}\ $ & \\
$\mathbf{\omega}_{2}=\frac{3}{5}\mathbf{\alpha}_{1}+\frac{6}{5}\mathbf{\alpha
}_{2}+\frac{4}{5}\mathbf{\alpha}_{3}+\frac{2}{5}\mathbf{\alpha}_{4}\ $ & \\
$\mathbf{\omega}_{3}=\frac{2}{5}\mathbf{\alpha}_{1}+\frac{4}{5}\mathbf{\alpha
}_{2}+\frac{6}{5}\mathbf{\alpha}_{3}+\frac{3}{5}\mathbf{\alpha}_{4}\ $ & \\
$\mathbf{\omega}_{4}=\frac{1}{5}\mathbf{\alpha}_{1}+\frac{2}{5}\mathbf{\alpha
}_{2}+\frac{3}{5}\mathbf{\alpha}_{3}+\frac{4}{5}\mathbf{\alpha}_{4}\ $ &
\end{tabular}
\  \  \  \  \label{120}%
\end{equation}
offering a unified way to parameterize the hyperdiamond $\mathbb{L}_{4}$; its
sublattices $\mathbb{A}$ and $\mathbb{B}$ and their reciprocal spaces.

\subsection{Supercell compactification}

First we study supercell $C_{\text{supercell}}$ compactification in
hyperdiamond; then we move to deal with the boundary conditions of fields on
supercell $C_{\text{supercell}}\equiv SC_{4}$.

\subsubsection{p-cycles in supercell and 4-torus homology}

The supercell $SC_{4}$ in hyperdiamond is 4-dimensional extensions of fig
\ref{PARA}; and is described by a hyper-parallelogram $\mathcal{P}_{4}$ with
periodic boundary conditions. Positions in supercell are parameterized by the
vectors $\mathbf{r=}\sum_{i}x^{i}\frac{\sqrt{2}}{2}\mathbf{\alpha}_{i}$ with
the real coordinates $x^{i}=\left(  x,y,z,\tau \right)  $ compactified as
\begin{equation}%
\begin{tabular}
[c]{llllll}%
$0\leq x^{i}<l_{i}$ & , & $l_{i}=aN_{i}$ & $,$ & $x^{i}\equiv x^{i}+l_{i}$ & .
\end{tabular}
\  \  \label{SC}%
\end{equation}
The 4D-parallelogram $\mathcal{P}_{4}$ is a real compact 4-cycle with the
homology class of a 4-torus $\mathbb{T}^{4}$; it has the following features:

\begin{itemize}
\item $\mathcal{P}_{4}$ has \emph{4} basis divisors $\Delta_{i}$ and \emph{4}
dual 1-cycles $E_{i}$ associated with the hyper-planes $x^{i}=0$ in
$\mathcal{P}_{4}$. These are:%
\[%
\begin{tabular}
[c]{l|l|l}
& \ 3-cycles \  \  & \ duals $=$ 1-cycles \  \  \\ \hline
$x=0$ \  \  & $\  \  \  \Delta_{x}$ & $\  \  \  \tilde{\Delta}_{x}=E_{x}$\\
$y=0$ & $\  \  \  \Delta_{y}$ & $\  \  \  \tilde{\Delta}_{y}=E_{y}$\\
$z=0$ & $\  \  \  \Delta_{z}$ & $\  \  \  \tilde{\Delta}_{z}=E_{z}$\\
$\tau=0$ & $\  \  \  \Delta_{\tau}$ & $\  \  \  \tilde{\Delta}_{\tau}=E_{\tau}%
$\\ \hline
\end{tabular}
\  \
\]
where $\Delta_{i}$ have the homology of a 3-torus $\mathbb{T}^{3}$ and $E_{i}
$ of a circle $\mathbb{S}^{1}$. Positions in $E_{1}$ and in $\Delta_{1} $ are
given by the vectors
\begin{equation}%
\begin{tabular}
[c]{llll}%
$E_{1}$ & : & $\mathbf{r}=x$ $\frac{\sqrt{2}}{2}\mathbf{\alpha}_{1}$ & \\
$\Delta_{1}$ & : & $\mathbf{r}=y$ $\frac{\sqrt{2}}{2}\mathbf{\alpha}_{2}+z$
$\frac{\sqrt{2}}{2}\mathbf{\alpha}_{3}+\tau$ $\frac{\sqrt{2}}{2}%
\mathbf{\alpha}_{4}$ &
\end{tabular}
\end{equation}
with the corresponding volume forms:%
\begin{equation}%
\begin{tabular}
[c]{llll}%
$E_{1}$ & : & $dx$ $\frac{\sqrt{2}}{2}\mathbf{\alpha}_{1}$ & \\
$\Delta_{1}$ & : & $\  \frac{\sqrt{2}}{4}dydzd\tau$ $\mathbf{\alpha}_{2}%
\wedge \mathbf{\alpha}_{3}\wedge \mathbf{\alpha}_{4}$ &
\end{tabular}
\end{equation}
Similar relations can be written down for the others $E_{i}$'s and $\Delta
_{i}$'s.

\item $\mathcal{P}_{4}$ has also six 2-cycles $C_{ij}$ and six dual ones
$\tilde{C}_{ij}$ given by 2D-parallelograms.%
\begin{equation}%
\begin{tabular}
[c]{l|l|l}
& \ 2-cycles \  & \ dual 2-cycles \  \\ \hline
$x=y=0$ $\  \  \ $ & $\  \  \ C_{xy}$ & $\  \  \  \tilde{C}_{xy}=C_{z\tau}$\\
$x=z=0$ & $\  \  \ C_{xz}$ & $\  \  \  \tilde{C}_{xz}=C_{y\tau}$\\
$x=\tau=0$ & $\  \  \ C_{x\tau}$ & $\  \  \  \tilde{C}_{x\tau}=C_{yz}$\\
$y=z=0$ & $\  \  \ C_{yz}$ & $\  \  \  \tilde{C}_{yz}=C_{x\tau}$\\
$y=\tau=0$ & $\  \  \ C_{y\tau}$ & $\  \  \  \tilde{C}_{y\tau}=C_{xz}$\\
$z=\tau=0$ & $\  \  \ C_{z\tau}$ & $\  \  \  \tilde{C}_{z\tau}=C_{xy}$\\ \hline
\end{tabular}
\end{equation}
These 2-cycles have the homology of 2-torii $\mathbb{T}^{2}$; they will be
used later on to compute the fluxes of the background field components going
through $C_{ij}$. These 2-cycles are needed for the derivation of the
topological index of the Dirac operator. Positions in these cycles are as
\begin{equation}%
\begin{tabular}
[c]{llll}%
$C_{12}$ & : & $\mathbf{r}=x\  \frac{\sqrt{2}}{2}\mathbf{\alpha}_{1}%
+y\  \frac{\sqrt{2}}{2}\mathbf{\alpha}_{2}$ & \\
$\tilde{C}_{12}$ & : & $\mathbf{r}=z\  \frac{\sqrt{2}}{2}\mathbf{\alpha}%
_{3}+\tau \  \frac{\sqrt{2}}{2}\mathbf{\alpha}_{4}$ &
\end{tabular}
\end{equation}
with volume forms like%
\begin{equation}%
\begin{tabular}
[c]{llll}%
$C_{12}$ & : & $\frac{1}{2}dxdy$ $\mathbf{\alpha}_{1}\wedge \mathbf{\alpha}%
_{2}$ & \\
$\Delta_{1}$ & : & $\frac{1}{2}dzd\tau$ $\mathbf{\alpha}_{2}\wedge
\mathbf{\alpha}_{3}\wedge \mathbf{\alpha}_{4}$ &
\end{tabular}
\end{equation}
Similar relations can be written down for the others.

\item the 4D-parallelogram $\mathcal{P}_{4}$ has $2^{4}$\ vertices, they read
in the P-frame as follows
\begin{equation}%
\begin{tabular}
[c]{llll}%
$\left(  l_{1}\epsilon_{1},\text{ }l_{2}\epsilon_{2},\text{ }l_{3}\epsilon
_{3},l_{4}\epsilon_{4}\right)  $ & , & $\epsilon_{i}=0,1$ & .
\end{tabular}
\end{equation}
with the $l_{i}$'s as in (\ref{SC}).
\end{itemize}

\  \  \newline Notice that as a 4-cycle, the 4D-parallelogram $\mathcal{P}_{4}$
and its p-cycles are generated by the 1-cycle basis $E_{1}$, $E_{2}$, $E_{3}$,
$E_{4} $. Notice also that hyperdiamond sites in $SC_{4}$ are given by
discrete position vectors $\mathbf{r}_{\mathbf{n}}$ belonging to two classes%
\[
\left \{  \mathbf{r}_{\mathbf{n}}^{\mathbb{A}}%
\begin{tabular}
[c]{l}%
\\
\end{tabular}
\right \}  \cup \left \{  \mathbf{r}_{\mathbf{n}}^{\mathbb{B}}%
\begin{tabular}
[c]{l}%
\\
\end{tabular}
\right \}
\]
respectively associated with the sublattice $\mathbb{A}$ and the sublattice
$\mathbb{B}$. Sites in $\mathbb{A}$ are given by the integer vectors
\begin{equation}%
\begin{tabular}
[c]{lll}%
$\mathbf{r}_{\mathbf{n}}^{\mathbb{A}}=n_{1}\mathbf{a}_{1}+n_{2}\mathbf{a}%
_{2}+n_{3}\mathbf{a}_{3}+n_{4}\mathbf{a}_{4}$ & , & $0\leq n_{i}\leq N_{i}-1$%
\end{tabular}
\end{equation}
with $\mathbf{a}_{i}$\ as in (\ref{41}) and $\mathbf{n}=\left(  n_{1}%
,n_{2},n_{3},n_{4}\right)  $. The sites of the sublattice $\mathbb{B}$ are
given by the globally shifted ones,
\begin{equation}%
\begin{tabular}
[c]{ll}%
$\mathbf{r}_{\mathbf{n}}^{\mathbb{B}}$ & $=\left(  n_{1}+\frac{1}{5}\right)
\mathbf{a}_{1}+\left(  n_{2}+\frac{2}{5}\right)  \mathbf{a}_{2}+\left(
n_{3}+\frac{3}{5}\right)  \mathbf{a}_{3}+\left(  n_{4}+\frac{4}{5}\right)
\mathbf{a}_{4}$%
\end{tabular}
\label{BA}%
\end{equation}
with shift vector $\mathbf{s}=\mathbf{r}_{n}^{\mathbb{B}}-\mathbf{r}%
_{n}^{\mathbb{A}}$ given by%
\begin{equation}%
\begin{tabular}
[c]{lll}%
$\mathbf{s}$ & $=\frac{1}{5}\mathbf{a}_{1}+\frac{2}{5}\mathbf{a}_{2}+\frac
{3}{5}\mathbf{a}_{3}+\frac{4}{5}\mathbf{a}_{4}$ & \\
& $=d\sqrt{\frac{5}{2}}\mathbf{\omega}^{4}$ &
\end{tabular}
\end{equation}
By substituting the vectors $\mathbf{a}_{i}$ by their expressions (\ref{SR}),
we obtain%
\begin{equation}
\mathbf{s}=\left(
\begin{array}
[c]{c}%
0\\
0\\
0\\
d
\end{array}
\right)
\end{equation}
Notice also that the total number of sites in the supercell (\ref{SC}) is
$2N_{1}N_{2}N_{3}N_{4}$; half of them belong to the sublattice $\mathbb{A}$
and the other half to $\mathbb{B}$.

\subsubsection{Boundary conditions}

To recover the hyperdiamond lattice from a given supercell, one uses periodic
boundary conditions on its borders as
\begin{equation}%
\begin{tabular}
[c]{lllll}%
$\mathbf{r}$ & $\equiv \mathbf{r}+N_{1}\mathbf{a}_{1}$ & , & $\mathbf{r}$ &
$\equiv \mathbf{r}+N_{3}\mathbf{a}_{3}$\\
$\mathbf{r}$ & $\equiv \mathbf{r}+N_{2}\mathbf{a}_{2}$ & , & $\mathbf{r}$ &
$\equiv \mathbf{r}+N_{4}\mathbf{a}_{4}$%
\end{tabular}
\end{equation}
or equivalently
\begin{equation}
\mathbf{r}\equiv \mathbf{r}+l_{i}\frac{\sqrt{2}}{2}\mathbf{\alpha}_{i}.
\end{equation}
To get the right spectrum of the Dirac operator, one has also to worry about
the boundary conditions on the gauge field $\mathcal{A}_{\mu}\left(
\mathbf{r}\right)  $ and the fermionic waves $\Psi \left(  \mathbf{r}\right)  $
living on supercell. Generally these boundary conditions close up to a gauge
transformation as follows%
\begin{equation}%
\begin{tabular}
[c]{lll}%
\textrm{F}$\left(  \mathbf{r}+\frac{l_{i}\sqrt{2}}{2}\mathbf{\alpha}%
_{i}\right)  $ & $=$ & \textrm{F}$^{\Omega_{i}}\left(  \mathbf{r}\right)  $%
\end{tabular}
\  \  \label{CD}%
\end{equation}
where \textrm{F }stands for $\mathcal{A}_{\mu}$, $\Psi$ and where $\Omega_{i}
$ some particular gauge transformation along the $\mathbf{\alpha}_{i}$-
direction. To deal with the conditions (\ref{CD}), we extend the trick of
\emph{Smit and Vink }\textrm{\cite{B2}} that we have developed in section 3 by
taking the gauge fields as
\begin{equation}%
\begin{tabular}
[c]{llll}%
$\mathcal{A}_{x}\left(  \mathbf{r}\right)  =0$ & , & $\mathcal{A}_{y}\left(
\mathbf{r}\right)  =-\mathcal{B}x$ & \\
$\mathcal{A}_{z}\left(  \mathbf{r}\right)  =0$ & , & $\mathcal{A}_{\tau
}\left(  \mathbf{r}\right)  =-\mathcal{E}z$ &
\end{tabular}
\  \  \label{GP}%
\end{equation}
with no dependence in the $y$ nor $\tau$\ variables. This particular gauge
choice allows amongst others the two following helpful features:\newline%
\textbf{(i)} it permits to use gauge invariance $\mathcal{A}_{i}^{\Omega
}\left(  \mathbf{r}\right)  =\mathcal{A}_{i}\left(  \mathbf{r}\right)
+\partial_{i}\lambda$ to ensure the periodicity of the components of the gauge
potential. This leads to a strong constraint on the y- and $\tau$- dependence
into the gauge parameter $\lambda$ which are no longer arbitrary since they
have to take the form
\begin{equation}
\lambda \left(  \mathbf{r}\right)  =\mathcal{B}l_{1}y+\mathcal{E}l_{3}%
\tau+\varphi \left(  x,z\right)  ,\qquad \Omega \left(  \mathbf{r}\right)
=e^{i\lambda \left(  \mathbf{r}\right)  }%
\end{equation}
with some integration\ function $\varphi \left(  x,z\right)  $ that we drop out
below as it is not of great importance for the following analysis. Moreover
periodicity of the gauge group element on supercell which has to obey
$\Omega \left(  \mathbf{r}+\frac{l_{i}\sqrt{2}}{2}\mathbf{\alpha}_{i}\right)
=\Omega \left(  \mathbf{r}\right)  $ leads to the quantization of the
background fields as follows%
\begin{equation}%
\begin{tabular}
[c]{lllll}%
$\mathcal{B}l_{1}l_{2}$ & $=2\pi Q_{B}$ & , & $Q_{B}\in \mathbb{Z}$ & \\
$\mathcal{E}l_{3}l_{4}$ & $=2\pi Q_{E}$ & , & $Q_{E}\in \mathbb{Z}$ &
\end{tabular}
\end{equation}
\textbf{(ii)} it permits also to put the boundary condition on the fermionic
waves into two classes: the two trivial ones%
\begin{equation}%
\begin{tabular}
[c]{ll}%
$\Psi \left(  x,y+l_{2},z,\tau \right)  $ & $=\Psi \left(  x,y,\tau,z\right)
\ $\\
$\Psi \left(  x,y,z,\tau+l_{4}\right)  $ & $=\Psi \left(  x,y,\tau,z\right)  \ $%
\end{tabular}
\  \  \label{F}%
\end{equation}
easy to deal with; and the two less trivial ones%
\begin{equation}%
\begin{tabular}
[c]{ll}%
$\Psi \left(  x+l_{1},y,z,\tau \right)  $ & $=\Psi^{\Omega_{1}}\left(
x,y,\tau,z\right)  \ $\\
$\Psi \left(  x,y,z+l_{3},\tau \right)  $ & $=\Psi^{\Omega_{3}}\left(
x,y,\tau,z\right)  \ $%
\end{tabular}
\end{equation}
where $\Omega_{1}$ and $\Omega_{3}$ are particular gauge transformations to be
given later on. \newline Eq(\ref{F}) allows to expand $\Psi \left(
x,y,z,\tau \right)  $ in Fourier modes as follows
\begin{equation}
\Psi \left(  x,y,z,\tau \right)  =%
{\displaystyle \sum \limits_{r\in \mathbb{Z}}}
e^{i\frac{2\pi r}{L_{2}}y}%
{\displaystyle \sum \limits_{s\in \mathbb{Z}}}
e^{i\frac{2\pi s}{l_{4}}\tau}\text{ }\tilde{\Psi}_{r,s}\left(  x,z\right)
\label{EX}%
\end{equation}
restricting the problem of computing wave functions to looking for the Fourier
modes $\tilde{\Psi}_{r,s}\left(  x,z\right)  $ whose dependence in x and z
variables of $\tilde{\Psi}_{r,s}\left(  x,z\right)  $ can be obtained by
solving the Dirac equation that we study below.

\section{Spectrum of Dirac operator}

Here, we compute the spectrum of the fermions on hyperdiamond. First we
consider the case of the continuum limit (unit cell). This is interesting to
compare with the case of lattice fermions using supercell compactification to
be considered later on.

\subsection{Dirac equation coupled to $\mathcal{F}_{\mu \nu} $: continuum case}

The Dirac equation describing the dynamics of four component Dirac fermions
$\Psi_{\alpha}^{a}\left(  \mathbf{x}\right)  $, with $SU_{C}\left(  3\right)
$ color symmetry in presence of background vector potentials $\mathcal{A}%
_{\mu}^{G}\left(  \mathbf{x}\right)  $, is given by,%
\begin{equation}
\sum_{\mu=1}^{4}\sum_{\beta=1}^{4}\sum_{b=1}^{3}i\left(  \gamma^{\mu}\right)
_{\alpha}^{\beta}\left[  \delta_{ab}\partial_{\mu}-i\left(  \mathcal{A}_{\mu
}^{G}\right)  _{ab}\right]  \Psi_{\beta}^{b}=\epsilon \Psi_{\alpha}^{b}
\label{S3}%
\end{equation}
where
\[
\Psi_{\alpha}^{a}=\left(
\begin{array}
[c]{c}%
\Psi_{\alpha}^{1}\\
\Psi_{\alpha}^{2}\\
\Psi_{\alpha}^{3}%
\end{array}
\right)  ,
\]
and%
\[%
\begin{tabular}
[c]{llll}%
$\Psi_{\alpha}^{a}=\left(
\begin{array}
[c]{c}%
\Psi_{L}^{a}\\
\Psi_{R}^{a}%
\end{array}
\right)  ,$ & $\Psi_{L}^{a}=\left(
\begin{array}
[c]{c}%
\psi_{1}^{a}\\
\psi_{2}^{a}%
\end{array}
\right)  ,$ & $\Psi_{R}^{a}=\left(
\begin{array}
[c]{c}%
\bar{\chi}_{1}^{a}\\
\bar{\chi}_{2}^{a}%
\end{array}
\right)  $ &
\end{tabular}
\
\]
and where $\mathcal{A}_{\mu}^{G}=\mathcal{A}_{\mu}^{U_{em}\left(  1\right)
}+\mathcal{A}_{\mu}^{SU_{c}\left(  3\right)  }$ with
\begin{equation}
\mathcal{A}_{\mu}^{U_{em}\left(  1\right)  }=\mathcal{A}_{\mu}^{em}%
Q_{em}\text{, \qquad}\mathcal{A}_{\mu}^{SU_{c}\left(  3\right)  }=%
{\displaystyle \sum \limits_{I=1}^{8}}
\mathcal{A}_{\mu}^{I}\mathcal{T}_{I}%
\end{equation}
with the $3\times3$ matrices $\mathcal{T}_{I}$ standing for the generators of
$SU_{c}\left(  3\right)  $. Below, we focus on the particular case where
\begin{equation}%
\begin{tabular}
[c]{llll}%
$\mathcal{A}_{\mu}^{U_{em}\left(  1\right)  }\neq0$ & , & $\mathcal{A}_{\mu
}^{SU_{c}\left(  3\right)  }=0$ & ,
\end{tabular}
\end{equation}
but keep in mind that such analysis can be straightforwardly extended to
include the other abelian components $\mathcal{A}_{\mu}^{SU_{c}\left(
3\right)  }\neq0$ leading to a non zero uniform field strength $\mathcal{F}%
_{\mu \nu}^{SU_{c}\left(  3\right)  }$ along the Cartan directions in the
$SU_{C}\left(  3\right)  $ Lie algebra. Notice that in (\ref{S3}) the number
$\epsilon$ captures deformations of the energy spectrum around the Dirac point
where live zero modes of QCD$_{4}$ fermions on lattice. Notice also that here
we have ignored the flavor symmetry which is associated with the zero modes in
the tight binding description of QCD on lattice; for details on implementation
of flavor symmetry in the case minimally doubled fermions see
\textrm{\cite{PSM,5A}}.

\begin{itemize}
\item \emph{Dirac equation in continuum}
\end{itemize}

By setting $A_{\mu}^{SU_{c}\left(  3\right)  }=0$ in eq(\ref{S3}), the Dirac
equation in the \emph{canonical frame} of the euclidian space $\mathbb{R}^{4}
$ reduces to its usual form,%
\begin{equation}
i\gamma^{\mu}\left(  \partial_{\mu}-iA_{\mu}\right)  \Psi=\epsilon \Psi.
\label{DIR}%
\end{equation}
In this equation, the $4\times4$ matrices $\gamma^{\mu}$ are euclidian Dirac
matrices obeying the usual \emph{4D} Clifford algebra $\gamma^{\mu}\gamma
^{\nu}+\gamma^{\nu}\gamma^{\mu}=2\delta^{\mu \nu}I_{4}$ with $I_{4}$ the
$4\times4$ identity matrix. These $\gamma^{\mu}$ matrices can be built by
using the Pauli ones $\tau^{i}$ and $\sigma^{i}$ of the group product
$SU_{L}\left(  2\right)  \times SU_{R}\left(  2\right)  \simeq SO\left(
4\right)  $,%
\begin{equation}%
\begin{tabular}
[c]{lll}%
$\varrho^{1}=\left(
\begin{array}
[c]{cc}%
0 & 1\\
1 & 0
\end{array}
\right)  ,$ & $\varrho^{2}=\left(
\begin{array}
[c]{cc}%
0 & -i\\
i & 0
\end{array}
\right)  ,$ & $\varrho^{3}=\left(
\begin{array}
[c]{cc}%
1 & 0\\
0 & -1
\end{array}
\right)  ,$%
\end{tabular}
\end{equation}
with $\varrho^{i}$ standing for both $\tau^{i}$ and $\sigma^{i}$. The
$\varrho^{1}$, $\varrho^{2}$ satisfy the 2D Clifford algebra $\varrho
^{i}\varrho^{j}+\varrho^{j}\varrho^{i}=2\delta^{ij}I_{2}$ and the $SU\left(
2\right)  $ symmetry bracket $\left[  \varrho^{1},\varrho^{2}\right]
=2i\varrho^{3}$. We have%
\begin{equation}%
\begin{tabular}
[c]{lll}%
$\gamma^{i}=\tau^{2}\otimes \sigma^{i}$ & , & $\gamma^{4}=\tau^{1}\otimes
\sigma^{4}$\\
$\gamma^{5}=\tau^{3}\otimes \sigma^{4}$ & , & $\gamma^{0}=\tau^{4}\otimes
\sigma^{4}$%
\end{tabular}
\  \  \  \label{GAM}%
\end{equation}
with $\tau^{4}$, $\sigma^{4}=I_{2}\equiv I$ and $\gamma^{5}=\gamma^{1}%
\gamma^{2}\gamma^{3}\gamma^{4}$. More explicitly
\begin{equation}%
\begin{tabular}
[c]{lll}%
$\gamma^{k}{\small =}\left(
\begin{array}
[c]{cc}%
{\small 0} & {\small -i\sigma}^{k}\\
{\small i\sigma}^{k} & {\small 0}%
\end{array}
\right)  $, & $\gamma^{4}{\small =}\left(
\begin{array}
[c]{cc}%
{\small 0} & {\small I}\\
{\small I} & {\small 0}%
\end{array}
\right)  ,$ & $\Upsilon^{0}{\small =}\left(
\begin{array}
[c]{cc}%
{\small I} & {\small 0}\\
{\small 0} & {\small I}%
\end{array}
\right)  $%
\end{tabular}
\end{equation}
and%
\begin{equation}%
\begin{tabular}
[c]{lll}%
$\gamma^{5}=\left(
\begin{array}
[c]{cccc}%
1 & 0 & 0 & 0\\
0 & 1 & 0 & 0\\
0 & 0 & -1 & 0\\
0 & 0 & 0 & -1
\end{array}
\right)  $ & , & $\tau^{3}\otimes \sigma^{3}=\left(
\begin{array}
[c]{cccc}%
1 & 0 & 0 & 0\\
0 & -1 & 0 & 0\\
0 & 0 & -1 & 0\\
0 & 0 & 0 & 1
\end{array}
\right)  $%
\end{tabular}
\end{equation}
The commutators $\gamma^{\left[  \mu \nu \right]  }$ give precisely the \emph{6}
generators of the spinorial representation of the $SO\left(  4\right)  $
symmetry,%
\begin{equation}%
\begin{tabular}
[c]{lll}%
$\gamma^{i}\gamma^{j}-\gamma^{j}\gamma^{i}$ & $=2i\varepsilon^{ijk}\left(
\tau^{2}\otimes \sigma^{k}\right)  $ & ,\\
$\gamma^{4}\gamma^{i}-\gamma^{i}\gamma^{4}$ & $=2i\varepsilon^{123}\left(
\tau^{3}\otimes \sigma^{i}\right)  $ & ,
\end{tabular}
\end{equation}
where $\varepsilon^{ijk}$ is the completely antisymmetric \emph{3D}
Levi-Civita tensor.\newline The Dirac operator (\ref{DIR}) involves also the
gauge potential $A_{\mu}$ defined up to the gauge symmetry transformations
\[%
\begin{tabular}
[c]{llll}%
$\Psi=e^{-i\lambda}\Psi$ & , & $A_{\mu}^{\Omega}\equiv A_{\mu}+\partial_{\mu
}\lambda$ & ,
\end{tabular}
\  \  \
\]
with $\lambda \left(  \mathbf{x}\right)  $ arbitrary gauge parameter. Since the
background field strength $F_{\mu \nu}=\partial_{\mu}A_{\nu}-\partial_{\nu
}A_{\mu}$ is a constant, the gauge potential $A_{\mu}$ reads as follows,
\begin{equation}%
\begin{tabular}
[c]{lll}%
$A_{\mu}=\frac{1}{2}F_{\mu \nu}x^{\nu}$ & , & $A_{\mu}^{\Omega}\equiv A_{\mu
}+\partial_{\mu}\lambda$%
\end{tabular}
\  \  \  \label{AM}%
\end{equation}
Notice that generally $F_{\mu \nu}$ has six real degrees of freedom, \emph{3}
magnetic $B_{i}$ and \emph{3} electric $E_{i}$:
\begin{equation}
F_{\mu \nu}=\left(
\begin{array}
[c]{cccc}%
0 & +B_{3} & -B_{2} & +E_{1}\\
-B_{3} & 0 & +B_{1} & +E_{2}\\
+B_{2} & -B_{1} & 0 & +E_{3}\\
-E_{1} & -E_{2} & -E_{3} & 0
\end{array}
\right)  . \label{GEN}%
\end{equation}

\begin{itemize}
\item \emph{Choice of the background fields}
\end{itemize}

Below we will think about this $F_{\mu \nu}$ as a sub-matrix of the following
$5\times5$ antisymmetric one,%
\begin{equation}
F_{MN}=\left(
\begin{array}
[c]{cc}%
F_{\mu \nu} & F_{\mu5}\\
F_{5\nu} & 0
\end{array}
\right)  ,
\end{equation}
where now the $F_{\mu5}$'s are the \emph{4} components of the electric field
in $5D$; and $F_{\mu \nu}$ are the \emph{6} components of the magnetic tensor.
We also make the two following useful choices:\newline \textbf{(i)} we restrict
the matrix $F_{\mu \nu}$ to the particular case,
\begin{equation}
F_{\mu \nu}=\left(
\begin{array}
[c]{cccc}%
0 & -B & 0 & 0\\
+B & 0 & 0 & 0\\
0 & 0 & 0 & -E\\
0 & 0 & +E & 0
\end{array}
\right)  \label{FAC}%
\end{equation}
allowing exact computations due to the decoupling of the left and right
sectors of fermions. Using the antisymmetric tensor $\varepsilon_{\mu \nu
\rho \sigma}$, we have%
\begin{equation}
F_{\mu \nu}=B\varepsilon_{\mu \nu34}+E\varepsilon_{12\mu \nu}.
\end{equation}
\textbf{(ii)} To deal with the gauge field, we can either use the symmetric
choice%
\begin{equation}%
\begin{tabular}
[c]{ll}%
$A_{1}=\frac{B}{2}y,$ & $A_{2}=-\frac{B}{2}x,$\\
$A_{3}=\frac{E}{2}\tau,$ & $A_{4}=-\frac{E}{2}z,$%
\end{tabular}
\  \  \label{C1}%
\end{equation}
or the Smit-Vink method
\begin{equation}%
\begin{tabular}
[c]{ll}%
$A_{1}=0,$ & $A_{2}=Bx$\\
$A_{3}=0,$ & $A_{4}=Ez$%
\end{tabular}
\  \  \label{MA}%
\end{equation}
In the next sub-subsection, we use the first choice as it allows to take
advantage of the symmetric role played by the components fields to solve the
Dirac equation in the continuum. Later on, we use the second choice to study
the Dirac operator of fermion on supercell; the gauge (\ref{MA}) is convenient
for the study the degeneracy of the zero modes of the Dirac operator and its
topological index.

\subsection{Spectrum in the gauge (\ref{C1})}

To get the spectrum of the Dirac operator (\ref{DIR}), notice that the
\emph{4} gauge covariant derivatives $D_{1},$ $D_{2},$ $D_{3},$ $D_{4}$
satisfy the generic commutation relations
\begin{equation}
\left[  D_{\mu},D_{\nu}\right]  =-iF_{\mu \nu},
\end{equation}
but because of the choice (\ref{FAC}) of the background fields, they reduce
to,%
\begin{equation}%
\begin{tabular}
[c]{lll}%
$\left[  D_{1},D_{2}\right]  $ & $=iB$ & $,$\\
$\left[  D_{3},D_{4}\right]  $ & $=iE$ & $,$%
\end{tabular}
\  \label{CR}%
\end{equation}
and all others vanishing.

\subsubsection{Deriving the wave functions}

The fermionic wave functions solving the Dirac equation are given by
representations of the algebra (\ref{CR}). By using (\ref{C1}), the four
covariant derivatives organize in $2+2$ capturing a complex structure as
follows,%
\begin{align}
&
\begin{tabular}
[c]{lll}%
$D_{1}-iD_{2}=\frac{2\partial}{\partial u}+\frac{QB}{2c}\bar{u}$ & $,$ &
$D_{1}+iD_{2}=\frac{2\partial}{\partial \bar{u}}-\frac{QB}{2c}u$\\
$D_{3}-iD_{4}=\frac{2\partial}{\partial v}+\frac{QE}{2c}\bar{v}$ & $,$ &
$D_{3}+iD_{4}=\frac{2\partial}{\partial \bar{v}}-\frac{QE}{2c}v$%
\end{tabular}
\label{RC}\\
& \nonumber
\end{align}
In getting these relations, we have used the explicit expressions
$D_{1}=\partial_{1}-i\frac{QB}{2c}y$, $D_{2}=\partial_{2}+i\frac{QB}{2c}x$
together with similar relations for $D_{3}$, $D_{4}$; and we have set%
\begin{equation}%
\begin{tabular}
[c]{llll}%
$u=x+iy$ & $,$ & $\frac{\partial}{\partial u}=\frac{1}{2}\left(
\frac{\partial}{\partial x}-i\frac{\partial}{\partial y}\right)  $ & ,\\
$v=z+i\tau$ & , & $\frac{\partial}{\partial v}=\frac{1}{2}\left(
\frac{\partial}{\partial z}-i\frac{\partial}{\partial \tau}\right)  $ & .
\end{tabular}
\  \  \label{CV}%
\end{equation}
The representations of eqs(\ref{CR}) depends on the sign of $B$ and $E$.
Setting
\begin{equation}%
\begin{tabular}
[c]{llll}%
$i\left(  D_{1}-iD_{2}\right)  =A^{-}\sqrt{2\left \vert B\right \vert }$ & , &
$i\left(  D_{1}+iD_{2}\right)  =A^{+}\sqrt{2\left \vert B\right \vert }$ & ,\\
$i\left(  D_{3}-iD_{4}\right)  =C^{-}\sqrt{2\left \vert E\right \vert }$ & , &
$i\left(  D_{3}+iD_{4}\right)  =C^{+}\sqrt{2\left \vert E\right \vert }$ & ,
\end{tabular}
\  \  \label{AAC}%
\end{equation}
the commutation relations (\ref{CR}) read also as
\begin{equation}%
\begin{tabular}
[c]{llll}%
$\left[  A^{-},A^{+}\right]  =\frac{B}{\left \vert B\right \vert }I$ & , &
$\left[  C^{-},C^{+}\right]  =\frac{E}{\left \vert E\right \vert }I$ & ,\\
$\left[  A^{-},C^{\pm}\right]  =0$ & , & $\left[  A^{+},C^{\pm}\right]  =0$ &
;
\end{tabular}
\  \  \label{COM}%
\end{equation}
These relations (\ref{COM}) show that the Dirac fermion in the background
field $F_{\mu \nu}$ (\ref{FAC}) describe a priori \emph{2} quantum harmonic
oscillators with oscillation frequencies
\begin{equation}
\mathrm{\nu}=\sqrt{2\left \vert B\right \vert }\qquad,\qquad \mathrm{\nu}%
^{\prime}=\sqrt{2\left \vert B\right \vert }. \label{OME}%
\end{equation}
The operators $A^{+}A^{-}$ and $C^{+}C^{-}$ , which a priori give the number
of energy excitations in $\mathrm{\nu}$ and $\mathrm{\nu}^{\prime}$ units
respectively, read in terms of the gauge covariant derivatives as follows
\begin{equation}%
\begin{tabular}
[c]{llll}%
$2\left \vert B\right \vert A^{+}A^{-}$ & = & $\left(  D_{1}\right)
^{2}+\left(  D_{2}\right)  ^{2}+i\left[  D_{1},D_{2}\right]  $ & \\
& = & $\left(  D_{1}\right)  ^{2}+\left(  D_{2}\right)  ^{2}-B$ & ,
\end{tabular}
\end{equation}
and similarly%
\begin{equation}%
\begin{tabular}
[c]{llll}%
$2\left \vert E\right \vert C^{+}C^{-}$ & = & $\left(  D_{3}\right)
^{2}+\left(  D_{4}\right)  ^{2}+i\left[  D_{3},D_{4}\right]  $ & \\
& = & $\left(  D_{3}\right)  ^{2}+\left(  D_{4}\right)  ^{2}-E$ & .
\end{tabular}
\end{equation}
Using the expression\textrm{\ }of the matrices $\gamma^{\mu}$, we can write
this matrix operator $H_{\alpha \beta}=i\gamma_{\alpha \beta}^{\mu}D_{\mu}$ in
terms of the "creation" operators $A^{+},$ $C^{+}$ and the "annihilation" ones
$A^{-},$ $C^{-}$ as follows:%
\begin{align}
H_{\alpha \beta}  &  =\frac{1}{i}\left(
\begin{array}
[c]{cccc}%
0 & 0 & -\mathrm{\nu}^{\prime}C^{+} & -\mathrm{\nu}A^{-}\\
0 & 0 & -\mathrm{\nu}A^{+} & \mathrm{\nu}^{\prime}C^{-}\\
\mathrm{\nu}^{\prime}C^{-} & \mathrm{\nu}A^{-} & 0 & 0\\
\mathrm{\nu}A^{+} & -\mathrm{\nu}^{\prime}C^{+} & 0 & 0
\end{array}
\right)  ,\label{DC}\\
& \nonumber
\end{align}
with $H^{\dagger}=H$. Moreover, using the commutation properties $C^{+}A^{\pm
}=A^{\pm}C^{+}$ and $C^{-}A^{\pm}=A^{\pm}C^{-}$, the squared hamiltonian
$H^{2}$ reads as follows,%
\begin{align*}
&  \left(
\begin{array}
[c]{cccc}%
{\small \nu}^{{\small 2}}{\small A}^{-}{\small A}^{+}{\small +\nu}%
^{\prime{\small 2}}{\small C}^{+}{\small C}^{-} & 0 & 0 & 0\\
0 & {\small \nu}^{{\small 2}}{\small A}^{+}{\small A}^{-}{\small +\nu
^{\prime2}C}^{-}{\small C}^{+} & 0 & 0\\
0 & 0 & {\small \nu}^{{\small 2}}{\small A}^{-}{\small A}^{+}{\small +\nu
^{\prime2}C}^{-}{\small C}^{+} & 0\\
0 & 0 & 0 & {\small \nu}^{{\small 2}}{\small A}^{+}{\small A}^{-}%
{\small +\nu^{\prime2}C}^{+}{\small C}^{-}%
\end{array}
\right) \\
&
\end{align*}
and, by using $\Psi=\left(  \mathrm{\psi}_{a},\mathrm{\bar{\xi}}_{\dot{a}%
}\right)  $, leads to
\begin{equation}%
\begin{tabular}
[c]{lll}%
$\mathcal{O\mathcal{O}}^{\dagger}\left(
\begin{array}
[c]{c}%
\psi_{1}\\
\psi_{2}%
\end{array}
\right)  $ & $=\epsilon^{2}\left(
\begin{array}
[c]{c}%
\psi_{1}\\
\psi_{2}%
\end{array}
\right)  $ & ,\\
&  & \\
$\mathcal{O}^{\dagger}\mathcal{O}\left(
\begin{array}
[c]{c}%
\bar{\xi}_{1}\\
\bar{\xi}_{2}%
\end{array}
\right)  $ & $=\epsilon^{2}\left(
\begin{array}
[c]{c}%
\bar{\xi}_{1}\\
\bar{\xi}_{2}%
\end{array}
\right)  $ & ,
\end{tabular}
\  \  \label{PS}%
\end{equation}
with%
\begin{align}
&
\begin{tabular}
[c]{lll}%
$\mathcal{OO}^{\dagger}$ & $=$ & $\left(
\begin{array}
[c]{cc}%
\mathrm{\nu}^{2}A^{-}A^{+}+\mathrm{\nu}^{\prime2}C^{+}C^{-} & 0\\
0 & \mathrm{\nu}^{2}A^{+}A^{-}+\mathrm{\nu}^{\prime2}C^{-}C^{+}%
\end{array}
\right)  $\\
&  & \\
$\mathcal{O}^{\dagger}\mathcal{O}$ & $=$ & $\left(
\begin{array}
[c]{cc}%
\mathrm{\nu}^{2}A^{-}A^{+}+\mathrm{\nu}^{\prime2}C^{-}C^{+} & 0\\
0 & \mathrm{\nu}^{2}A^{+}A^{-}+\mathrm{\nu}^{\prime2}C^{+}C^{-}%
\end{array}
\right)  $%
\end{tabular}
\label{21}\\
& \nonumber
\end{align}
showing that the two Weyl spinors $\psi_{a}$ and $\bar{\xi}_{\dot{a}}$ can be
treated separately. Putting the expression of $\mathcal{OO}^{\dagger}$ and
$\mathcal{O}^{\dagger}\mathcal{O}$ back into eq(\ref{PS}), we obtain
\begin{equation}%
\begin{tabular}
[c]{lll}%
$\left(  \mathrm{\nu}^{\prime2}C^{+}C^{-}+\mathrm{\nu}^{2}A^{-}A^{+}\right)
\psi_{1}$ & $=\epsilon^{2}\psi_{1}$ & ,\\
$\left(  \mathrm{\nu}^{\prime2}C^{-}C^{+}+\mathrm{\nu}^{2}A^{+}A^{-}\right)
\psi_{2}$ & $=\epsilon^{2}\psi_{2}$ & ,\\
$\left(  \mathrm{\nu}^{\prime2}C^{-}C^{+}+\mathrm{\nu}^{2}A^{-}A^{+}\right)
\bar{\xi}_{1}$ & $=\epsilon^{2}\bar{\xi}_{1}$ & ,\\
$\left(  \mathrm{\nu}^{\prime2}C^{+}C^{-}+\mathrm{\nu}^{2}A^{+}A^{-}\right)
\bar{\xi}_{2}$ & $=\epsilon^{2}\bar{\xi}_{2}$ & .
\end{tabular}
\end{equation}
The solutions of these equations depend on the sign of the background fields
$B$ and $E$. In the case $\frac{B}{\left \vert B\right \vert }>0$, $\frac
{E}{\left \vert E\right \vert }>0$, we have:%
\begin{equation}
\left(
\begin{array}
[c]{c}%
\psi_{1}\\
\psi_{2}\\
\bar{\xi}_{1}\\
\bar{\xi}_{2}%
\end{array}
\right)  =\left(
\begin{array}
[c]{c}%
\mathbf{\Theta}_{n-1,\text{ }m}\text{ \  \ }\\
\Theta_{n,\text{ }m-1}\text{ \ }\\
\Theta_{n-1,\text{ }m-1}\\
\Theta_{n,\text{ }m}\text{ \  \  \  \ }%
\end{array}
\right)
\end{equation}
with
\begin{equation}
\Theta_{n,m}\left(  u,\bar{u},v,\bar{v}\right)  =\theta_{n}\left(  u,\bar
{u}\right)  \times \theta_{m}^{\prime}\left(  v,\bar{v}\right)  ,\qquad
n,m\geq0,
\end{equation}
and
\begin{equation}%
\begin{tabular}
[c]{ll}%
$A^{-}\times \theta_{0}\left(  u,\bar{u}\right)  =\left(  \frac{2\partial
}{\partial u}+\frac{QB}{2c}\bar{u}\right)  \theta_{0}\left(  u,\bar{u}\right)
=0$ & \\
$C^{-}\times \theta_{0}^{\prime}\left(  v,\bar{v}\right)  =\left(
\frac{2\partial}{\partial v}+\frac{QE}{2c}\bar{v}\right)  \theta_{0}\left(
v,\bar{v}\right)  =0$ &
\end{tabular}
\end{equation}
solved as follows%
\begin{equation}%
\begin{tabular}
[c]{llll}%
$\theta_{0}\left(  u,\bar{u}\right)  =\mathcal{N}_{0}\left(  \bar{u}\right)
e^{-\frac{QB}{4}u\bar{u}}$ & , & $\theta_{0}^{\prime}\left(  v,\bar{v}\right)
=\mathcal{N}_{0}^{\prime}\left(  \bar{v}\right)  e^{-\frac{QE}{4}v\bar{v}}$ &
,
\end{tabular}
\  \  \label{NN}%
\end{equation}
where the complex functions $\mathcal{N}_{0}\left(  \bar{u}\right)  $ and
$\mathcal{N}_{0}^{\prime}\left(  \bar{v}\right)  $ are anti-holomorphic
functions. The excited waves functions $\theta_{n}\left(  u,\bar{u}\right)  $
and $\theta_{m}^{\prime}\left(  v,\bar{v}\right)  $ are obtained by applying
the creation operators.
\begin{equation}%
\begin{tabular}
[c]{llll}%
$\theta_{n}\left(  u,\bar{u}\right)  $ & $=$ & $\frac{1}{n!}\left(
\frac{2\partial}{\partial \bar{u}}-\frac{QB}{2c}u\right)  ^{n}\theta_{0}\left(
u,\bar{u}\right)  $ & ,\\
$\theta_{n}^{\prime}\left(  v,\bar{v}\right)  $ & $=$ & $\frac{1}{n!}\left(
\frac{2\partial}{\partial \bar{v}}-\frac{QE}{2c}v\right)  ^{n}\theta
_{0}^{\prime}\left(  v,\bar{v}\right)  $ & .
\end{tabular}
\  \  \label{W}%
\end{equation}

\subsubsection{Zero modes and topological index}

The zero modes of the Dirac operator depend on the sign of the background
fields $B$ and $E$ since the algebra of the commutation relations (\ref{COM})
depend on $\frac{B}{\left \vert B\right \vert }$ and $\frac{E}{\left \vert
E\right \vert }$ as given below:%
\begin{equation}%
\begin{tabular}
[c]{llll}%
$\left[  A^{-},A^{+}\right]  =\frac{B}{\left \vert B\right \vert }I$ & , &
$\left[  C^{-},C^{+}\right]  =\frac{E}{\left \vert E\right \vert }I$ & .
\end{tabular}
\end{equation}
\ We have the \emph{4} following possibilities:

\textbf{(a) }Case $\frac{B}{\left \vert B\right \vert }>0$, $\frac{E}{\left \vert
E\right \vert }>0$ \newline In this situation, the $A^{+},$ $C^{+}$ are
creation operators and $A^{-},$ $C^{-}$ annihilation ones. So the zero mode is
given by%
\begin{equation}
\Psi_{\epsilon=0}=\left(
\begin{array}
[c]{c}%
0\text{\ }\\
0\\
0\\
\mathbf{\Theta}_{0,0}%
\end{array}
\right)  \label{ZE}%
\end{equation}
with the chirality property
\begin{equation}%
\begin{tabular}
[c]{ll}%
$\left(  \sigma^{3}\otimes \tau^{3}\right)  \Psi_{\epsilon=0}$ & $=+\Psi
_{\varepsilon=0}$\\
$\  \  \  \  \  \  \  \  \  \gamma^{5}\Psi_{\epsilon=0}$ & $=-\Psi_{\varepsilon=0}$%
\end{tabular}
\end{equation}
and by setting $Tr\left(  \bar{\Psi}_{\epsilon=0}\Psi_{\varepsilon=0}\right)
=1$, we have
\begin{equation}%
\begin{tabular}
[c]{ll}%
$Tr\left[  \bar{\Psi}_{\epsilon=0}\left(  \sigma^{3}\otimes \tau^{3}\right)
\Psi_{\varepsilon=0}\right]  $ & $=+1$\\
$\  \  \  \  \  \  \  \  \  \  \ Tr\left[  \bar{\Psi}_{\epsilon=0}\gamma^{5}%
\Psi_{\varepsilon=0}\right]  $ & $=-1$%
\end{tabular}
\end{equation}

\textbf{(b) }Case $\frac{B}{\left \vert B\right \vert }<0$, $\frac{E}{\left \vert
E\right \vert }<0$ \newline In this case $A^{+},$ $C^{+}$ are annihilation
operators and $A^{-},$ $C^{-}$ creation ones so that the zero mode is%
\begin{equation}%
\begin{tabular}
[c]{lll}%
$\Psi_{\epsilon=0}$ & $=\left(
\begin{array}
[c]{c}%
\mathbf{\Theta}_{0,0}\text{\ }\\
0\\
0\\
0
\end{array}
\right)  $ & ,
\end{tabular}
\end{equation}
and%
\begin{equation}%
\begin{tabular}
[c]{lll}%
$\left(  \sigma^{3}\otimes \tau^{3}\right)  \Psi_{\epsilon=0}$ & $=+\Psi
_{\varepsilon=0}$ & \\
$\  \  \  \  \  \  \  \  \  \  \gamma^{5}\Psi_{\epsilon=0}$ & $=+\Psi_{\varepsilon=0}$
&
\end{tabular}
\end{equation}
as well as%
\begin{equation}%
\begin{tabular}
[c]{ll}%
$Tr\left[  \bar{\Psi}_{\epsilon=0}\left(  \sigma^{3}\otimes \tau^{3}\right)
\Psi_{\varepsilon=0}\right]  $ & $=+1$\\
$\  \  \  \  \  \  \  \  \  \  \ Tr\left[  \bar{\Psi}_{\epsilon=0}\gamma^{5}%
\Psi_{\varepsilon=0}\right]  $ & $=+1$%
\end{tabular}
\end{equation}

\textbf{(c) }Case $\frac{B}{\left \vert B\right \vert }>0$, $\frac{E}{\left \vert
E\right \vert }<0$\newline Here $A^{+},$ $C^{-}$ are creation operators and
$A^{-},$ $C^{-}$ annihilations. The zero mode is given by%
\begin{equation}
\Psi_{\epsilon=0}=\left(
\begin{array}
[c]{c}%
0\text{\ }\\
0\\
\mathbf{\Theta}_{0,0}\\
0
\end{array}
\right)
\end{equation}
with%
\begin{equation}%
\begin{tabular}
[c]{ll}%
$\left(  \sigma^{3}\otimes \tau^{3}\right)  \Psi_{\epsilon=0}$ & $=-\Psi
_{\varepsilon=0}$\\
$\  \  \  \  \  \  \  \  \  \gamma^{5}\Psi_{\epsilon=0}$ & $=-\Psi_{\varepsilon=0}$%
\end{tabular}
\end{equation}
and%
\begin{equation}%
\begin{tabular}
[c]{ll}%
$Tr\left[  \bar{\Psi}_{\epsilon=0}\left(  \sigma^{3}\otimes \tau^{3}\right)
\Psi_{\varepsilon=0}\right]  $ & $=1$\\
$\  \  \  \  \  \  \  \  \  \  \ Tr\left[  \bar{\Psi}_{\epsilon=0}\gamma^{5}%
\Psi_{\varepsilon=0}\right]  $ & $=1$%
\end{tabular}
\end{equation}

\textbf{(d) }Case $\frac{B}{\left \vert B\right \vert }<0$, $\frac{E}{\left \vert
E\right \vert }>0$\newline In this case, the zero mode reads as%
\begin{equation}
\Psi_{\epsilon=0}=\left(
\begin{array}
[c]{c}%
0\text{\ }\\
\mathbf{\Theta}_{0,0}\\
0\\
0
\end{array}
\right)
\end{equation}
with%
\begin{equation}%
\begin{tabular}
[c]{ll}%
$\left(  \sigma^{3}\otimes \tau^{3}\right)  \Psi_{\epsilon=0}$ & $=-\Psi
_{\varepsilon=0}$\\
$\  \  \  \  \  \  \  \  \  \  \  \gamma^{5}\Psi_{\epsilon=0}$ & $=+\Psi_{\varepsilon=0}
$%
\end{tabular}
\end{equation}
and%
\begin{equation}%
\begin{tabular}
[c]{ll}%
$Tr\left[  \bar{\Psi}_{\epsilon=0}\left(  \sigma^{3}\otimes \tau^{3}\right)
\Psi_{\varepsilon=0}\right]  $ & $=+1$\\
$\  \  \  \  \  \  \  \  \  \  \ Tr\left[  \bar{\Psi}_{\epsilon=0}\gamma^{5}%
\Psi_{\varepsilon=0}\right]  $ & $=-1$%
\end{tabular}
\end{equation}

\  \  \  \  \newline From the above analysis on the chirality of zero modes, it
follows that the chirality operator that satisfies the Atiyah-Singer theorem
is $\sigma^{3}\otimes \tau^{3}$ since%
\begin{equation}%
\begin{tabular}
[c]{ll}%
$Tr\left[  \bar{\Psi}_{\epsilon=0}\left(  \sigma^{3}\otimes \tau^{3}\right)
\Psi_{\varepsilon=0}\right]  $ & $=1$%
\end{tabular}
\end{equation}
This feature can be explained as due to the factorization of the $SO\left(
4\right)  $ symmetry of $\mathbb{R}^{4}$ in terms of the product of
$SU_{L}\left(  2\right)  \times SU_{R}\left(  2\right)  $. The relation that
involves $\gamma^{5}$ reads as $Tr\left[  \bar{\Psi}_{\epsilon=0}\gamma
^{5}\Psi_{\varepsilon=0}\right]  =\frac{E}{\left \vert E\right \vert }$.

\section{Solving Dirac equation on supercell}

To study the spectrum of the Dirac equation in the supercell compactification
of the hyperdiamond, we extend the\textrm{\ }result of sub-section 5.2
concerning the continuum to the case of the 4D lattice. There, we have used
the canonical frame of $\mathbb{R}^{4}$ with the local coordinates $X^{\mu
}=\left(  X,Y,Z,T\right)  $; which will be used also in the case of
lattice.\newline In the C-frame, the \emph{4} gauge covariant derivatives
$D_{\mu}=$ $\left(  D_{1},D_{2},D_{3},D_{4}\right)  $ satisfy the usual
commutation relations giving the components of the U$\left(  1\right)  $ gauge
curvature%
\begin{equation}
\left[  D_{\mu},D_{\nu}\right]  =-iF_{\mu \nu}. \label{FD}%
\end{equation}
By choosing the background fields as in (\ref{FAC}), these commutation
relations factorize into two decoupled Heisenberg algebras as follows,%
\begin{equation}%
\begin{tabular}
[c]{lll}%
$\left[  D_{1},D_{2}\right]  =iB$ & , & $\left[  D_{1},D_{3}\right]  =\left[
D_{1},D_{4}\right]  =0$\\
$\left[  D_{3},D_{4}\right]  =iE$ & , & $\left[  D_{2},D_{3}\right]  =\left[
D_{2},D_{4}\right]  =0$%
\end{tabular}
\end{equation}
To work out explicit solutions of these equations, one may used either the
potential vector (\ref{C1}) or (\ref{MA}). In this section, we use
eq(\ref{MA}) leading to the following gauge covariant derivatives $D_{\mu
}=\frac{\partial}{\partial X^{\mu}}-iA_{\mu}$
\begin{equation}%
\begin{tabular}
[c]{lll}%
$D_{1}=\frac{\partial}{\partial X}$ & $,$ & $D_{2}=\frac{\partial}{\partial
Y}-iBX$\\
$D_{3}=\frac{\partial}{\partial Z}$ & $,$ & $D_{4}=\frac{\partial}{\partial
T}-iEZ$%
\end{tabular}
\  \  \  \label{GD}%
\end{equation}
This choice breaks the symmetric role played by the variables $\left(
X,Y\right)  $ and $\left(  Z,T\right)  $; but is suitable to deal with the
boundary conditions of the fields on supercell.\newline Notice that
eqs(\ref{FD}) and (\ref{GD}) can be also expressed in the P- frame with
positions as $x^{i}=\left(  x,y,z,\tau \right)  $. The passage between C- and
P- frames is given by the transformations%
\begin{equation}%
\begin{tabular}
[c]{lllll}%
$X^{\mu}=\alpha_{i}^{\mu}x^{i}$ & , & $x^{i}=\omega_{\mu}^{i}X^{\mu}$ & , &
$\alpha_{i}^{\mu}\omega_{\mu}^{j}=\delta_{i}^{i}.$%
\end{tabular}
\  \  \  \label{CC}%
\end{equation}
Similar relations can be written down for the other the objects; for instance
the potential vector $A_{\mu}$ and the gauge covariant derivatives $D_{\mu}$
in C-frame are related to their homologue $\mathcal{A}_{i}$ and $\mathcal{D}%
_{i}$ in the P-frame as%
\begin{equation}%
\begin{tabular}
[c]{lll}%
$A_{\mu}=\omega_{\mu}^{i}\mathcal{A}_{i}$ & $,$ & $\  \ D_{\mu}=\omega_{\mu
}^{i}\mathcal{D}_{i}$\\
$\mathcal{A}_{i}=\alpha_{i}^{\mu}A_{\mu}$ & $,$ & $\  \  \mathcal{D}_{i}%
=\alpha_{i}^{\mu}D_{\mu}$%
\end{tabular}
\end{equation}

\subsection{Computing the fluxes of background fields}

The flux of the background fields through hyperdiamond supercell is a scalar
quantity and is frame independent. This flux give the total topological charge
inside the supercell $SC_{4}$; it controls the chirality of the ground state
and allows to determine the topological index of the Dirac operator in the
background fields $B$ and $E$. \newline To compute the flux, one can either
use the C-frame or the P-frame; in fact it is frame independent. To see this
property recall that in the C-frame the gauge curvature is given by $F_{\mu
\nu}$ and in the P-frame is $\mathcal{F}_{ij}$:%
\begin{equation}%
\begin{tabular}
[c]{llll}%
$\left[  D_{\mu},D_{\nu}\right]  =-iF_{\mu \nu}$ & , & $\left[  \mathcal{D}%
_{i},\mathcal{D}_{j}\right]  =-i\mathcal{F}_{ij}$ & .
\end{tabular}
\end{equation}
These two tensors are related as
\begin{equation}%
\begin{tabular}
[c]{ll}%
$\mathcal{F}_{ij}$ & $=\alpha_{i}^{\mu}\alpha_{j}^{\nu}F_{\mu \nu}=\frac{1}%
{2}\left(  \alpha_{i}^{\mu}\alpha_{j}^{\nu}-\alpha_{j}^{\mu}\alpha_{i}^{\nu
}\right)  F_{\mu \nu}$\\
$F_{\mu \nu}$ & $=\omega_{\mu}^{i}\omega_{\nu}^{j}\mathcal{F}_{ij}=\frac{1}%
{2}\left(  \omega_{\mu}^{i}\omega_{\nu}^{j}-\omega_{\nu}^{i}\omega_{\mu}%
^{j}\right)  \mathcal{F}_{ij}$%
\end{tabular}
\  \  \label{2F}%
\end{equation}
with
\begin{equation}%
\begin{tabular}
[c]{lll}%
$\mathbf{\alpha}_{i}\wedge \mathbf{\alpha}_{j}$ & $=$ & $\left(  \alpha
_{i}^{\mu}\alpha_{j}^{\nu}-\alpha_{j}^{\mu}\alpha_{i}^{\nu}\right)  $\\
$\mathbf{\omega}^{i}\wedge \mathbf{\omega}^{j}$ & $=$ & $\left(  \omega_{\mu
}^{i}\omega_{\nu}^{j}-\omega_{\nu}^{i}\omega_{\mu}^{j}\right)  $%
\end{tabular}
\end{equation}
The corresponding gauge invariant 2-form field strengths are then given by%
\begin{equation}%
\begin{tabular}
[c]{lll}%
$\mathcal{F}$ & $=\frac{1}{2}dx^{i}\wedge dx^{j}\mathcal{F}_{ij}$ & \\
$F$ & $=\frac{1}{2}dX^{\mu}\wedge dX^{\nu}F_{\mu \nu}$ &
\end{tabular}
\end{equation}
and are equal $\mathcal{F}=F$ since they are frame independent; thanks to
$\alpha_{i}^{\mu}\omega_{\mu}^{j}=\delta_{i}^{j}$. Moreover, because of the
choice
\begin{equation}
F_{\mu \nu}=B\varepsilon_{\mu \nu34}+E\varepsilon_{12\mu \nu}%
\end{equation}
that we have used in this paper to work out explicit solutions of the Dirac
equation, the 2-form $F$ reduces to the simple expression%
\begin{equation}%
\begin{tabular}
[c]{lll}%
$F$ & $=BdX\wedge dY$ $+$ $EdZ\wedge dT$ &
\end{tabular}
\  \  \label{BE}%
\end{equation}
By using the coordinate change (\ref{CC}) to the P-frame, it can be also
written as%
\begin{equation}%
\begin{tabular}
[c]{ll}%
$F$ & $=\left(  B\alpha_{i}^{1}\alpha_{j}^{2}+E\alpha_{i}^{3}\alpha_{j}%
^{4}\right)  dx^{i}\wedge dx^{j}$%
\end{tabular}
\end{equation}
where $F$ takes a general expression in the basis $dx^{i}\wedge dx^{j}$. With
these relations at hand, we can compute the flux of the background fields
through the various p-cycles of the supercell. We will do the calculations the
C-frame. \newline The total topological charge $Q_{tot}$ of the background
field within the supercell is given by the integration of the 4-form $F\wedge
F$ over the supercell,%
\begin{equation}%
\begin{tabular}
[c]{llll}%
$\frac{1}{\left(  2\pi \right)  ^{2}}%
{\displaystyle \int \nolimits_{SC_{4}}}
\frac{1}{2}F\wedge F=Q_{tot}$ & , & $\  \  \ Q_{tot}\in \mathbb{Z}$ & ,
\end{tabular}
\label{FF}%
\end{equation}
Substituting (\ref{BE}) back into (\ref{FF}), we obtain a quantization
condition on the background fields given by
\begin{equation}
B\times E=\frac{\left(  2\pi \right)  ^{2}}{L_{1}L_{2}L_{3}L_{4}}Q_{tot}
\label{QT}%
\end{equation}
One can also compute the fluxes
\begin{equation}
\frac{1}{2\pi}\int_{C_{2}}F
\end{equation}
of the field strength $F$ through the 2-cycles $C_{ij}\equiv C_{2}$ of the
supercell; this gives extra quantization conditions. Because of the choice
(\ref{FAC}), non trivial fluxes are indeed given by the 2-cycles $C_{12}$ and
$C_{34}$. Using the relation $F=dA$ to map the integration over the 2-cycles
$C_{ij}$ to circulation around its boundaries $\partial C_{ij}$, we end with%
\begin{equation}%
\begin{tabular}
[c]{lll}%
$\frac{1}{2\pi}%
{\displaystyle \int \nolimits_{C_{12}}}
F$ & $=\frac{1}{2\pi}%
{\displaystyle \oint \nolimits_{\partial C_{12}}}
A$ & $=Q_{B}$%
\end{tabular}
\end{equation}
and%
\begin{equation}%
\begin{tabular}
[c]{lll}%
$\frac{1}{2\pi}%
{\displaystyle \int \nolimits_{C_{34}}}
F$ & $=\frac{1}{2\pi}%
{\displaystyle \oint \nolimits_{\partial C_{34}}}
A$ & $=Q_{E}$%
\end{tabular}
\end{equation}
giving two extra quantization conditions; one on the background field $B$ and
the other on $E$. More precisely, we have:
\begin{equation}%
\begin{tabular}
[c]{lll}%
$B=\frac{2\pi}{L_{1}L_{2}}Q_{B}$ & , & $Q_{B}\in \mathbb{Z}$\\
&  & \\
$E=\frac{2\pi}{L_{3}L_{4}}Q_{E}$ & , & $Q_{E}\in \mathbb{Z}$%
\end{tabular}
\  \  \label{EB}%
\end{equation}
Comparing (\ref{EB}) with (\ref{QT}) we get the following relation between the
topological charges%
\begin{equation}
Q_{tot}=Q_{B}Q_{E}. \label{B}%
\end{equation}

\subsection{Dirac equation on 4D- supercell}

The euclidian Dirac equation on supercell is given by the 4-dimensional
extension of the Dirac equation on honeycomb lattice. In addition to periodic
background potentials, this equation involves four component fermionic waves
with boundary conditions described by the $SO\left(  4\right)  $ spinor
$\Psi=\left(  \psi_{a},\bar{\xi}_{\dot{a}}\right)  $ that we want to determine below.

\subsubsection{The hamiltonian}

The euclidian Dirac equation is%
\begin{equation}
H\left(
\begin{array}
[c]{c}%
\psi_{a}\\
\bar{\xi}_{\dot{a}}%
\end{array}
\right)  =\epsilon \left(
\begin{array}
[c]{c}%
\psi_{a}\\
\bar{\xi}_{\dot{a}}%
\end{array}
\right)  \label{HEQ}%
\end{equation}
where the hamiltonian $H$ reads in the canonical frame as $H=\gamma^{\mu
}D_{\mu}$ or equivalently in the P-frame%
\begin{equation}
H=\gamma^{\mu}D_{\mu}=\gamma^{\mu}\omega_{\mu}^{i}\mathcal{D}_{i}=\Gamma
^{i}\mathcal{D}_{i}%
\end{equation}
with $\Gamma^{i}=\gamma^{\mu}\omega_{\mu}^{i}$. The solution of (\ref{HEQ}) in
continuum has been worked out in the previous section; below we want to extend
these results to the lattice case where boundary conditions put strong
constraints on the solutions. To that purpose, let us start by collecting
useful tools. First the hamiltonian $H$ has the form%
\begin{equation}%
\begin{tabular}
[c]{ll}%
$H=\left(
\begin{array}
[c]{cc}%
0 & D\\
\bar{D} & 0
\end{array}
\right)  $ & ,
\end{tabular}
\end{equation}
with%
\begin{equation}%
\begin{tabular}
[c]{ll}%
$D=\left(
\begin{array}
[c]{cc}%
\sqrt{\left \vert E\right \vert }C^{-} & -i\sqrt{\left \vert B\right \vert }%
A^{-}\\
-i\sqrt{\left \vert B\right \vert }A^{+} & \sqrt{\left \vert E\right \vert }C^{+}%
\end{array}
\right)  $ & ,
\end{tabular}
\end{equation}
and
\begin{equation}
\bar{D}=\left(
\begin{array}
[c]{cc}%
\sqrt{\left \vert E\right \vert }C^{+} & i\sqrt{\left \vert B\right \vert }A^{-}\\
i\sqrt{\left \vert B\right \vert }A^{+} & \sqrt{\left \vert E\right \vert }C^{-}%
\end{array}
\right)
\end{equation}
and where we have set%
\begin{equation}%
\begin{tabular}
[c]{llll}%
$A^{-}=\frac{1}{\sqrt{\left \vert B\right \vert }}\left(  D_{1}-iD_{2}\right)  $
& , & $C^{-}=\frac{1}{\sqrt{\left \vert E\right \vert }}\left(  D_{4}%
-iD_{3}\right)  $ & \\
$A^{+}=\frac{1}{\sqrt{\left \vert B\right \vert }}\left(  D_{1}+iD_{2}\right)  $
& , & $C^{+}=\frac{1}{\sqrt{\left \vert E\right \vert }}\left(  D_{4}%
+iD_{3}\right)  $ &
\end{tabular}
\end{equation}
Second the operators $A^{\pm}$ and $C^{\pm}$\ obey the commutation relations%
\begin{equation}%
\begin{tabular}
[c]{lll}%
$\left[  A^{-},A^{+}\right]  $ & $=-\frac{B}{\left \vert B\right \vert }I$ & ,\\
$\left[  C^{-},C^{+}\right]  $ & $=-\frac{E}{\left \vert E\right \vert }I$ & ,
\end{tabular}
\  \  \label{AC}%
\end{equation}
and all other vanishing. Notice that these commutation relations have a
remarkable dependence on the sign of the background fields B and E. Third, the
squared hamiltonian $H^{2}$ has the diagonal form
\[
H^{2}=\left(
\begin{array}
[c]{cc}%
D\bar{D} & 0\\
0 & \bar{D}D
\end{array}
\right)
\]
with%
\begin{equation}
D\bar{D}=\left(
\begin{array}
[c]{cc}%
\left \vert B\right \vert A^{-}A^{+}+\left \vert E\right \vert C^{-}C^{+} & 0\\
0 & \left \vert B\right \vert A^{+}A^{-}+\left \vert E\right \vert C^{+}C^{-}%
\end{array}
\right)
\end{equation}
and%
\begin{equation}
D\bar{D}=\left(
\begin{array}
[c]{cc}%
\left \vert B\right \vert A^{-}A^{+}+\left \vert E\right \vert C^{+}C^{-} & 0\\
0 & \left \vert B\right \vert A^{+}A^{-}+\left \vert E\right \vert C^{-}C^{+}%
\end{array}
\right)
\end{equation}
involving the four possible quadratic combinations of $A^{\pm}$ and $C^{\pm}$
namely $A^{-}A^{+}$, $A^{+}A^{-}$, $C^{-}C^{+}$ and $C^{+}C^{-}$.

\subsubsection{The solutions and Index$\left(  D\right)  $}

The solutions of (\ref{HEQ}) on supercell are representations of the algebra
(\ref{AC}) that have to satisfy moreover the boundary conditions on the
fields. These conditions are quite similar to those studied in the case of 2D
honeycomb; so we omit here the lengthy technical details and just give the
results. \newline There are 4 classes of solutions of (\ref{HEQ}) depending on
the signs of $\frac{B}{\left \vert B\right \vert }$ and $\frac{E}{\left \vert
E\right \vert }$. They are obtained as follows: First, expand the wave
$\Psi \left(  \mathbf{x}\right)  $ on the periodic supercell in Fourier series
as
\begin{equation}%
\begin{tabular}
[c]{lll}%
$\Psi \left(  \mathbf{x}\right)  $ & $=%
{\displaystyle \sum \limits_{l,q}}
e^{i\left(  \frac{2l\pi}{L_{2}}y+\frac{2q\pi}{L_{4}}\tau \right)  }\Psi
_{l,q}\left(  x,z\right)  $ &
\end{tabular}
\end{equation}
This expansion follows from the periodicity of $\Psi \left(  \mathbf{x}\right)
$ along the y- and $\tau$- axis. Second, solve the non trivial boundary
conditions along the x- and z- axes by following the method used in the case
of 2D honeycomb which lead to eqs(\ref{SH}-\ref{HS}). As in the present case
we have to deal with the 2 variables x and z, we write the $\Psi_{l,q}\left(
x,z\right)  $ modes like
\begin{equation}
\Psi_{l,q}\left(  x,z\right)  =\Phi \left(  \xi_{l},\zeta_{q}\right)
\end{equation}
with%
\begin{equation}%
\begin{tabular}
[c]{llll}%
$\xi_{l}=x+\frac{l}{Q_{B}}L_{1}$ & , & $\zeta_{q}=z+\frac{q}{Q_{E}}L_{3}$ &
\end{tabular}
\end{equation}
The next step is to determine the function $\Phi \left(  \xi,\zeta \right)  $;
this is a Dirac spinor which we set as%
\begin{equation}
\Phi \left(  \xi,\zeta \right)  =\left(
\begin{array}
[c]{c}%
\phi \left(  \xi,\zeta \right) \\
\varrho \left(  \xi,\zeta \right) \\
\varphi \left(  \xi,\zeta \right) \\
\chi \left(  \xi,\zeta \right)
\end{array}
\right)
\end{equation}
with components obeying the following coupled equations
\begin{equation}%
\begin{tabular}
[c]{lll}%
$\left(  \left \vert B\right \vert A^{-}A^{+}+\left \vert E\right \vert C^{-}%
C^{+}\right)  \phi$ & $=$ & $\epsilon^{2}\phi$\\
$\left(  \left \vert B\right \vert A^{+}A^{-}+\left \vert E\right \vert C^{+}%
C^{-}\right)  \varrho$ & $=$ & $\epsilon^{2}\varrho$\\
$\left(  \left \vert B\right \vert A^{-}A^{+}+\left \vert E\right \vert C^{+}%
C^{-}\right)  \varphi$ & $=$ & $\epsilon^{2}\varphi$\\
$\left(  \left \vert B\right \vert A^{+}A^{-}+\left \vert E\right \vert C^{-}%
C^{+}\right)  \chi$ & $=$ & $\epsilon^{2}\chi$%
\end{tabular}
\end{equation}
The operators $A^{\pm}$ and $C^{\pm}$ satisfy the algebra (\ref{AC}) and show
that solutions for $\Phi \left(  \xi,\zeta \right)  $ depend on the sign of the
background fields. These solutions are as follows:

\begin{itemize}
\item case $\frac{B}{\left \vert B\right \vert }>0$, $\frac{E}{\left \vert
E\right \vert }>0$
\end{itemize}

In this case, which corresponds also to a positive topological charge
$Q_{tot}$, the algebra (\ref{AC}) reads as follows%
\begin{equation}%
\begin{tabular}
[c]{lll}%
$\left[  A^{+},A^{-}\right]  $ & $=1$ & ,\\
$\left[  C^{+},C^{-}\right]  $ & $=1$ & ,
\end{tabular}
\end{equation}
showing $A^{-}$, $C^{-}$ are creation operators and $A^{+}$, $C^{+}$ are
annihilation ones. Using general results on quantum harmonic oscillators and
the relations%
\[%
\begin{tabular}
[c]{llll}%
$\mathcal{N}_{A}=A^{-}A^{+}$ & , & $A^{+}A^{-}=A^{-}A^{+}+1$ & \\
$\mathcal{N}_{C}=C^{-}C^{+}$ & , & $C^{+}C^{-}=C^{-}C^{+}+1$ &
\end{tabular}
\  \
\]
where $\mathcal{N}$ stands for the number operator, it is not difficult to see
that the $\epsilon$ energies are discrete as%
\begin{equation}
\epsilon_{_{n,m}}^{2}=n\left \vert B\right \vert +m\left \vert E\right \vert
,\qquad n,m\in \mathbb{N}%
\end{equation}
and the corresponding wave functions $\Phi_{_{n,m}}$ are given by
\begin{equation}
\Phi_{_{_{n,m}}}\left(  \xi,\zeta \right)  =\left(
\begin{array}
[c]{c}%
\Theta_{_{n,m}}\text{ \  \  \  \ }\\
\Theta_{_{n-1,m-1}}\\
\Theta_{_{n,m-1}}\text{ \  \ }\\
\Theta_{_{n-1,m}}\text{ \  \ }%
\end{array}
\right)
\end{equation}
where%
\[
\Theta_{n,m}\left(  \xi,\zeta \right)  =\theta_{n}\left(  \xi \right)
\times \vartheta_{m}\left(  \zeta \right)
\]
and%
\begin{equation}%
\begin{tabular}
[c]{llll}%
$\theta_{n}\left(  \xi \right)  $ & $=\frac{1}{n!}\left(  A^{-}\right)
^{n}\theta_{0}\left(  \xi \right)  $ & , & $A^{+}\theta_{0}\left(  \xi \right)
=0$\\
&  &  & \\
$\vartheta_{m}\left(  \zeta \right)  $ & $=\frac{1}{m!}\left(  C^{-}\right)
^{m}\vartheta_{0}\left(  \zeta \right)  $ & , & $C^{-}\vartheta_{0}\left(
\zeta \right)  =0$%
\end{tabular}
\end{equation}
with%
\begin{equation}%
\begin{tabular}
[c]{llll}%
$\theta_{0}\left(  \xi \right)  =\mathcal{N}_{0}e^{-\frac{\left \vert
B\right \vert }{2}\xi^{2}}$ & , & $\vartheta_{0}\left(  \zeta \right)
=\mathcal{N}_{0}e^{-\frac{\left \vert E\right \vert }{2}\zeta^{2}}$ & .
\end{tabular}
\end{equation}
Notice that because $\theta_{-1}\left(  \xi \right)  =\vartheta_{-1}\left(
\zeta \right)  =0,$ the ground state has only one component as%
\begin{equation}
\Phi_{_{0,0}}=\left(
\begin{array}
[c]{c}%
\Theta_{_{0,0}}\\
0\\
0\\
0
\end{array}
\right)  .
\end{equation}
Moreover since the degeneracy of $\theta_{0}$ and $\vartheta_{0}$ are
respectively $\left \vert Q_{B}\right \vert $ and $\left \vert Q_{E}\right \vert
$; it follows that the degree of degeneracy of $\Phi_{0,0} $ is precisely the
total topological charge
\begin{equation}
\left \vert Q_{B}Q_{E}\right \vert =\left \vert Q_{tot}\right \vert
\end{equation}
in agreement with Atiyah-Singer theorem in 4-dimensional Dirac theory.

\begin{itemize}
\item case $\frac{B}{\left \vert B\right \vert }>0$, $\frac{E}{\left \vert
E\right \vert }<0$
\end{itemize}

This case corresponds to $Q_{tot}<0$; the algebra (\ref{AC}) reduces to%
\begin{equation}%
\begin{tabular}
[c]{lll}%
$\left[  A^{+},A^{-}\right]  $ & $=1$ & ,\\
$\left[  C^{-},C^{+}\right]  $ & $=1$ & .
\end{tabular}
\end{equation}
It shows that $A^{-}$, $C^{+}$ are the creation operators and $A^{+}$, $C^{-}
$ are the annihilation ones. The energies $\epsilon_{n,m}^{2}$ are same as
above but the fermionic wave are like%
\begin{equation}
\Phi_{_{n,m}}\left(  \xi,\zeta \right)  =\left(
\begin{array}
[c]{c}%
\Theta_{_{n,m-1}}\text{\  \  \ }\\
\Theta_{_{n-1,m}}\text{ \  \ }\\
\Theta_{_{n,m}}\text{ \  \  \  \  \ }\\
\Theta_{_{n-1,m-1}}%
\end{array}
\right)
\end{equation}
Here also the ground state has one component given by%
\begin{equation}
\Phi_{_{0,0}}=\left(
\begin{array}
[c]{c}%
0\\
0\\
\Theta_{_{0,0}}\\
0
\end{array}
\right)
\end{equation}
it has the same degree of degeneracy as\ in the previous case.

\begin{itemize}
\item case $\frac{B}{\left \vert B\right \vert }<0$, $\frac{E}{\left \vert
E\right \vert }>0$, with $Q_{tot}<0$
\end{itemize}

The algebra (\ref{AC}) becomes%
\begin{equation}%
\begin{tabular}
[c]{lll}%
$\left[  A^{-},A^{+}\right]  $ & $=1$ & ,\\
$\left[  C^{+},C^{-}\right]  $ & $=1$ & .
\end{tabular}
\end{equation}
Here $A^{+}$, $C^{-}$ are creation operators and $A^{-}$, $C^{+}$ are
annihilations. The fermionic waves are as follows:%
\begin{equation}
\Phi_{_{n,m}}\left(  \xi,\zeta \right)  =\left(
\begin{array}
[c]{c}%
\Theta_{_{n-1,m}}\text{ \  \ }\\
\Theta_{_{n,m-1}}\text{ \  \ }\\
\Theta_{_{n-1,m-1}}\\
\Theta_{_{n,m}}\text{ \  \  \  \ }%
\end{array}
\right)
\end{equation}
with ground state as:%
\begin{equation}
\Phi_{_{0,0}}=\left(
\begin{array}
[c]{c}%
0\\
0\\
0\\
\Theta_{_{0,0}}%
\end{array}
\right)
\end{equation}

\begin{itemize}
\item case $\frac{B}{\left \vert B\right \vert }<0$, $\frac{E}{\left \vert
E\right \vert }<0$, $Q_{tot}>0$
\end{itemize}

The commutation relations (\ref{AC}) are as
\begin{equation}%
\begin{tabular}
[c]{lll}%
$\left[  A^{-},A^{+}\right]  $ & $=+1$ & ,\\
$\left[  C^{-},C^{+}\right]  $ & $=+1$ & ,
\end{tabular}
\end{equation}
with $A^{+}$, $C^{+}$ the creations operators and $A^{-}$, $C^{-}$ the
annihilators. The fermionic waves are also different from the previous ones;
they are as%
\begin{equation}
\Phi_{_{n,m}}\left(  \xi,\zeta \right)  =\left(
\begin{array}
[c]{c}%
\Theta_{_{n-1,m-1}}\\
\Theta_{_{n,m}}\text{ \  \  \  \ }\\
\Theta_{_{n-1,m}}\text{ \  \ }\\
\Theta_{_{n,m-1}}\text{ \  \ }%
\end{array}
\right)
\end{equation}
with ground state like%
\begin{equation}
\Phi_{_{0,0}}=\left(
\begin{array}
[c]{c}%
0\\
\Theta_{_{0,0}}\\
0\\
0
\end{array}
\right)
\end{equation}
The index of the Dirac operator is given by\textrm{\ }$Tr\left[  \bar{\Psi
}_{\epsilon=0}\left(  \sigma^{3}\otimes \tau^{3}\right)  \Psi_{\varepsilon
=0}\right]  =Q_{tot}$\textrm{.}

\section{Conclusion and comments}

In this paper, we have studied topological aspects of fermions on a family of
2N-dimensional lattices in presence of background fields with special focus on
the 2 leading crystals namely the graphene and the 4D hyperdiamond of
\emph{QCD}$_{4}$. With the results obtained by our explicit study, we have now
an exact answer on the population of the ground state of fermions on lattices
in presence of uniform background fields. For example, in the case of graphene
in a strong magnetic field, we find that the chiral anomaly is behind the
observed anomalous in the filling factor $\nu_{gra}=4\left(  n+\frac{1}%
{2}\right)  $ of the integer quantum Hall effect. This means that the ground
state of graphene with $\nu_{gra}^{\epsilon=0}=4\times \frac{1}{2}=2$ is
occupied\textrm{ either by positive chiral states or negative ones} depending
on the sign of the magnetic field $B$. The same statement can be made for
light quarks of \emph{QCD}$_{4}$ on hyperdiamond and more generally fermions
on higher even- dimensional honeycombs. In \emph{QCD}$_{4}$ on lattice with a
Dirac fermion (say the quark u) in the background fields $B$ and $E$; the
filling factor reads as $\nu_{{\small QCD}}=\# \left(  n+\frac{1}{2}\right)
\left(  m+\frac{1}{2}\right)  $ there are 4 possible configurations for the
population of the ground state depending on the signs of $B$ and $E$; in the
case of u quark these are $u_{\uparrow L},$ $u_{\downarrow L},$ $u_{\uparrow
R},$ $u_{\downarrow R}$. \newline To exhibit this behavior, recall that in
honeycomb the magnetic field $B$ appears in the \emph{2} gauge covariant
derivatives $D_{1}$ and $D_{2}$ whose curvature can, up on using the scaling
(\ref{EH}-\ref{HE}), be put into the remarkable form
\[
\left[  A^{-},A^{+}\right]  =\frac{B}{\left \vert B\right \vert }I
\]
This is a typical Heisenberg algebra; but with \emph{two sectors} depending on
the sign of the magnetic field $B$. For positive $B$, the operators $A^{+}$
and $A^{-}$ are respectively the creation operator and the annihilation one;
but for negative $B$, this property gets reversed since now $A^{-}$ plays the
role of the creation operator and $A^{+}$ the role of the annihilation one.
Almost the same thing happens for quarks in \emph{QCD}$_{4}$ and fermions on
higher dimensional lattices. For example, in the case of fermions on 4D
hyperdiamond, we have \emph{4} gauge covariant derivatives $D_{\mu}=\left(
D_{1},D_{2},D_{3},D_{4}\right)  $ obeying the general commutation relations
\[%
\begin{tabular}
[c]{ll}%
$\left[  D_{\mu},D_{\nu}\right]  $ & $=-iF_{\mu \nu}$\\
$\left[  D_{\mu},F_{\nu \rho}\right]  $ & $=0$%
\end{tabular}
\  \  \  \  \  \  \  \
\]
where $F_{\mu \nu}$ is a C-number capturing in general \emph{6} degrees of
freedom (provided $<F_{\mu \nu}^{{\small QCD}}>$ $=0$). A careful analysis of
this algebra shows that it describes two interacting quantum harmonic
oscillators. However by choosing the gauge field strength as $F_{\mu \nu}%
=B_{1}\varepsilon_{\mu \nu34}+B_{2}\varepsilon_{12\mu \nu}$, the above algebra
reduces to 2 uncoupled Heisenberg algebras%
\[
\left[  A_{i}^{-},A_{j}^{+}\right]  =\frac{B_{i}}{\left \vert B_{i}\right \vert
}\delta_{ij},
\]
but with $2^{2}$ sectors according to the signs of $B_{1}$ and $B_{2}$. This
result extends straightforwardly to the case of fermions on 2N-dimensional
honeycombs with background fields. There, the field strength $F_{\mu \nu}$ has
generally $N\left(  2N-1\right)  $ moduli and so the corresponding algebra of
the gauge covariant derivatives describes $N$ interacting quantum harmonic
oscillators. By choosing the field strength as
\[
F_{\mu \nu}=B_{1}\varepsilon_{\mu \nu34....2N}+B_{2}\varepsilon_{12\mu
\nu56....2N}+...+B_{N}\varepsilon_{12....\mu \nu},
\]
the algebra of the covariant derivatives gets reduced to N uncoupled
Heisenberg ones as above and has $2^{N}$ sectors depending on the signs of the
$B_{i}$'s. In this generic case, the total topological charge $Q_{tot}$ of the
background fields within the compact supercell $SC_{2N}$ of the 2N-dimensional
honeycomb as well as the partial fluxes through the 2-cycle basis $C_{i}$ of
the supercell are given by%
\[
Q_{tot}=\frac{1}{N!\left(  2\pi \right)  ^{N}}%
{\displaystyle \int \nolimits_{SC_{2N}}}
F\wedge...\wedge F,\qquad Q_{i}=\frac{1}{2\pi}%
{\displaystyle \int \nolimits_{Ci}}
F
\]
leading to the relation $Q_{tot}=%
{\displaystyle \prod \nolimits_{i=1}^{N}}
Q_{i}$ which can be proved by thinking about the supercell $SC_{2N}$ as given
by the product of those 2-cycles $C_{i}$ of $SC_{2N}$ with no intersection;
i.e: $C_{i}\cap C_{j}=\emptyset$. The computation of these charges for the
case $N=2$ was done in section 6; see eqs(\ref{EB}-\ref{B}); they can be
easily extended to higher dimensions; in particular for the case fermions on
the 6-dimensional honeycomb. By taking $F_{\mu \nu}=$ $B_{1}\varepsilon_{\mu
\nu3456}$ $+$ $B_{2}\varepsilon_{12\mu \nu56}+$ $B_{3}\varepsilon_{1234\mu \nu}%
$, one ends with $Q_{tot}=Q_{1}Q_{2}Q_{3}$. \newline In the end, we would like
to add that our explicit approach gives as well a unified group theoretical
description of fermions in both graphene and \emph{QCD}$_{4}$. The
construction of \textrm{\cite{4C}} turns out to be intimately related with the
weight lattice of $SU\left(  3\right)  $ we have given in section 2 and 3; the
primitive compactification used in \textrm{\cite{4C}} has also to do with the
simple roots basis of the SU$\left(  3\right)  $. The latter is a hidden
symmetry of the 2D honeycomb; it allows many explicit calculations in graphene
and moreover draws the path to follow to get the extension of results on
graphene to fermions on 2N-dimensional honeycombs where the job is done by the
hidden $SU\left(  2N+1\right)  $ symmetry.

\section{Appendix: Lattice calculations}

In this appendix, we give some details on the lattice calculations used in
this study. These computations, which have been understood in the paper; are
based on the method of refs \textrm{\cite{2B,B2}}; and constitute an extension
of results, obtained in ref \textrm{\cite{4C}} concerning graphene, to the
case of QCD on 4D hyperdiamond.\ For completeness, we describe below the 2
following useful things:\newline$\left(  1\right)  $ review briefly the main
lines of tight binding model for graphene; a quite similar analysis is valid
for QCD$_{4}$ on hyperdiamond (see section 4). We also take this opportunity
to develop further the link between the electronic properties of graphene and
$SU\left(  3\right)  $ representations. This group theoretical approach
extends directly to $4D$ lattice QCD on hyperdiamond. There, the role of
$SU\left(  3\right)  $ is played by $SU\left(  5\right)  $.\newline$\left(
2\right)  $ give a comment on the reason behind solving Dirac equation with
periodicity conditions on the wave functions. This is important for two
things: $\left(  i\right)  $ working out the wave functions with explicit
dependence on lattice parameters, the quantized topological charge \emph{Q};
and too particularly the exact determination of the degree $g_{m}$ of
degeneracies of the states of energy $E_{m}$ as required by the computation of
eq(\ref{01}). $\left(  ii\right)  $ performing numerical calculations as done
in \textrm{\cite{4C}} for $30\times30$ lattice to check analytic predictions
on the Atiyah-Singer theorem, the chiral anomaly and related issues.

\subsection{Case of graphene: a brief review and link with $SU\left(
\emph{3}\right)  $}

On the 2D honeycomb lattice, made by the superposition of two sublattices
$\mathbb{A}$ and $\mathbb{B}$ as depicted by figs \ref{PC}-\ref{ARA}, the
tight binding hamiltonian describing the hopping of electrons of graphene to
first nearest neighbors, in presence of a magnetic background field
$\mathcal{B}$, reads as%
\begin{equation}
H=-t%
{\displaystyle \sum \limits_{\mathbf{r}_{m}\in \mathbb{A}}}
\left(
{\displaystyle \sum \limits_{l=0}^{2}}
a_{\mathbf{r}_{m}}^{-}\text{ }\mathcal{U}_{\mathbf{r}_{m},\mathbf{v}_{l}%
}\text{ }b_{\mathbf{r}_{m}+\mathbf{v}_{l}}^{+}+%
{\displaystyle \sum \limits_{l=0}^{2}}
b_{\mathbf{r}_{m}}^{-}\text{ }\mathcal{U}_{\mathbf{r}_{m},\mathbf{v}_{l}%
}^{\ast}\text{ }a_{\mathbf{r}_{m}-\mathbf{v}_{l}}^{+}\right)
\end{equation}
Here $t$ is the hop energy; $a_{\mathbf{r}_{m}}^{\pm}$ and $b_{\mathbf{r}%
_{m}+\mathbf{v}_{l}}^{\pm}$ are the fermionic waves respectively associated
with $\mathbb{A}$ and $\mathbb{B}$ sublattices and satisfying the usual
anticommutation relations. $\mathcal{U}_{\mathbf{r}_{m},\mathbf{v}_{l}}$ is
the link field given by%
\begin{equation}
\mathcal{U}_{\mathbf{r}_{m},\mathbf{v}_{l}}=e^{i\mathbf{v}_{l}.\mathcal{A}%
\left(  \mathbf{r}_{m}\right)  }=e^{iv_{l}^{\mu}\mathcal{A}_{\mu}\left(
\mathbf{r}_{m}\right)  }%
\end{equation}
with $\mathcal{A}_{\mu}\left(  x,y\right)  $ the potential potential of the
external magnetic field. This is a $u\left(  1\right)  $-valued gauge
connection emerging from the site $\mathbf{r_{n}}$ and lying along the
$\mathbf{v}_{l}$ direction. Notice that if switching of the external
$\mathcal{B}$, the field $\mathcal{U}_{\mathbf{r}_{m},\mathbf{v}_{l}}$ reduces
to identity; and then $H$ describes a tight binding model of hopping free
electrons. We also have the following relations%
\begin{equation}%
\begin{tabular}
[c]{lll}%
$\mathbf{v}_{1}.\mathcal{A}$ & $=$ & $+\frac{2}{3}\mathcal{A}_{1}+\frac{1}%
{3}\mathcal{A}_{2}$\\
$\mathbf{v}_{2}.\mathcal{A}$ & $=$ & $-\frac{1}{3}\mathcal{A}_{1}+\frac{1}%
{3}\mathcal{A}_{2}$\\
$\mathbf{v}_{0}.\mathcal{A}$ & $=$ & $-\frac{1}{3}\mathcal{A}_{1}-\frac{2}%
{3}\mathcal{A}_{2}$%
\end{tabular}
\end{equation}
where $\mathcal{A}_{i}$ stands for $\alpha_{i}^{\mu}\mathcal{A}_{\mu}\left(
\mathbf{r}\right)  $ with $\mathbf{\alpha}_{1},$ $\mathbf{\alpha}_{2}$ the
vectors generating the sublattices $\mathbb{A}$ and $\mathbb{B}$ of the
honeycomb. In our present study, we have used the method of
\textrm{\cite{2B,B2} }to deal with the gauge potential\textrm{\ }%
$\mathcal{A}_{\mu}\left(  \mathbf{r}\right)  $\textrm{\ }by working
with\textrm{\ }the helpful gauge choice
\begin{equation}%
\begin{tabular}
[c]{lll}%
$\mathcal{A}_{1}=\alpha_{1}^{\mu}\mathcal{A}_{\mu}\left(  \mathbf{r}\right)
=\mathbf{0}$ & , & $\  \  \  \  \  \  \mathcal{A}_{2}=\alpha_{2}^{\mu}%
\mathcal{A}_{\mu}\left(  \mathbf{r}\right)  =-By$%
\end{tabular}
\end{equation}
Recall that on honeycomb each pi-electron of a carbon atom, say a fermion
$a_{\mathbf{r}_{n}}$ of the sublattice $\mathbb{A}$, has \emph{3} first
nearest atom neighbors of $\mathbb{B}$-type with fermions $b_{\mathbf{r}%
_{n}+\mathbf{v}_{0}},$ $b_{\mathbf{r}_{n}+\mathbf{v}_{1}},b_{\mathbf{r}%
_{n}+\mathbf{v}_{2}}$. The vectors $\mathbf{v}_{0},$\textbf{\ }$\mathbf{v}%
_{1},$\textbf{\ }$\mathbf{v}_{2}$ with entries $\mathbf{v}_{l}=\left(
v_{l}^{\mu}\right)  $ parameterize the relative positions of $b_{\mathbf{r}%
_{n}+\mathbf{v}_{l}}$ with respect to $a_{\mathbf{r}_{n}}$ and satisfy some
remarkable features; in particular the following:\newline$\left(  a\right)  $
they obey the vector constraint relation
\begin{equation}
\mathbf{v}_{0}+\mathbf{v}_{1}+\mathbf{v}_{2}=0 \label{BT3}%
\end{equation}
that captures physical information on the phases of wave functions.\newline%
$\left(  b\right)  $ they play a central role in dealing with lattice
calculations as they encode the hopping to nearest neighbors. In particular,
they allow to build the analogue of the usual derivative term of the Dirac
hamiltonian in continuum; the lattice derivative turns out to be captured by
some combination of the phases $e^{i\mathbf{k.v}_{l}}$. \newline$\left(
c\right)  $ in our approach, the $\mathbf{v}_{l}$'s are interpreted as the 3
weight vectors of the fundamental representation of the $SU\left(  3\right)  $
group. This is a remarkable observation that allows to simplify drastically
the lattice calculations.\ To get the point, set%
\begin{equation}%
\begin{tabular}
[c]{lllll}%
$\mathbf{v}_{0}=d$ $\mathbf{\lambda}_{0}$ & $,$ & $\mathbf{v}_{1}=d$
$\mathbf{\lambda}_{1}$ & , & $\mathbf{v}_{2}=d$ $\mathbf{\lambda}_{2}$%
\end{tabular}
\end{equation}
with $d\simeq1.42\mathring{A}$ the length of the carbon-carbon bond in
graphene; then put back into (\ref{BT3}), we end with a well known group
theoretical identity namely
\begin{equation}
\mathbf{\lambda}_{0}+\mathbf{\lambda}_{1}+\mathbf{\lambda}_{2}=0 \label{man}%
\end{equation}
describing precisely the constraint equation on the weight vectors of the
$SU\left(  3\right)  $ fundamental representation. \newline Recall by the way
that for $SU\left(  3\right)  $, the weight vectors of the complex 3-dimension
representation can be written in two basic manners: either in terms of the
fundamental weights $\mathbf{\omega}_{1}$ and $\mathbf{\omega}_{2}$ of
$SU\left(  3\right)  $ as follows
\begin{equation}%
\begin{tabular}
[c]{lllll}%
$\mathbf{\lambda}_{1}=+\mathbf{\omega}_{1}$ & , & $\mathbf{\lambda}%
_{2}=-\mathbf{\omega}_{1}+\mathbf{\omega}_{2}$ & , & $\mathbf{\lambda}%
_{0}=-\mathbf{\omega}_{2}$%
\end{tabular}
\label{la}%
\end{equation}
or equivalently\ in terms of the two simple roots $\mathbf{\alpha}_{1}$ and
$\mathbf{\alpha}_{2}$ of $SU\left(  3\right)  $ like:%
\begin{equation}%
\begin{tabular}
[c]{ll}%
$\mathbf{\lambda}_{1}$ & $=+\frac{2}{3}\mathbf{\alpha}_{1}+\frac{1}%
{3}\mathbf{\alpha}_{2}$\\
$\mathbf{\lambda}_{2}$ & $=-\frac{1}{3}\mathbf{\alpha}_{1}+\frac{1}%
{3}\mathbf{\alpha}_{2}$\\
$\mathbf{\lambda}_{0}$ & $=-\frac{1}{3}\mathbf{\alpha}_{1}-\frac{2}%
{3}\mathbf{\alpha}_{2}$%
\end{tabular}
\end{equation}
satisfying manifestly (\ref{man}). Obviously the fundamental weights
$\mathbf{\omega}_{1}$, $\mathbf{\omega}_{2}$ and the simple roots
$\mathbf{\alpha}_{1}$, $\mathbf{\alpha}_{2}$ are linked by the duality
relations $\mathbf{\alpha}_{i}.\mathbf{\omega}_{j}=\delta_{ij}$ which lead to
$\mathbf{\alpha}_{1}=2\mathbf{\omega}_{1}-\mathbf{\omega}_{2},$
$\mathbf{\alpha}_{2}=2\mathbf{\omega}_{2}-\mathbf{\omega}_{1}$. We also have
the helpful relations%
\begin{equation}%
\begin{tabular}
[c]{lll}%
$\  \  \  \  \  \mathbf{\alpha}_{1}$ & $=$ & $\mathbf{\lambda}_{1}-\mathbf{\lambda
}_{2}$\\
$\  \  \  \  \  \mathbf{\alpha}_{2}$ & $=$ & $\mathbf{\lambda}_{2}-\mathbf{\lambda
}_{0}$\\
$\mathbf{\alpha}_{1}+\mathbf{\alpha}_{2}$ & $=$ & $\mathbf{\lambda}%
_{1}-\mathbf{\lambda}_{0}$%
\end{tabular}
\  \  \  \  \  \label{al}%
\end{equation}
in dealing with lattice calculations. This $SU\left(  3\right)  $ group
theoretical analysis has been shown to extend straightforwardly to the
4-dimensional hyperdiamond of section 4 where the role of $SU\left(  3\right)
$ is played by $SU\left(  5\right)  $; see eqs(\ref{41})-(\ref{120}). \newline
If switching off the interaction between the electrons of graphene with the
external magnetic field $\mathcal{B}$; and then performing the Fourier
transform of the local fields, we can put the tight binding hamiltonian
$H|_{\mathcal{B}=0}$ into the form $\sum_{\mathbf{k}}H_{\mathbf{k}}$ with wave
vectors $\mathbf{k}=\left(  k_{x},k_{y}\right)  $ and%
\begin{equation}%
\begin{tabular}
[c]{ll}%
$H_{\mathbf{k}}=$ & $\left(  a_{\mathbf{k}}^{-},b_{\mathbf{k}}^{-}\right)
\left(
\begin{array}
[c]{cc}%
0 & \varepsilon_{\mathbf{k}}\\
\varepsilon_{\mathbf{k}}^{\ast} & 0
\end{array}
\right)  \left(
\begin{array}
[c]{c}%
a_{\mathbf{k}}^{+}\\
b_{\mathbf{k}}^{+}%
\end{array}
\right)  $%
\end{tabular}
\  \  \  \  \  \  \  \  \label{hk}%
\end{equation}
with $\varepsilon_{\mathbf{k}}=\sum_{l}e^{id\mathbf{k.\lambda}_{l}}$ or more
explicitly%
\begin{equation}
\varepsilon_{\mathbf{k}}=e^{id\mathbf{k.\lambda}_{0}}+e^{id\mathbf{k.\lambda
}_{1}}+e^{id\mathbf{k.\lambda}_{2}}.
\end{equation}
Moreover, using eqs(\ref{la}-\ref{al}), we can also put the above relation
into the remarkable form%
\begin{equation}
\varepsilon_{\mathbf{k}}=e^{-\frac{i}{3}d\mathbf{k.}\left(  \mathbf{\alpha
}_{1}-\mathbf{\alpha}_{2}\right)  }\left[  1+e^{id\mathbf{k.\alpha}_{1}%
}+e^{-id\mathbf{k.\alpha}_{2}}\right]  \label{ep}%
\end{equation}
which is manifestly invariant under translations in the reciprocal lattice;
that is under the shifts%
\begin{equation}%
\begin{tabular}
[c]{lllll}%
$\mathbf{k}$ & $\mathbf{\rightarrow}$ & $\mathbf{k+}\frac{2\pi}{d}\left(
n_{1}\mathbf{\omega}_{1}+n_{2}\mathbf{\omega}_{2}\right)  $ & $,$ &
$n_{1},n_{2}\in \mathbb{Z}$%
\end{tabular}
\  \  \  \  \  \  \label{EP}%
\end{equation}
The diagonalization of the hamiltonian mode $H_{\mathbf{k}}$ (\ref{hk}) leads
to the energy dispersion relations%
\begin{equation}
E_{\pm}=\pm2t\sqrt{\frac{3}{4}+\frac{1}{2}\cos d\mathbf{k.\alpha}_{1}+\frac
{1}{2}\cos d\mathbf{k.\alpha}_{2}+\frac{1}{2}\cos d\mathbf{k.\alpha}_{3}}%
\end{equation}
where we have set $\mathbf{\alpha}_{3}=\mathbf{\alpha}_{1}+\mathbf{\alpha}%
_{2}$. \newline Notice that zero modes of the hamiltonian $H_{\mathbf{k}}$,
obtained by solving the vanishing condition $\varepsilon_{\mathbf{k}}=0$, are
immediately learnt from (\ref{ep}) and are given, modulo translations in the
reciprocal lattice, by the two following Dirac points%
\begin{equation}
\mathbf{k}_{\pm}=\mathbf{\pm}\frac{2\pi}{3d}\left(  \mathbf{\omega}%
_{1}+\mathbf{\omega}_{2}\right)  =\pm \mathbf{k}_{F} \label{kf}%
\end{equation}
Notice moreover that near these zeros, say for $\mathbf{k}_{+}=\mathbf{q}%
+\mathbf{k}_{F}$ with small $q$, the dispersion energy relation is linear in
$\mathbf{q}$ and the physics of the electron is mainly described by a free
2-dimensional Dirac theory in continuum with the periodicity property
(\ref{EP}) of the reciprocal lattice. \newline By switching on the interaction
between the electrons and $\mathcal{B}$, the previous energy dispersion
relation $\varepsilon_{\mathbf{k}}=\sum_{l}e^{id\mathbf{\lambda}_{l}^{\mu
}k_{\mu}}$ gets modified; it is given by a complicated expression which is
obtained by substituting the wave vector $k_{\mu}$ by the gauge covariant
quantity $\mathcal{K}_{\mu}=k_{\mu}\mathbf{-}\mathcal{A}_{\mu}$. Furthermore,
using the fact that the external magnetic field is constant, we end, after
some straightforward algebra, with the following covariant derivative in
reciprocal space
\begin{equation}
\mathcal{K}_{\mu}=k_{\mu}-i\frac{\mathcal{B}}{2}\varepsilon_{\mu \nu}%
\frac{\partial}{\partial k_{\nu}}=\frac{\mathcal{B}}{2i}\varepsilon_{\mu \nu
}\left(  \frac{\partial}{\partial k_{\nu}}+\frac{2i}{B}\varepsilon^{\nu \rho
}k_{\rho}\right)
\end{equation}
leading in turns to $\varepsilon \left(  \mathcal{K}\right)  =\sum
_{l}e^{id\mathbf{\lambda}_{l}^{\mu}\mathcal{K}_{\mu}}$. Near the zeros
(\ref{kf}), $\varepsilon \left(  \mathcal{K}\right)  $ leads therefore to the
Dirac equation in continuum in presence of a background field $\mathcal{B}$.
This equation reads in the real space as,
\begin{equation}
i\gamma^{\mu}\left(  \frac{\partial}{\partial x^{\mu}}-i\mathcal{A}_{\mu
}\right)  \Psi \left(  x,y\right)  =0,\qquad \Psi=\left(
\begin{array}
[c]{c}%
\psi \\
\chi
\end{array}
\right)
\end{equation}
with the periodicity property%
\begin{equation}%
\begin{tabular}
[c]{lllll}%
$\mathbf{r}$ & $\mathbf{\rightarrow}$ & $\mathbf{r}+d\frac{\sqrt{3}}{2}\left(
n_{1}\mathbf{\alpha}_{1}+n_{2}\mathbf{\alpha}_{2}\right)  $ & $,$ &
$n_{1},n_{2}\in \mathbb{Z}$%
\end{tabular}
\end{equation}
In section 3, we have worked out the solutions $\Psi \left(  x,y\right)  $ near
the Dirac points that obey periodicity properties eq(\ref{PER}) and
(\ref{m}-\ref{SV}). It was shown that the ground state is chiral and depends
on the sign of $\frac{\mathcal{B}}{\left \vert \mathcal{B}\right \vert }$. In
the case $\frac{\mathcal{B}}{\left \vert \mathcal{B}\right \vert }=-1$ for
instance, the wave functions are given by eqs(\ref{A}), (\ref{AA}) and
(\ref{AAA}) where, in addition to the background field $\mathcal{B}$; the
dependence on the lattice parameters, the topological charge and the
degeneracy of the eigenvalues are manifestly exhibited.

\subsection{Numerical study}

\begin{figure}[ptbh]
\begin{center}
\hspace{0cm} \includegraphics[width=12cm]{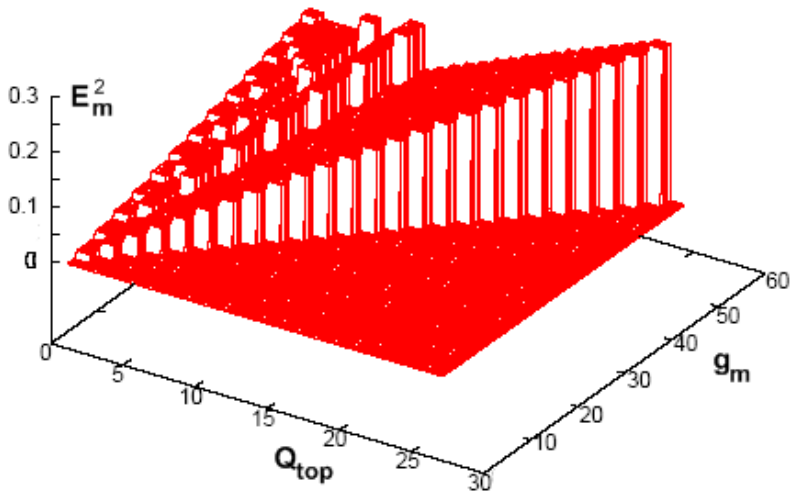}
\end{center}
\par
\vspace{-0.8cm}\caption{{\protect \small plot of E}$_{m}^{2}$
{\protect \small as a function of the topological charge }$\left \vert
{\protect \small Q}_{top}\right \vert $ {\protect \small and the degree of
degeneracy g}$_{m}$ {\protect \small on a 30}$\times${\protect \small 30 lattice
\cite{4C}.}}%
\label{EQ}%
\end{figure}

In order to analyze the spectrum of the Dirac operator $D$ of graphene in
various gauge field backgrounds $\mathcal{B}=\frac{2\pi Q}{L_{1}L_{2}}$, one
diagonalizes the matrix $D^{2}$ by using a subspace iteration technique as
well as Chebyshev polynomial iteration to accelerate the convergence of the
$E_{m}^{2}$ eigenvalues. Following \textrm{\cite{4C}}; see also
\textrm{\cite{16}} for technical details, the plot the \emph{60} smallest
eigenvalues $E_{m}^{2}$ ($1\leq m\leq60$) reveals a very good agreement
between the numerical results and the analytic predictions as shown by fig
\ref{EQ}. These energies are calculated on a $N_{1}\times N_{2}$ lattice
supercell ($N_{1}=N_{2}=30$) with primitive boundary conditions and for values
of topological charge $Q$ varying between $1$ and $26.$

\  \  \  \newline Recall that the analytic prediction for the eigenvalues of the
energy spectrum leads to $E_{m}^{2}=\frac{2\sqrt{3}}{3}m\left \vert
\mathcal{B}\right \vert $ with $m$ a positive integer. Lattice calculations
show also that the degeneracy pattern (\ref{PER}) of these eigenvalues is as
follows%
\begin{equation}
g_{m}=\left \{
\begin{array}
[c]{ccc}%
\left \vert Q\right \vert  & for\text{ } & m=0\\
2\left \vert Q\right \vert  & for & m>0
\end{array}
\right.
\end{equation}
By substituting the background field $\mathcal{B}$ by its expression in terms
of the topological charge $Q$ and the area of the supercell, we can rewrite
the energy spectrum like,
\begin{equation}
E_{m}^{2}=\frac{4\pi \sqrt{3}}{3L_{1}L_{2}}\left \vert Q\right \vert m
\end{equation}
As shown on the plot, the states with $E_{0}^{2}=0$, that form a triangle on
fig \ref{EQ}, corresponds precisely to the zero modes of the Dirac operator
with degree of degeneracy $g_{m}$ growing linearly with $Q$ in complete
agreement with the index theorem.

\begin{acknowledgement}
: \newline The authors thank the Hassan II Academy of Science and Technology
where part of this work has been done. E.H.S thanks the Moroccan Center for
Scientific Research and Technology; Project ref URAC09, for support.
\end{acknowledgement}

\end{document}